\documentclass[onecolumn,11pt,letterpaper]{usetex-onecol}

\usepackage{times}
\usepackage{breakurl}
\usepackage{color}
\usepackage{enumitem}
\usepackage{graphicx}
\usepackage{xspace}
\usepackage{epstopdf}
\usepackage{url}
\usepackage{cite}
\usepackage[colorlinks=true,linkcolor=black,citecolor=red,pdfborder={0 0 0}]{hyperref}
\usepackage{breakurl}
\usepackage{enumitem}
\usepackage{microtype}

\newcommand\HASHKV{\textsf{HashKV}\xspace}
\renewcommand{\paragraph}[1]{\smallskip\noindent {\bf #1}}

\setlist{nolistsep}
\begin{document}

\pagestyle{plain}

\title{Enabling Efficient Updates in KV Storage via Hashing:\\
Design and Performance Evaluation\thanks{An earlier version of this paper
appeared at the 2018 USENIX Annual Technical Conference \cite{chan18}.  In
this extended version, we conduct a comprehensive performance evaluation
study of \HASHKV to validate its design effectiveness in different
aspects.  We also demonstrate that \HASHKV can be integrated with other KV
stores to improve their respective performance.}}

\author{Yongkun Li$^\dagger$, Helen H. W. Chan$^\ddagger$, 
Patrick P. C. Lee$^\ddagger$, and Yinlong Xu$^\dagger$\\
$^\dagger$University of Science and Technology of China\\
$^\ddagger$The Chinese University of Hong Kong}

\maketitle

\begin{abstract}
Persistent key-value (KV) stores mostly build on the Log-Structured Merge
(LSM) tree for high write performance, yet the LSM-tree suffers from the
inherently high I/O amplification.  KV separation mitigates I/O
amplification by storing only keys in the LSM-tree and values in separate
storage.  However, the current KV separation design remains inefficient under
update-intensive workloads due to its high garbage collection (GC) overhead in
value storage.  We propose \HASHKV, which aims for high update performance
atop KV separation under update-intensive workloads.  \HASHKV uses
{\em hash-based data grouping}, which deterministically maps values to storage
space so as to make both updates and GC efficient.  We further relax the
restriction of such deterministic mappings via simple but useful design
extensions.  We extensively evaluate various design aspects of \HASHKV.  We
show that \HASHKV achieves 4.6$\times$ update throughput and 53.4\% less write
traffic compared to the current KV separation design.  In addition, we
demonstrate that we can integrate the design of \HASHKV with state-of-the-art
KV stores and improve their respective performance. 
\end{abstract}

\section{Introduction}
\label{sec:intro}

Persistent key-value (KV) stores are an integral part of modern large-scale
storage infrastructures for storing massive structured data
(e.g., \cite{Chang06,DeCandia07,Beaver10,Lai15}).
While many real-world KV storage workloads are read-intensive (e.g., the
Get-Update request ratio can reach 30$\times$ in Facebook's Memcached
workloads \cite{Atikoglu12}), {\em update-intensive} workloads are also
dominant in many storage scenarios, including online transaction processing
\cite{tpcc} and enterprise servers \cite{Kavalanekar08}.  Field studies show
that the amount of write requests becomes more significant in modern
enterprise workloads.  For example, Yahoo! reports that its low-latency
workloads increasingly move from reads to writes \cite{Sears12}; Baidu reports
that the read-write request ratio of a cloud storage workload is 2.78$\times$
\cite{Lai15}; Microsoft reports that read-write traffic ratio of a 3-month
OneDrive workload is 2.3$\times$ \cite{Chen17}.

Modern KV stores optimize the performance of writes (including inserts and
updates) using the Log-Structured Merge (LSM) tree \cite{Oneil96}. Its
idea is to transform updates into sequential writes through a log-structured
(append-only) design \cite{Rosenblum92}, while supporting efficient queries
including individual key lookups and range scans. In a nutshell, the
LSM-tree buffers written KV pairs and flushes them into a multi-level tree,
in which each node is a fixed-size file containing sorted KV pairs and their
metadata.  It stores the recently written KV pairs at higher tree levels,
and merges them with lower tree levels via {\em compaction}. The LSM-tree
design not only improves write performance by avoiding small random updates
(which are also harmful to the endurance of solid-state drives (SSDs)
\cite{Agrawal08,Min12}), but also improves range scan performance by keeping
sorted KV pairs in each node.
	
However, the LSM-tree incurs high I/O amplification in both writes and reads.
As the LSM-tree receives more writes of KV pairs, it will trigger frequent
compaction operations, leading to tremendous extra I/Os due to rewrites across
levels.  Such write amplification can reach a factor of at least
50$\times$ \cite{Wu15,Lu16}, which is detrimental to both write performance
and the endurance of SSDs \cite{Agrawal08,Min12}.  Also, as the LSM-tree grows
in size, reading the KV pairs at lower levels incurs many disk accesses.
Such read amplification can reach a factor of over 300$\times$ \cite{Lu16},
leading to low read performance.

In order to mitigate the compaction overhead, many research efforts focus on
optimizing LSM-tree indexing (\S\ref{sec:related}).  One approach is
{\em KV separation} from WiscKey \cite{Lu16}, in which keys and metadata
are still stored in the LSM-tree, while values are separately stored in an
append-only circular log.  The main idea of KV separation is to reduce the
LSM-tree size, while preserving the indexing feature of the LSM-tree for
efficient inserts/updates, individual key lookups, and range scans.

In this work, we argue that KV separation itself still cannot fully achieve
high performance under update-intensive workloads.  The root cause is that the
circular log for value storage needs frequent garbage collection (GC) to
reclaim the space from the KV pairs that are deleted or superseded by new
updates.  However, the GC overhead is actually expensive due to two
constraints of the circular log.  First, the circular log maintains a strict
GC order, as it always performs GC at the beginning of the log where the least
recently written KV pairs are located.  This can incur a large amount of
unnecessary data relocation (e.g., when the least recently written KV pairs
remain valid).  Second, the GC operation needs to query the LSM-tree to check
the validity of each KV pair.  These queries have high latencies, especially
when the LSM-tree becomes sizable under large workloads.

We propose \HASHKV, a high-performance KV store tailored for update-intensive
workloads.  \HASHKV builds on KV separation and uses a novel {\em hash-based
data grouping} design for value storage.  Its idea is to divide value storage
into fixed-size partitions and deterministically map the value of each written
KV pair to a partition by hashing its key.  Hash-based data grouping supports
lightweight updates due to deterministic mapping.  More importantly, it
significantly mitigates GC overhead, since each GC operation not only has the
flexibility to select a partition to reclaim space, but also eliminates the
queries to the LSM-tree for checking the validity of KV pairs.

On the other hand, the deterministic nature of hash-based data
grouping restricts where KV pairs are stored.  Thus, we propose three novel
design extensions to relax the restriction of hash-based data grouping:
(i) {\em dynamic reserved space allocation}, which dynamically allocates
reserved space for extra writes if their original hash partitions are full
given the size limit;
(ii) {\em hotness awareness}, which separates the storage of hot and cold KV
pairs to improve GC efficiency as inspired by existing SSD designs
\cite{Min12,Lee13}; and
(iii) {\em selective KV separation}, which keeps small-size KV pairs in
entirety in the LSM-tree to simplify lookups.

We implement our \HASHKV prototype atop LevelDB \cite{LevelDB}, and show via
testbed experiments that \HASHKV achieves 4.6$\times$ throughput and 53.4\%
less write traffic compared to the circular log design in WiscKey under
update-intensive workloads.  Also, \HASHKV generally achieves higher
throughput and significantly less write traffic compared to modern KV stores,
such as LevelDB and RocksDB \cite{RocksDB}, in various cases.

Our work makes a case of augmenting KV separation with a new value management
design.  \HASHKV currently targets commodity flash-based SSDs
under update-intensive workloads.  Its hash-based data
grouping design mitigates GC overhead and incurs less write traffic, thereby
improving not only the update performance but also the endurance of SSDs
\cite{Agrawal08,Min12}.  Note that \HASHKV incurs random writes due to
hashing, yet our implementation can feasibly mitigate the random access
overhead (\S\ref{subsec:impl}) since SSDs have a closer performance gap
between random and sequential writes compared to hard disks
\cite{Papagiannis16,Lu16}.  While \HASHKV builds on LevelDB by default for the
key and metadata management, it can also build on other KV stores that have
more efficient LSM-tree designs (e.g.,
\cite{Sears12,Shetty13,Wu15,Yao17,Yue17,Raju17}).  To demonstrate, we
replace LevelDB with RocksDB \cite{RocksDB}, HyperLevelDB
\cite{Hyperleveldb}, and PebblesDB \cite{Raju17} and show that \HASHKV
improves their respective performance via KV separation and more efficient
value management.

The source code of \HASHKV is available at:
\href{http://adslab.cse.cuhk.edu.hk/software/hashkv}%
{\bf http://adslab.cse.cuhk.edu.hk/software/hashkv}.

\section{Motivation}
\label{sec:motivation}

We use LevelDB \cite{LevelDB} as a representative example to explain the write
and read amplification problems of LSM-tree-based KV stores.  We show how KV
separation \cite{Lu16} mitigates both write and read amplifications, yet it
still cannot fully achieve efficient updates.

\subsection{LevelDB}
\label{subsec:leveldb}

\begin{figure}[!t]
\centering
\includegraphics[width=3.6in]{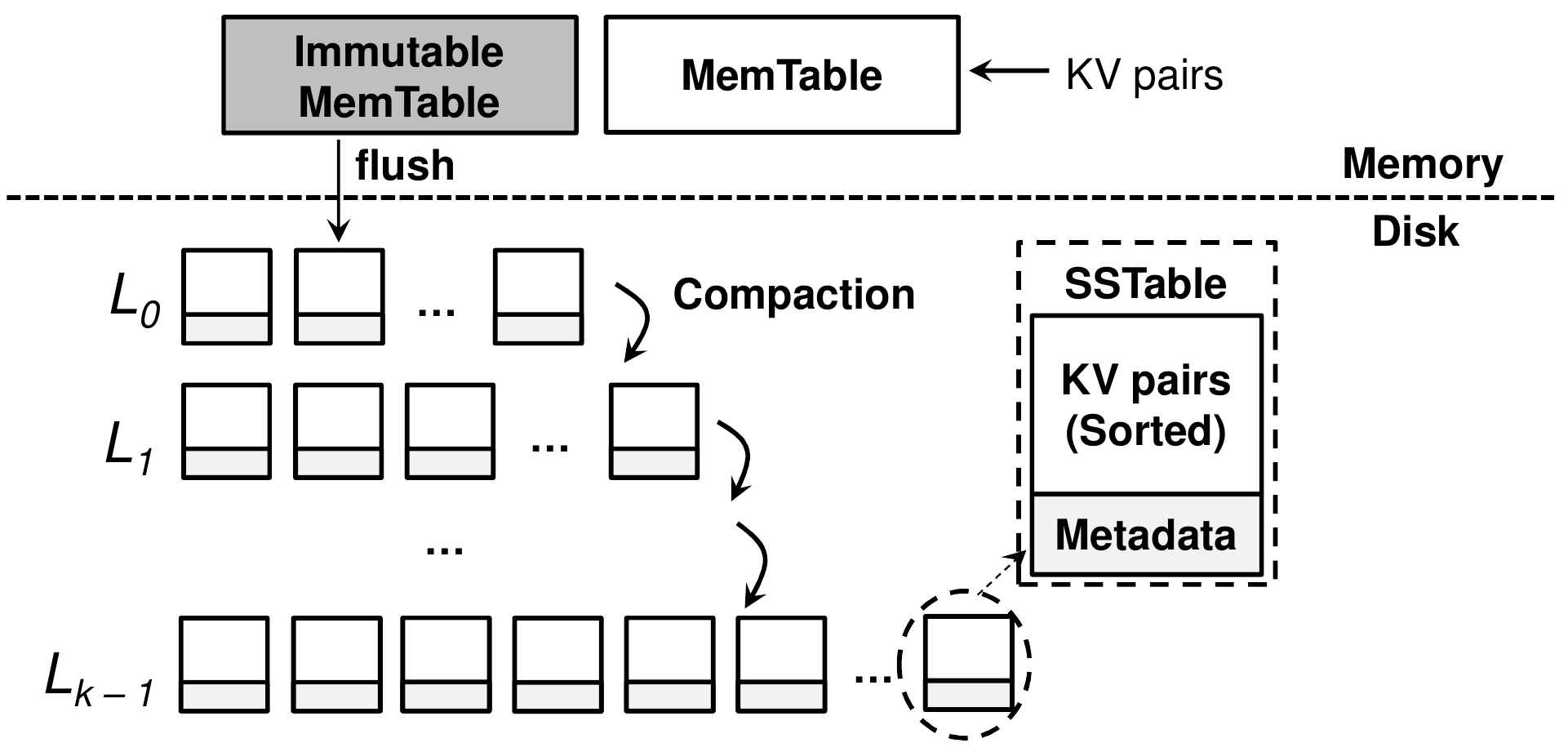}
\caption{Data organization in LevelDB.}
\label{fig:leveldb}
\end{figure}

LevelDB organizes KV pairs based on the LSM-tree \cite{Oneil96}, which
transforms small random writes into sequential writes and hence maintains
high write performance.  Figure~\ref{fig:leveldb} illustrates the data
organization in LevelDB.  It divides the storage space into $k$ levels (where
$k > 1$) denoted by $L_0$, $L_1$, $\cdots$, $L_{k-1}$.
It configures the capacity of each level $L_i$ to be a multiple (e.g.,
10$\times$) of that of its upper level $L_{i-1}$ (where $1\le i\le k-1$).

For inserts or updates of KV pairs, LevelDB first stores the new KV pairs in a
fixed-size in-memory buffer called {\em MemTable}, which uses a skip-list to
keep all buffered KV pairs sorted by keys.  When the MemTable is
full, LevelDB makes it {\em immutable} and flushes it to disk in level
$L_0$ as a file called {\em SSTable}.  Each SSTable has a size of around 
2\,MiB and is also immutable. It stores indexing metadata, a Bloom filter (for
quickly checking if a KV pair exists in the SSTable), and all sorted KV pairs.

If $L_0$ is full, LevelDB flushes and merges the KV pairs in $L_0$ into $L_1$
via {\em compaction}; similarly, if $L_1$ is full, LevelDB flushes and merges
the KV pairs in $L_1$ into $L_2$, and so on.  The compaction process comprises
three steps.  First, it reads out KV pairs in both $L_i$ and $L_{i+1}$ into
memory (where $i\ge 0$).  Second, it sorts the {\em valid} KV pairs 
(i.e., the KV pairs that are newly inserted or updated) by keys and
reorganizes them into SSTables.  It
also discards all {\em invalid} KV pairs (i.e., the KV pairs that are deleted
or superseded by new updates).  Finally, it writes back all SSTables with
valid KV pairs to $L_{i+1}$. Note that all KV pairs in each level, except
$L_0$, are sorted by keys.  In $L_0$, LevelDB only keeps KV pairs sorted
within each SSTable, but not across SSTables.  This improves performance of
flushing KV pairs from the MemTable to disk.

To perform a key lookup, LevelDB searches from $L_0$ to $L_{k-1}$ and returns
the first associated value found. In $L_0$, LevelDB searches all SSTables.
In each level between $L_1$ and $L_{k-1}$, LevelDB first identifies a
candidate SSTable and checks its Bloom filter to determine if the KV pair
exists. If so, LevelDB reads the candidate SSTable and searches for the KV
pair; otherwise, it directly searches the lower levels.  

\paragraph{Limitations:} LevelDB achieves high random write performance
via the LSM-tree-based design, but suffers from both write and read
amplifications.  First, the compaction process inevitably incurs extra
reads and writes.  In the worst case, to merge one SSTable from $L_{i-1}$ to
$L_i$, it reads and sorts 10 SSTables, and writes back all SSTables.  Prior
studies show that LevelDB can have an overall write amplification of at least
50$\times$ \cite{Wu15,Lu16}, since it may trigger more than one compaction to
move a KV pair down multiple levels under large workloads.

Also, a lookup operation may search multiple levels for a KV pair and
incur multiple disk accesses.  The reason is that the search in each level
needs to read the indexing metadata and the Bloom filter in the associated
SSTable. Although the Bloom filter is used, it may introduce false positives.
In this case, a lookup may still unnecessarily read an SSTable from disk even
though the KV pair actually does not exist in the SSTable.  Thus, each lookup
typically incurs multiple disk accesses.  Such read amplification further
aggravates under large workloads, as the LSM-tree builds up in levels.
Measurements show that the read amplification can reach over 300$\times$ in the
worst case \cite{Lu16}.

\subsection{KV Separation}
\label{subsec:kvsep}

KV separation, proposed by WiscKey \cite{Lu16}, decouples the management of
keys and values to mitigate both write and
read amplifications.  The rationale is that storing values in the LSM-tree
is unnecessary for indexing.  Thus, WiscKey stores only keys and metadata
(e.g., key/value sizes, value locations, etc.) in the LSM-tree, while
storing values in a separate append-only circular log called {\em vLog}.
KV separation effectively mitigates write and read
amplifications of LevelDB as it significantly reduces the size of the
LSM-tree, and hence both compaction and lookup overheads.

Since vLog follows the log-structured design \cite{Rosenblum92}, it is
critical for KV separation to achieve lightweight {\em garbage collection
(GC)} in vLog, i.e., to reclaim the free space from invalid values with
limited overhead.  Specifically, WiscKey tracks the {\em vLog head} and the
{\em vLog tail},
which correspond to the end and the beginning of vLog, respectively.  It
always inserts new values to the vLog head. When it performs a GC operation,
it reads a chunk of KV pairs from the vLog tail.  It first queries the
LSM-tree to see if each KV pair is valid.  It then discards the values of
invalid KV pairs, and writes back the valid values to the vLog head.  It
finally updates the LSM-tree for the latest locations of the valid values.
To support efficient LSM-tree queries during GC, WiscKey also stores the
associated key and metadata together with the value in vLog.  Note that vLog
is often over-provisioned with extra reserved space to mitigate GC overhead.

\paragraph{Limitations:}  While KV separation reduces compaction
and lookup overheads, we argue that it suffers from the substantial GC
overhead in vLog.  Also, the GC overhead becomes more severe if the
reserved space is limited.  The reasons are two-fold.

First, vLog can only reclaim space from its vLog tail due to its circular log
design. This constraint may incur unnecessary data movements.  In particular, 
real-world KV storage often exhibits strong locality \cite{Atikoglu12}, in
which a small portion of {\em hot} KV pairs are frequently updated, while the
remaining {\em cold} KV pairs receive only few or even no updates.
Maintaining a strict sequential order in vLog inevitably relocates cold KV
pairs many times and increases GC overhead.

Also, each GC operation queries the LSM-tree to check the validity of
each KV pair in the chunk at the vLog tail. Since the keys of the KV
pairs may be scattered across the entire LSM-tree, the query overhead is
high and increases the latency of the GC operation. Even though KV separation
has already reduced the size of the LSM-tree, the LSM-tree is still
sizable under large workloads, and this aggravates the query cost. 

\begin{figure}[!t]
\centering
\includegraphics[width=3.6in]{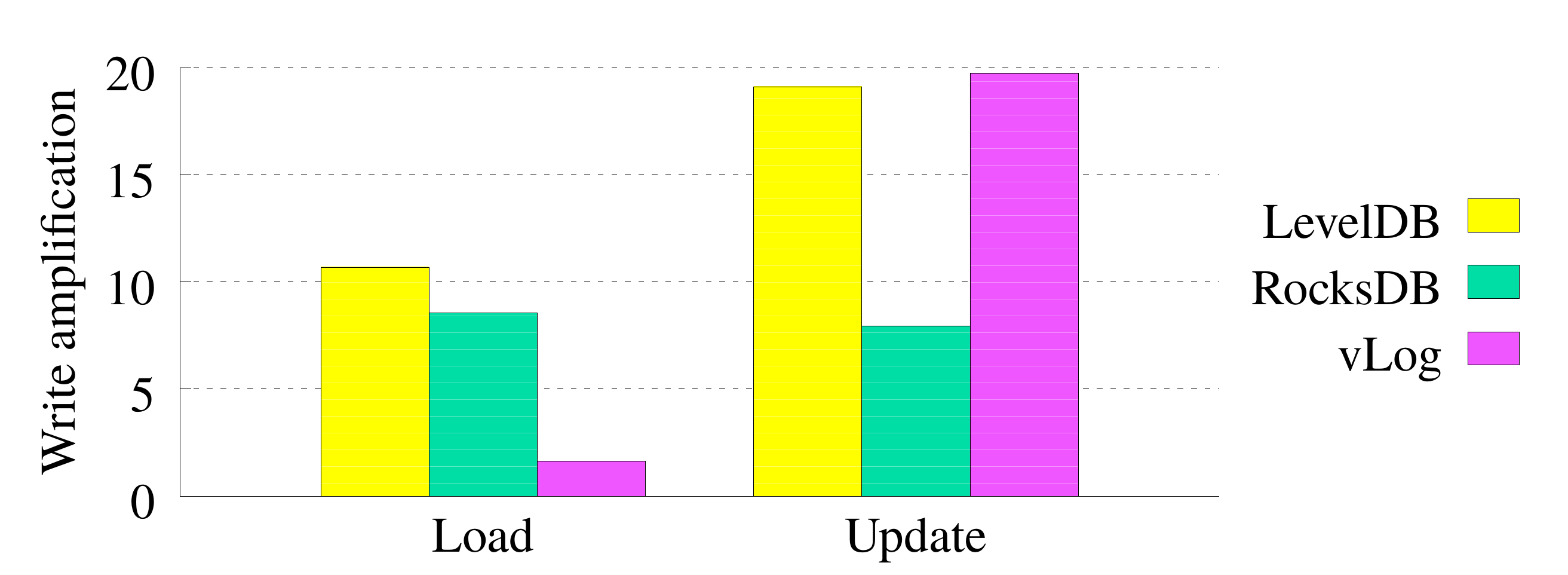}
\caption{Write amplifications of LevelDB, RocksDB, and vLog in the load and
update phases.}
\label{fig:wa_vlog_motivation}
\end{figure}

To validate the limitations of KV separation, we implement a KV store
prototype based on vLog (\S\ref{subsec:impl}) and evaluate its write
amplification. We consider two phases: load and update.  In the load phase, we
insert 40\,GiB of 1-KiB KV pairs into vLog that is initially empty.  Each
KV pair comprises 8-B metadata (including the key/value size fields and
reserved information), a 24-B key, and a 992-B value.  We insert the KV pairs
based on the default {\em hashed} order in YCSB \cite{Cooper10}, such that the
keys are the hashed outputs of some seed numbers; in other words, the
keys of the inserted KV pairs appear in random order.
In the update phase, we generate skewed access patterns (as
observed in real-world KV storage workloads \cite{Atikoglu12}); specifically,
we issue 40\,GiB of updates to the existing KV pairs based on a heavy-tailed
Zipf distribution with a Zipfian constant of 0.99 (the default Zipfian
constant in YCSB \cite{Cooper10}).
We provision 40\,GiB of space for vLog, and an additional 30\% (i.e., 12\,GiB)
of reserved space.  We also disable the write cache in our prototype
(\S\ref{subsec:storage}).  Figure~\ref{fig:wa_vlog_motivation} shows the write
amplification results of vLog in the load and update phases, in terms of the
ratio of the total device write size to the actual write size due to inserts
or updates.  For comparison, we also consider two modern KV stores, LevelDB
\cite{LevelDB} and RocksDB \cite{RocksDB}, based on their default parameters.
In the load phase, vLog has sufficient space to hold all KV pairs and does not
trigger GC, so its write amplification is only 1.6$\times$ due to KV
separation.  However, in the update phase, the updates fill up the reserved
space and start to trigger GC.  We see that vLog has a write amplification of
19.7$\times$, which is close to LevelDB (19.1$\times$) and higher than RocksDB
(7.9$\times$).  

To mitigate GC overhead in vLog, one approach is to partition vLog into
segments and choose the best candidate segments for GC 
(e.g., based on the cost-benefit policy or its variants
\cite{Rosenblum92,Matthews97,Rumble14}).  However, the hot and cold KV pairs
can still be mixed together in vLog, so the chosen segments for GC may still
contain cold KV pairs that are unnecessarily moved.

To address the mixture of hot and cold data, a better approach is to perform
{\em hot-cold data grouping} as in SSD designs \cite{Min12,Lee13}, in which
we separate the storage of hot and cold KV pairs into two regions and apply
GC to each region individually (more GC operations are expected to be
applied to the storage region for hot KV pairs).  However, the direct
implementation of hot-cold data grouping inevitably increases the update
latency in KV separation.  As a KV pair may be stored in either hot or cold
regions, each update needs to first query the LSM-tree for the exact storage
location of the KV pair.  Thus, a key motivation of our work is to enable
hotness awareness without LSM-tree lookups.

\section{\HASHKV Design}
\label{sec:design}

\HASHKV is a persistent KV store that specifically targets update-intensive
workloads.  It improves the management of value storage atop KV
separation to achieve high update performance.
It supports standard KV operations: {\tt PUT} (i.e., writing a KV pair),
{\tt GET} (i.e., retrieving the value of a key), {\tt DELETE} (i.e., deleting
a KV pair), and {\tt SCAN} (i.e., retrieving the values of a range of keys).

\subsection{Main Idea}
\label{subsec:idea}

\HASHKV follows KV separation \cite{Lu16} by storing only keys and metadata in
the LSM-tree for indexing KV pairs, while storing values in a separate area
called the {\em value store}.  Atop KV separation, \HASHKV introduces
several core design elements to achieve efficient value storage management.

\begin{itemize}[leftmargin=*]
\item
{\bf Hash-based data grouping:}  Recall that vLog incurs substantial
GC overhead in value storage.  Instead, \HASHKV maps values into fixed-size
partitions in the value store by hashing the associated keys.  This design
achieves: (i) {\em partition isolation}, in which all versions of value updates
associated with the same key must be written to the same partition, and (ii)
{\em deterministic grouping}, in which the partition where a value should be
stored is determined by hashing.  We leverage this design to achieve
flexible and lightweight GC.

\item
{\bf Dynamic reserved space allocation:}  Since we map values into
fixed-size partitions, one challenge is that a partition may receive more
updates than it can hold.  \HASHKV allows a partition to grow
{\em dynamically} beyond its size limit by allocating fractions of reserved
space in the value store to hold the extra updates.

\item
{\bf Hotness awareness:}  Due to deterministic grouping, a partition may be
filled with the values from a mix of hot and cold KV pairs, in which case a
GC operation unnecessarily reads and writes back the values of cold KV pairs.
\HASHKV uses a {\em tagging} approach to relocate the values of cold KV pairs
to a different storage area and separate the hot and cold KV pairs, so that we
can apply GC to hot KV pairs only and avoid re-copying cold KV pairs.

\item
{\bf Selective KV separation:}  \HASHKV differentiates KV pairs by their value
sizes, such that the small-size KV pairs can be directly stored in the
LSM-tree without KV separation.  This saves the overhead of accessing both the
LSM-tree and the value store for small-size KV pairs, while the compaction
overhead of storing the small-size KV pairs in the LSM-tree is limited.
\end{itemize}

\noindent
{\bf Remarks:}  \HASHKV maintains a single LSM-tree for indexing (instead of
hash-partitioning the LSM-tree as in the value store) to preserve the ordering
of keys and the range scan performance.  Since hash-based data grouping
spreads KV pairs across the value store, it incurs random writes; in
contrast, vLog maintains sequential writes with a log-structured storage
layout.  Our \HASHKV prototype (\S\ref{subsec:impl}) exploits both
multi-threading and batch writes to limit random write overhead.

\subsection{Storage Management}
\label{subsec:storage}

\begin{figure}[!t]
\centering
\includegraphics[width=5in]{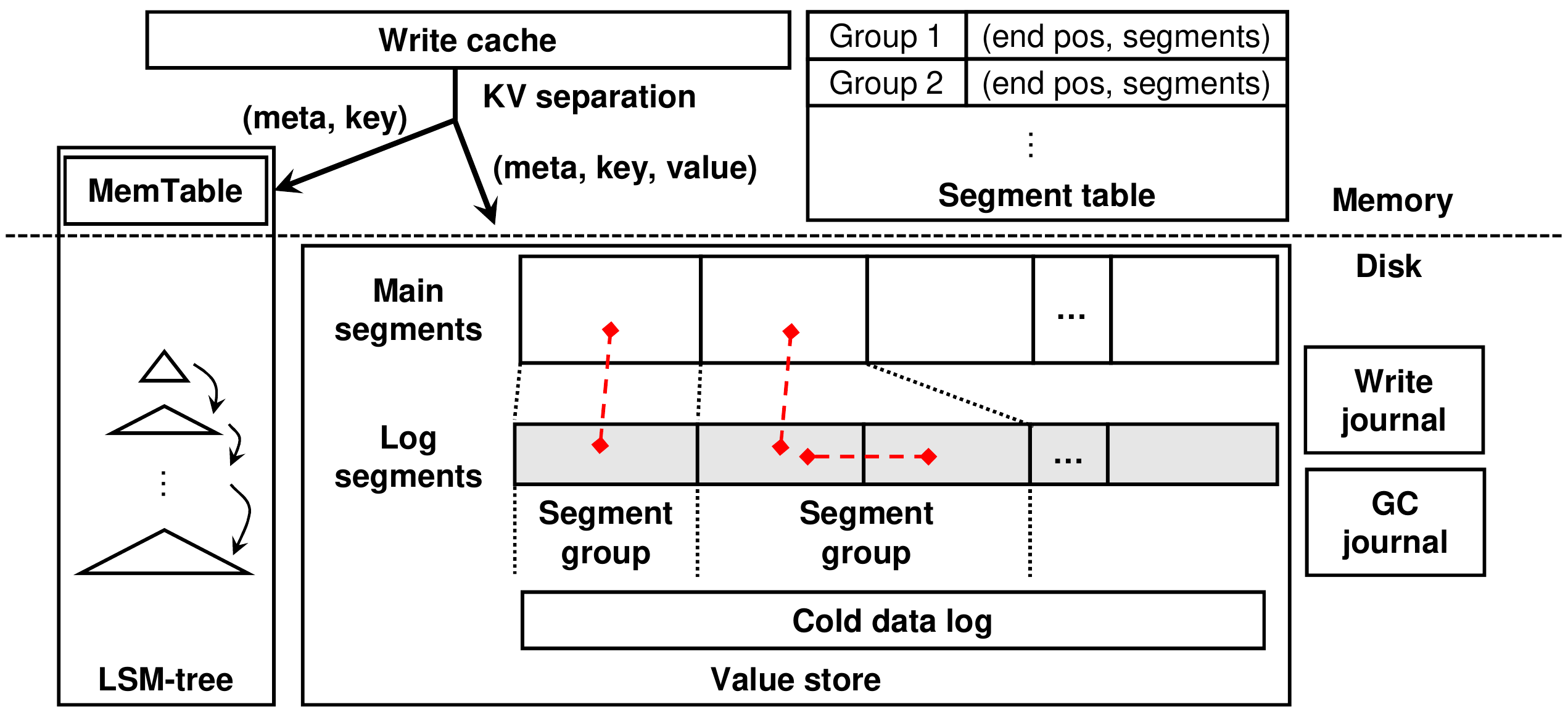}
\caption{\HASHKV architecture.
Each segment group consists of one main segment and a variable
number (zero or more) of log segments; the log segments (in shaded color) are
provisioned in reserved space.  For example, the segments that are linked
together belong to the same segment group.}
\label{fig:arch}
\end{figure}

Figure~\ref{fig:arch} depicts the architecture of \HASHKV. It divides the
logical address space of the value store into fixed-size units called {\em
main segments}.  Also, it over-provisions a fixed portion of reserved space,
which is again divided into fixed-size units called {\em log segments}.
Note that the sizes of main segments and log segments are configurable
and may differ; by default, we set them as 64\,MiB and 1\,MiB,
respectively.

For each insert or update of a KV pair, \HASHKV hashes its key into one of
the main segments. If the main segment is not full, \HASHKV stores the value
in a log-structured manner by appending the value to the end of the main
segment; on the other hand, if the main segment is full, \HASHKV dynamically
allocates a free log segment to store the extra values in a log-structured
manner. Again, it further allocates additional free log segments if the
current log segment is full.
We collectively call a main segment and all
its associated log segments (if any) a {\em segment group}, as shown in
Figure~\ref{fig:arch}.  Also, \HASHKV updates
the LSM-tree for the latest value location.  To keep track of the storage
status of the segment groups and segments, \HASHKV uses a global in-memory
{\em segment table} to store the current end position of each segment group
for subsequent inserts or updates, as well as the list of log segments
associated with each segment group.  Our design ensures that each insert or
update can be directly mapped to the correct write position without issuing
LSM-tree lookups on the write path, thereby achieving high write performance.
Also, the updates of the values associated with the same key must go to the
same segment group, and this simplifies GC.  For fault tolerance, \HASHKV
checkpoints the segment table to persistent storage.

To facilitate GC, \HASHKV also stores the key and metadata (e.g., key/value
sizes) together with the value for each KV pair in the value store as in
WiscKey \cite{Lu16} (Figure~\ref{fig:arch}). This enables a GC operation
to quickly identify the key associated with a value when it scans the value
store. However, our GC design inherently differs from vLog used by WiscKey
(\S\ref{subsec:gc}).

To improve write performance, \HASHKV holds an in-memory
{\em write cache} to store the recently written KV pairs, at the expense of
degrading reliability.  If the key of a new KV pair to be written is found in
the write cache, \HASHKV directly updates the value of the cached key
in-place without issuing the writes to the LSM-tree and the value
store. It can also return the KV pairs from the write cache for reads.  If the
write cache is full, \HASHKV flushes all the cached KV pairs to the LSM-tree
and the value store.  Note that the write cache is an optional component and
can be disabled for reliability concerns.

\HASHKV supports hotness awareness by keeping cold values in a separate
{\em cold data log} (\S\ref{subsec:hotness}).  It also addresses crash
consistency by tracking the updates in both {\em write journal} and {\em GC
journal} (\S\ref{subsec:crash}).  Note that the sizes of the cold data
log and the journals are configurable.

\subsection{Garbage Collection (GC)}
\label{subsec:gc}

\HASHKV necessitates GC to reclaim the space occupied by invalid KV pairs in
the value store.  In \HASHKV, GC operates in units of segment groups, and is
triggered when the free log segments in the reserved space are running out.
At a high level, a GC operation first selects a candidate segment group and
identifies all valid KV pairs in the group. It then writes back all valid KV
pairs to the main segment, or additional log segments if needed, in a
log-structured manner. It also releases any unused log segments that can be
later used by other segment groups. Finally, it updates the latest value
locations in the LSM-tree. Here, the GC operation needs to address two issues:
(i) which segment group should be selected for GC; and (ii) how the GC
operation quickly identifies the valid KV pairs in the selected segment group.

Unlike vLog, which requires the GC operation to follow a strict sequential
order, \HASHKV can flexibly choose which segment group to perform GC.  By
default, it adopts a {\em greedy} approach and selects the segment group with
the largest amount of writes.  The rationale is that the selected segment
group typically holds the hot KV pairs that have many updates and hence has a
large amount of writes.  Thus, selecting this segment group for GC likely
reclaims the most free space.  To realize the greedy approach, \HASHKV tracks
the amount of writes for each segment group in the in-memory segment table
(\S\ref{subsec:storage}), and uses a {\em heap} to quickly identify which
segment group receives the largest amount of writes.

\HASHKV can adopt other approaches of choosing a segment group for GC.  We
implement two additional approaches, namely the {\em cost-benefit algorithm
(CBA)} \cite{Rumble14} and the {\em greedy random algorithm (GRA)}
\cite{Li15}.  CBA takes into account both the age of a segment group (the time
since its last GC operation) and the fraction of valid data in the segment
group when determining which segment group is chosen for GC.
Intuitively, CBA prefers to choose a segment group with a larger age (i.e., it
is more stable) and a lower fraction of valid data (i.e., it has more free
space reclaimed) for GC.  GRA mixes both the greedy and random approaches,
such that it first selects the top-$d$ segment groups that receive the
largest amounts of writes for some configurable parameter $d$, followed by
randomly selecting a segment group among the top-$d$ ones.  Note that when $d
= 1$, GRA reduces to the greedy approach; when $d$ is at least the total
number of segment groups, GRA reduces to the random approach.

To check the validity of KV pairs in the selected segment group, \HASHKV
sequentially scans the KV pairs in the segment group without querying the
LSM-tree (note that it also checks the write cache for any latest KV pairs in
the segment group).  Since \HASHKV writes the KV pairs to the segment group in a
log-structured manner, the KV pairs must be sequentially placed according to
their order of being updated.  For a KV pair that has multiple versions of
updates, the version that is nearest to the end of the segment group must be
the latest one and correspond to the valid KV pair, while other versions are
invalid.  Thus, the running time for each GC operation only depends on
the size of the segment group that needs to be scanned.  In contrast, the GC
operation in vLog reads a chunk of KV pairs from the vLog tail
(\S\ref{subsec:kvsep}). It queries the LSM-tree (based on the keys stored
along with the values) for the latest storage location of each KV pair in
order to check if the KV pair is valid \cite{Lu16}.  The overhead of querying
the LSM-tree becomes substantial under large workloads.

During a GC operation on a segment group, \HASHKV constructs a temporary
in-memory hash table (indexed by keys) to buffer the addresses of the valid KV
pairs being found in the segment group.
As the key and address sizes are generally small and the number of KV pairs
in a segment group is limited, the hash table has limited size and can be
entirely stored in memory.
\HASHKV blocks incoming writes during a GC operation, so that KV
pairs remain intact when being relocated to different segments.

\subsection{Hotness Awareness}
\label{subsec:hotness}

Hot-cold data separation improves GC performance in log-structured storage
(e.g., SSDs \cite{Min12,Lee13}).  In fact, the current hash-based data
grouping design realizes some form of hot-cold data separation, since the
updates of the hot KV pairs must be hashed to the same segment group and our
current GC policy always chooses the segment group that is likely to store the
hot KV pairs (\S\ref{subsec:gc}).  However, it is inevitable that some
cold KV pairs are hashed to the segment group selected for GC, leading to
unnecessary data rewrites.  Thus, a challenge is to fully realize hot-cold
data separation to further improve GC performance.

\begin{figure}[!t]
\centering
\includegraphics[width=2.8in]{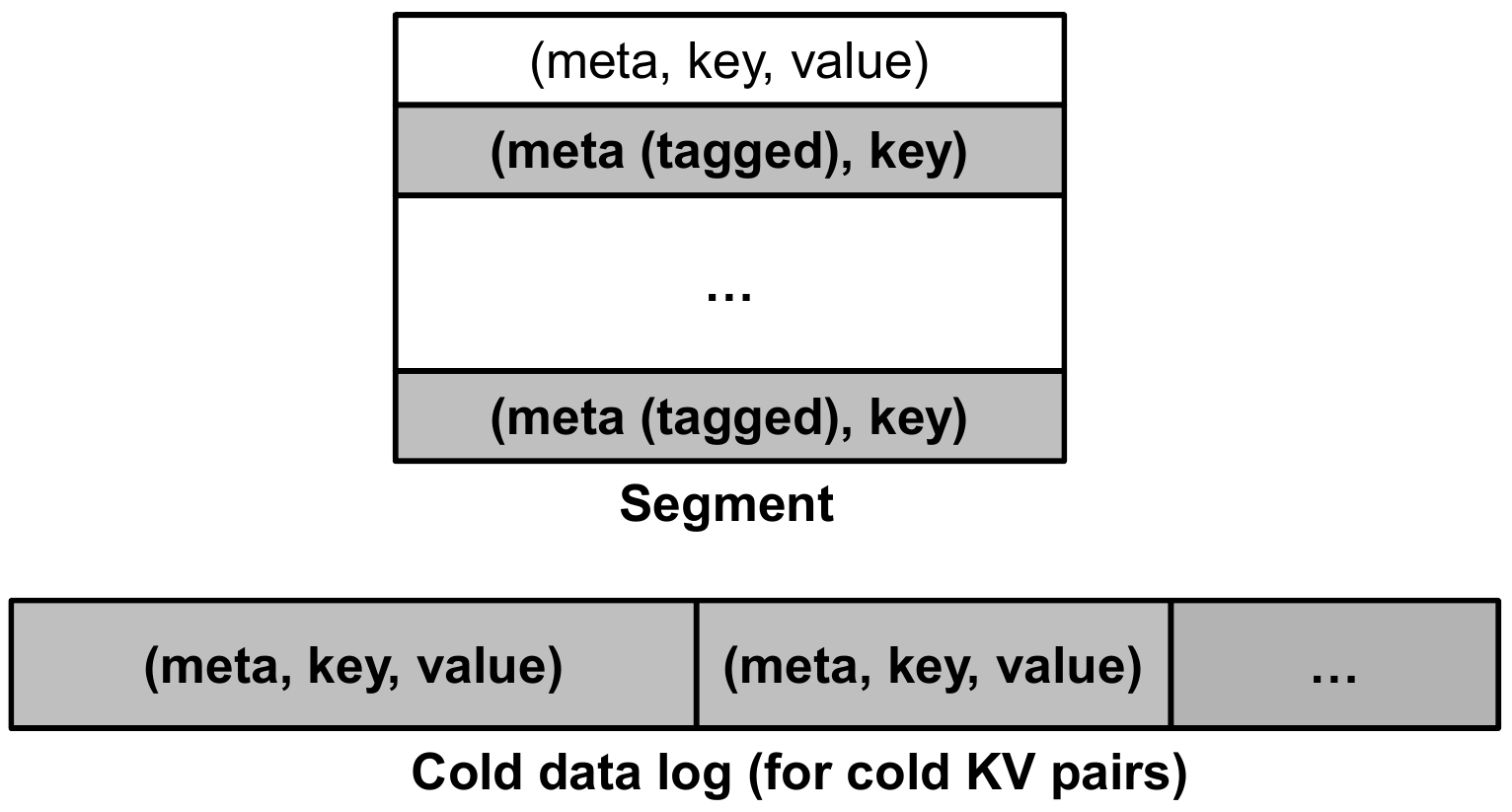}
\caption{Tagging in \HASHKV. Note that the locations of the cold
KV pairs in the cold data log are stored in the LSM-tree.}
\label{fig:tagging}
\vspace{-6pt}
\end{figure}

\HASHKV relaxes the restriction of hash-based data grouping via a {\em
tagging} approach (Figure~\ref{fig:tagging}).  Specifically, when \HASHKV
performs a GC operation on a segment group, it classifies each KV pair in the
segment group as hot or cold.  Currently, we treat the KV pairs that are
updated at least once since their last inserts as hot, or cold otherwise (more
accurate hot-cold data identification approaches \cite{Hsieh06} can be used).
For the hot KV pairs, \HASHKV still writes back their latest versions to the
same segment group via hashing.  However, for the cold KV pairs, it now writes
their values to a separate storage area, and keeps their key and metadata only
(i.e., without values) in the segment group.  In addition, it adds a {\em tag}
in the metadata of each cold KV pair to indicate its presence in the segment
group.  Thus, if a cold KV pair is not updated, a GC operation only rewrites
its key and metadata without rewriting its value, thereby saving the rewrite
overhead.  On the other hand, if a cold KV pair is later updated, we know
directly from the tag (without querying the LSM-tree) that the cold KV pair
has already been stored, so that we can treat it as hot based on our
classification policy; also, the tagged KV pair will become invalid and can be
reclaimed in the next GC operation. Finally, at the end of a GC operation,
\HASHKV updates the latest value locations in the LSM-tree, such that the
locations of the cold KV pairs point to the separate storage area.

With tagging, \HASHKV avoids storing the values of cold KV pairs in the
segment group and rewriting them during GC.  Also, tagging is only triggered
during GC, and does not add extra overhead to the write path.  Currently, we
implement the separate storage area for cold KV pairs as an append-only log
(called the {\em cold data log}) in the value store, and perform GC on the
cold data log as in vLog.  The cold data log can also be put in secondary
storage with a larger capacity (e.g., hard disks) if the cold KV pairs are
rarely accessed.

\subsection{Selective KV Separation}
\label{subsec:selective_kv_sep}

\HASHKV supports workloads with general value sizes.  Our rationale is that
KV separation reduces compaction overhead especially for large-size KV pairs,
yet its benefits for small-size KV pairs are limited, and it incurs extra
overhead of accessing both the LSM-tree and the value store.  Thus, we propose
{\em selective} KV separation, in which we still apply KV separation to KV
pairs with large value sizes, while storing KV pairs with small value sizes in
{\em entirety} in the LSM-tree.  A key challenge of selective KV separation
is to choose the KV pair size threshold of differentiating between
small-size and large-size KV pairs (assuming that the key size remains
fixed).

\begin{figure}[!t]
\centering
\includegraphics[width=3.6in]{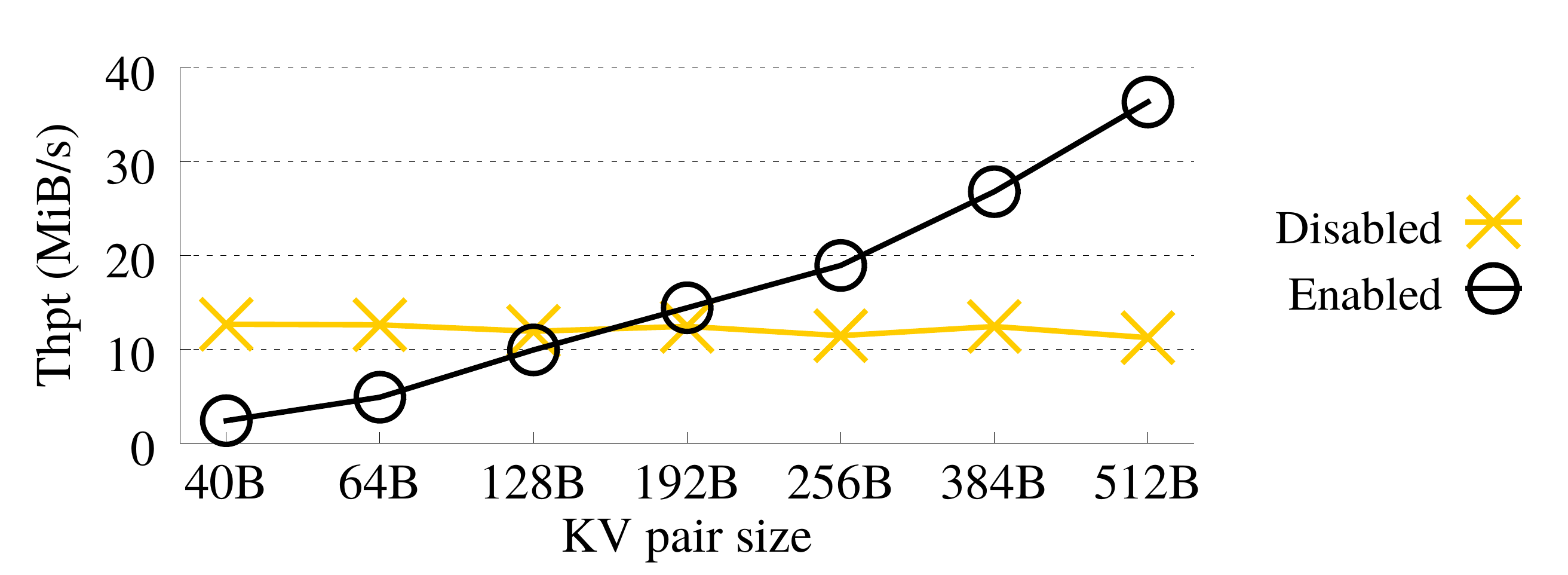}
\caption{Update throughput versus the KV pair size with and without KV
separation.}
\label{fig:kvsep_kv_thp}
\end{figure}

Here, we propose a simple search method to choose the appropriate threshold.
Our idea is to benchmark the update performance of different KV pair sizes
with and without KV separation on the platform where we deploy \HASHKV.  As a
case study, we conduct performance benchmarking based on our testbed
(\S\ref{subsec:exp_setup}) and update-intensive workloads
(\S\ref{subsec:exp_perf}).  We first load 40\,GiB of KV pairs into \HASHKV,
which is initially empty.  We then repeatedly issue three phases of 40\,GiB
updates to obtain a stable update throughput. If KV separation is disabled,
all updates are directly stored in LevelDB (on which \HASHKV manages keys and
metadata); otherwise if KV separation is enabled, we allocate 30\% of
reserved space to \HASHKV for value management.  Figure~\ref{fig:kvsep_kv_thp}
shows the update throughput for the last phase of 40\,GiB updates versus the
KV pair size; here, we fix the key size as 24\,B and the metadata size as
8\,B.  When the KV pair size is small, the overhead of accessing both the
LSM-tree and the value store in KV separation is significant, so enabling KV
separation has worse performance.  When the KV pair size increases, KV
separation improves performance, and it shows performance gains when the KV
pair size is between 128\,B and 192\,B.  Thus, we choose 192\,B as our default
threshold for selective KV separation.  Interestingly, even though we choose
our threshold based on update-intensive workloads, our search method is fairly
robust and the threshold also works well for other types of workloads
(see \S\ref{subsec:exp_ycsb} for details).

\subsection{Range Scans}
\label{subsec:range}

One critical reason of using the LSM-tree for indexing is its efficient
support of
range scans. Since the LSM-tree stores and sorts KV pairs by keys, it can
return the values of a range of keys via sequential reads.  However, KV
separation now stores values in separate storage space, so it incurs extra
reads of values.  In \HASHKV, the values are scattered across different
segment groups, so range scans will trigger many random reads that degrade
performance.  \HASHKV currently leverages the {\em read-ahead} mechanism to
speed up range scans by prefetching values into the page cache.
For each scan request, \HASHKV iterates over the range of sorted keys in the
LSM-tree, and issues a read-ahead request to each value (via
\texttt{posix\_fadvise}). It then reads all values and returns the sorted KV
pairs.  Note that Wisckey \cite{Lu16} fetches values in the background while
iterating over keys.

\subsection{Crash Consistency}
\label{subsec:crash}

Crashes can occur while \HASHKV issues writes to persistent storage.  \HASHKV
addresses crash consistency based on {\em metadata journaling} and focuses on
two aspects: (i) flushing the write cache and (ii) GC operations.

Flushing the write cache involves writing the KV pairs to the value store and
updating metadata in the LSM-tree.  \HASHKV maintains a {\em write journal}
to track each flushing operation.  It performs the following steps when
flushing the write cache: (i) flushing the cached KV pairs to the value store;
(ii) appending metadata updates to the write journal; (iii) writing a commit
record to the journal end; (iv) updating keys and metadata in the LSM-tree;
and (v) marking the flush operation free in the journal (the freed journaling
records can be recycled later).
If a crash occurs after step (iii) completes, \HASHKV replays the updates in
the write journal and ensures that the LSM-tree and the value store are
consistent.

Handling crash consistency in GC operations is different, as they may
overwrite existing valid KV pairs.  Thus, we also need to protect
existing valid KV pairs against crashes during GC.   \HASHKV
maintains a {\em GC journal} to track each GC operation.
It performs the following steps after identifying all valid KV pairs during a
GC operation: (i) appending the valid KV pairs that are overwritten as well as
metadata updates to the GC journal; (ii) writing all valid KV pairs back to
the segment group; (iii) updating the metadata in the LSM-tree; and (iv)
marking the GC operation free in the journal.

\subsection{Implementation Details}
\label{subsec:impl}

We prototype \HASHKV in C++ on Linux.  For key and metadata
management, \HASHKV uses LevelDB v1.20 \cite{LevelDB} by default, yet it can
also leverage other LSM-tree-based KV stores to improve their respective
performance (\S\ref{subsec:exp_ycsb}).  Our prototype contains around
6.7K~lines of code (without LevelDB).

\paragraph{Storage organization:}
We currently deploy \HASHKV on the Linux Ext4 file system, on which we
run both LevelDB and the value store.  In particular,
\HASHKV manages the value store as a large file.  It partitions the value
store file into two regions, one for main segments and another for log
segments, according to the pre-configured segment sizes.  All segments are
aligned in the value store file, such that the start offset of each main
(resp. log) segment is a multiple of the main (resp. log) segment size.  If
hotness awareness is enabled (\S\ref{subsec:hotness}), \HASHKV adds a
separate region in the value store file for the cold data log.  Also, to
address crash consistency (\S\ref{subsec:crash}), \HASHKV uses separate
files to store both write and GC journals.

\paragraph{Multi-threading:} \HASHKV implements multi-threading via
{\tt threadpool} \cite{threadpool} to boost I/O performance when flushing KV
pairs in the write cache to different segments (\S\ref{subsec:storage}) and
retrieving segments from segment groups in parallel during GC
(\S\ref{subsec:gc}).

To mitigate random write overhead due to deterministic grouping
(\S\ref{subsec:idea}), \HASHKV implements batch writes. When \HASHKV flushes
KV pairs in the write cache, it first identifies and buffers a number of KV
pairs that are hashed to the same segment group in a {\em batch}, and then
issues a sequential write (via a thread) to flush the batch.  A larger batch
size reduces random write overhead, yet it also degrades parallelism.
Currently, we configure a batch write threshold, such that after adding a KV
pair into a batch, if the batch size reaches or exceeds the batch size
threshold, the batch will be flushed; in other words, \HASHKV directly flushes
a KV pair if its size is larger than the batch write threshold.
Note that WiscKey \cite{Lu16} also uses a write buffer to batch KV pairs into
sequential writes. \HASHKV further issues parallel writes via multi-threading
using a configurable batch size threshold.

\section{Evaluation}
\label{sec:evaluation}

We compare via testbed experiments \HASHKV with several
state-of-the-art KV stores: LevelDB (v1.20) \cite{LevelDB}, RocksDB (v5.8)
\cite{RocksDB}, HyperLevelDB \cite{Hyperleveldb}, PebblesDB \cite{Raju17}, and
our own vLog implementation for KV separation based on WiscKey \cite{Lu16}.
Note that BadgerDB \cite{badgerdb} is an open-source implementation of KV
separation based on WiscKey and is written in Golang.  However, it currently
supports manual compaction only, and we do not include BadgerDB in our
evaluation. 

For fair comparison, we build a unified framework to integrate each
state-of-the-art KV store and \HASHKV.  Specifically, we buffer all written KV
pairs in the write cache and flush them when the write cache is full.  For
LevelDB, RocksDB, HyperLevelDB, and PebblesDB, we flush all KV pairs in
entirety to them; for vLog and \HASHKV, we flush keys and metadata to LevelDB,
and values (together with keys and metadata) to the value store.  We address
the following questions:

\begin{itemize}[leftmargin=*]
\item
How is the update performance of \HASHKV compared to other KV stores under
update-intensive workloads?  (Experiment~1)
\item
How do the reserved space size and RAID configurations affect the update
performance of \HASHKV?  (Experiments~2 and 3)
\item
What is the performance of \HASHKV in updates and range scans for different KV
pair sizes? (Experiments~4 and 5)
\item 
What are the performance gains of hotness awareness and selective KV
separation?  (Experiments~6 and 7)
\item
How do different KV pair size thresholds affect the performance gains of
selective KV separation? (Experiment~8)
\item
How do different GC approaches affect the update performance of \HASHKV?
(Experiment~9)
\item 
How does the crash consistency mechanism affect the update performance of
\HASHKV? (Experiment~10)
\item
What is the performance of \HASHKV compared to other KV stores under YCSB core 
workloads? (Experiments~11 and 12)
\item
What is the performance of \HASHKV when it builds on other KV stores?
(Experiment~13)
\item
What is the storage distribution of the value store of \HASHKV?
(Experiment~14)
\item
How do parameter configurations (e.g., main segment size, log segment size,
and write cache size) affect the update performance of \HASHKV?
(Experiment~15)
\end{itemize}

\subsection{Setup}
\label{subsec:exp_setup}

\noindent
{\bf Testbed:} We conduct our experiments on a machine running Ubuntu 14.04~LTS
with Linux kernel 3.13.0. The machine is equipped with a quad-core Xeon
E3-1240v2, 16\,GiB RAM, and seven Plextor M5 Pro 128\,GiB SSDs.  One SSD
is attached to the motherboard as the OS drive, while the remaining six SSDs
are attached to the LSI SAS 9201-16i host bus adapter to form an SSD RAID
volume for high I/O performance.  Specifically, we create a software RAID
volume using \texttt{mdadm} \cite{mdadm} atop the six SSDs, with a chunk
size of 4\,KiB.  We run each KV store on the SSD RAID volume.

\paragraph{Default setup:} For LevelDB, RocksDB, HyperLevelDB, and
PebblesDB, we use their default parameters.  We allow them to use all
available capacity in our SSD RAID volume, so that their major overheads come
from read and write amplifications in the LSM-tree management.

For vLog, we configure it to read 64\,MiB from the vLog tail 
(\S\ref{subsec:kvsep}) in each GC operation. For \HASHKV, we set the main
segment size as 64\,MiB and the log segment size as 1\,MiB.  Both vLog and
\HASHKV are configured with 40\,GiB of storage space and
over-provisioned with 30\% (i.e., 12\,GiB) of reserved space,
while their key and metadata storage in LevelDB can use all available storage
space.  Also, we do not limit the sizes of the cold data log and the
journals.
Here, we provision the storage space of vLog and \HASHKV to be close to the
actual KV store sizes of LevelDB and RocksDB based on our evaluation
(Experiment~1).

We mount the SSD RAID volume under RAID-0 (no fault tolerance) by
default to maximize performance. All KV stores run in asynchronous mode and
are equipped with a write cache of size 64\,MiB.  For \HASHKV, we set the
batch write threshold (\S\ref{subsec:impl}) to 4\,KiB, and configure 32
and 8 threads for write cache flushing and segment retrieval in GC,
respectively.  We disable selective KV separation, hotness awareness, and
crash consistency in \HASHKV by default, except when we evaluate them.

\subsection{Performance Comparison}
\label{subsec:exp_perf}

We compare the performance of different KV stores under update-intensive
workloads.  Specifically, we generate workloads using YCSB \cite{Cooper10},
and fix the size of each KV pair as 1\,KiB, which consists of 8-B
metadata (including the key/value size fields and reserved information), a
24-B key, and a 992-B value.  We assume that each KV store is initially
empty. We first load 40\,GiB of KV pairs (or 42\,M inserts) into each KV store
(call it Phase~P0). We then repeatedly issue 40\,GiB of updates over the
existing 40\,GiB of KV pairs {\em three} times (call them Phases~P1, P2, and
P3), accounting for 120\,GiB or 126\,M updates in total. Updates in each phase
follow a heavy-tailed Zipf distribution with a Zipfian constant of 0.99. We
issue the requests to each KV store as fast as possible to stress-test its
performance.

Note that vLog and \HASHKV do not trigger GC in Phase~P0.  In Phase~P1, when
the reserved space becomes full after 12\,GiB of updates, both systems start
to trigger GC; in both Phases~P2 and P3, updates are issued to the fully
filled value store and will trigger GC frequently. We include both Phases~P2
and P3 to ensure that the update performance is stable.


\begin{figure}[!t]
\centering
\begin{tabular}{cc}
\multicolumn{2}{c}{
\includegraphics[width=4.5in]{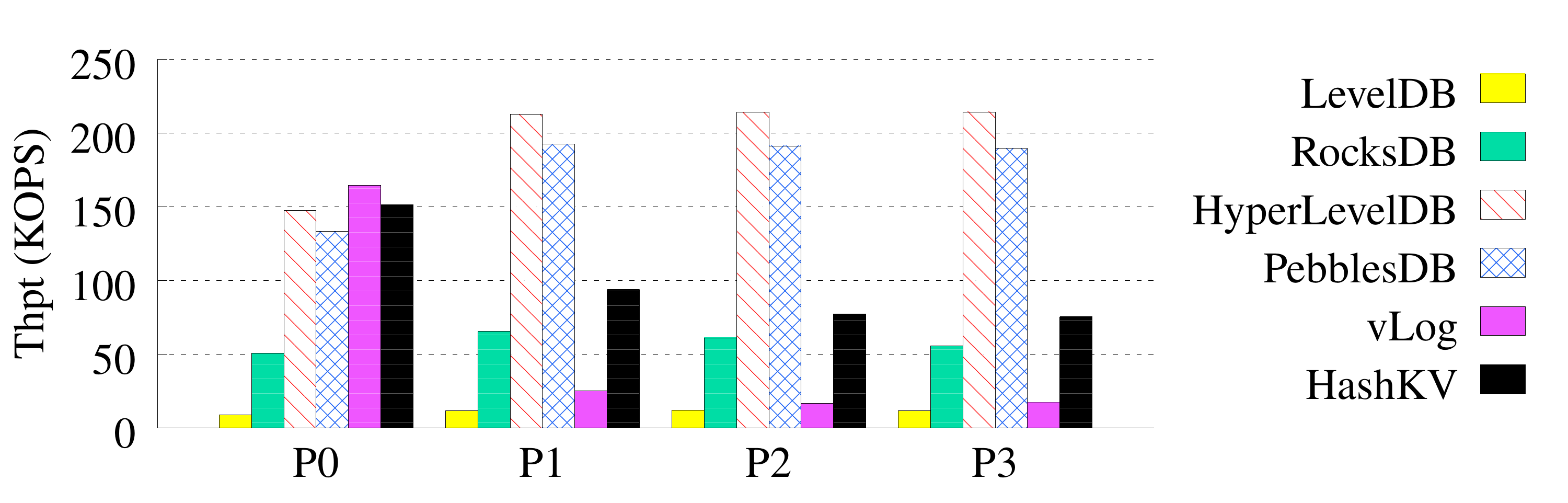}
} \\
\multicolumn{2}{c}{
\mbox{\small (a) Throughput (load phase: P0, update phases: P1 - P3)}
} \\
\includegraphics[width=2.1in]{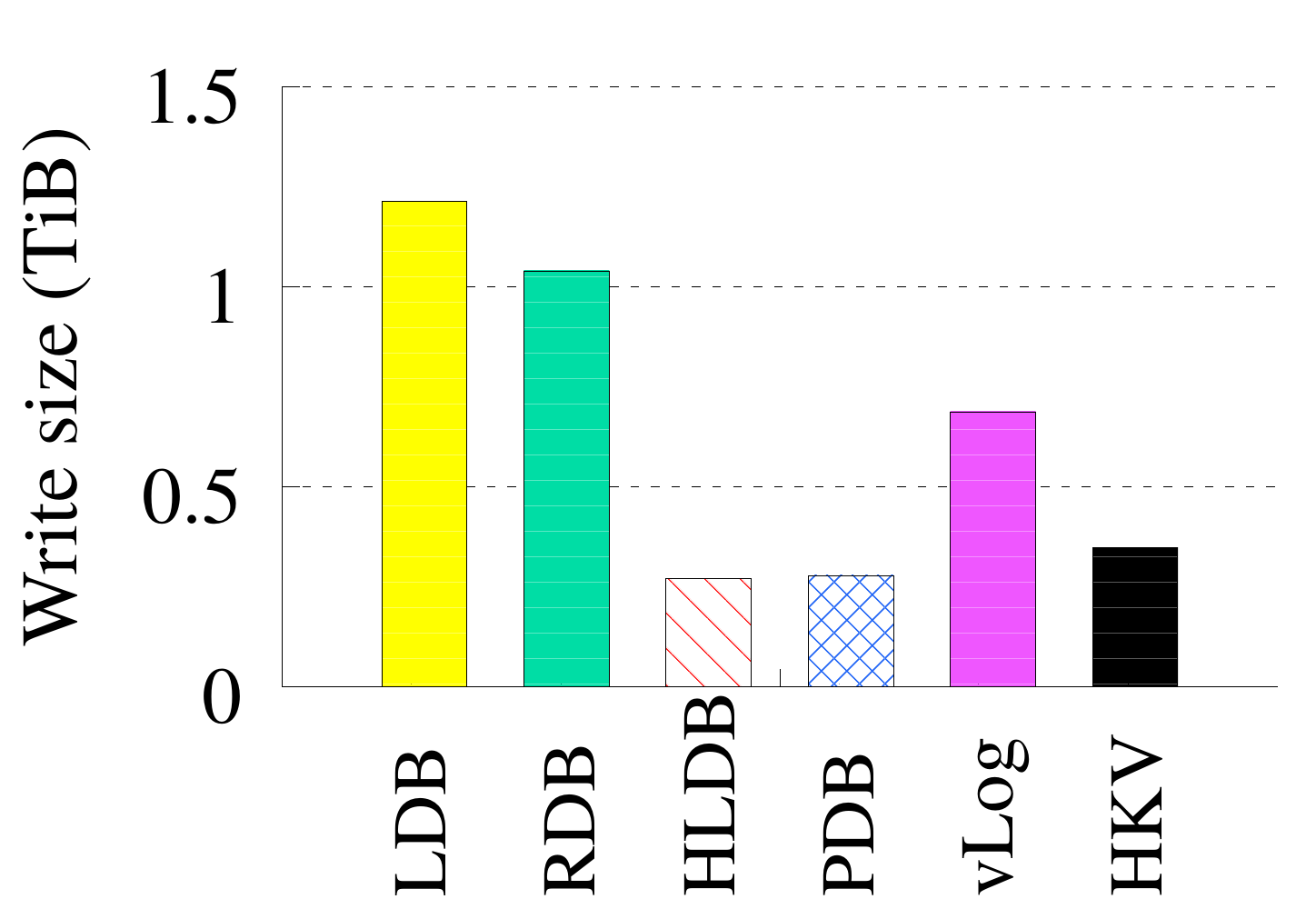} &
\includegraphics[width=2.1in]{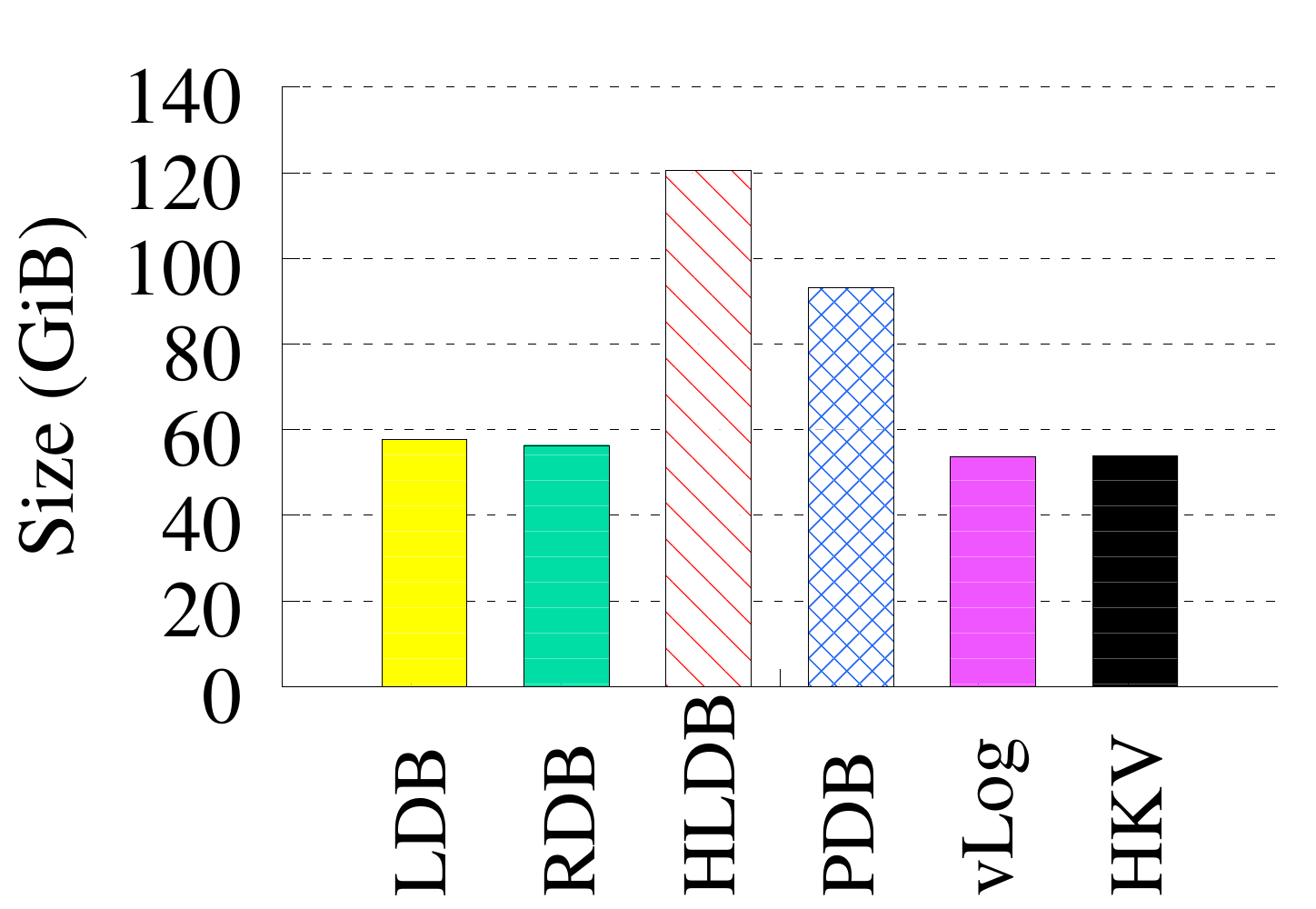}\\
\mbox{\small (b) Total write size} &
\mbox{\small (c) KV store size} \\
\multicolumn{2}{c}{
\includegraphics[width=4.5in]{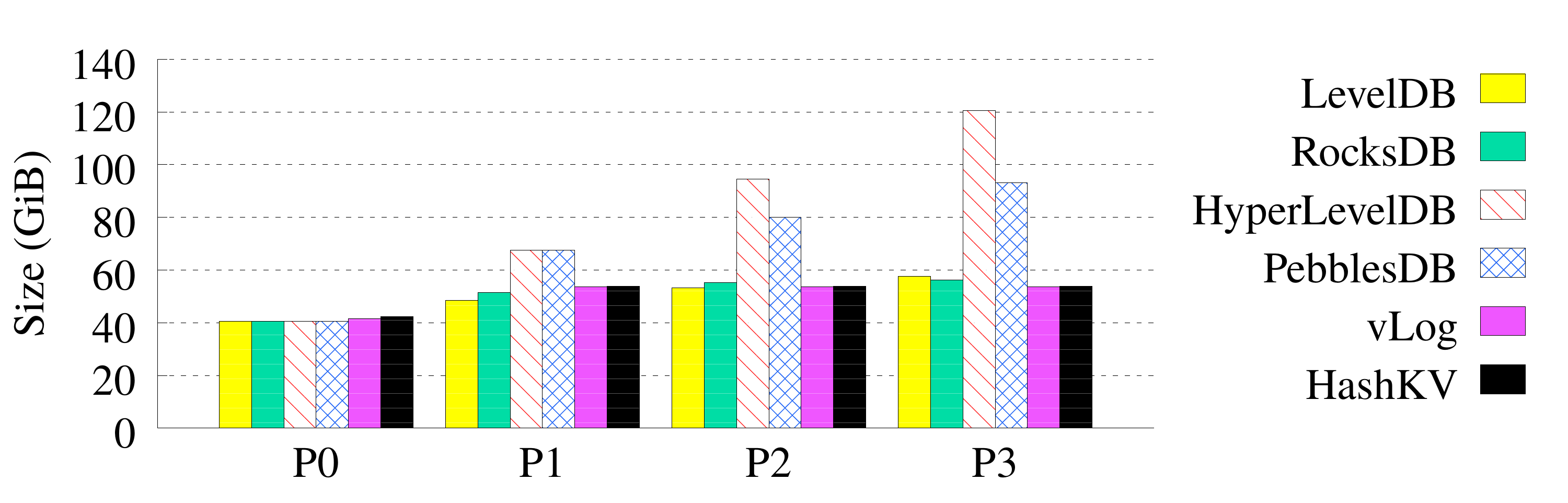}
}\\
\multicolumn{2}{c}{
\mbox{\small (d) KV store size at the end of each phase}
}\\
\multicolumn{2}{c}{
\includegraphics[width=4.5in]{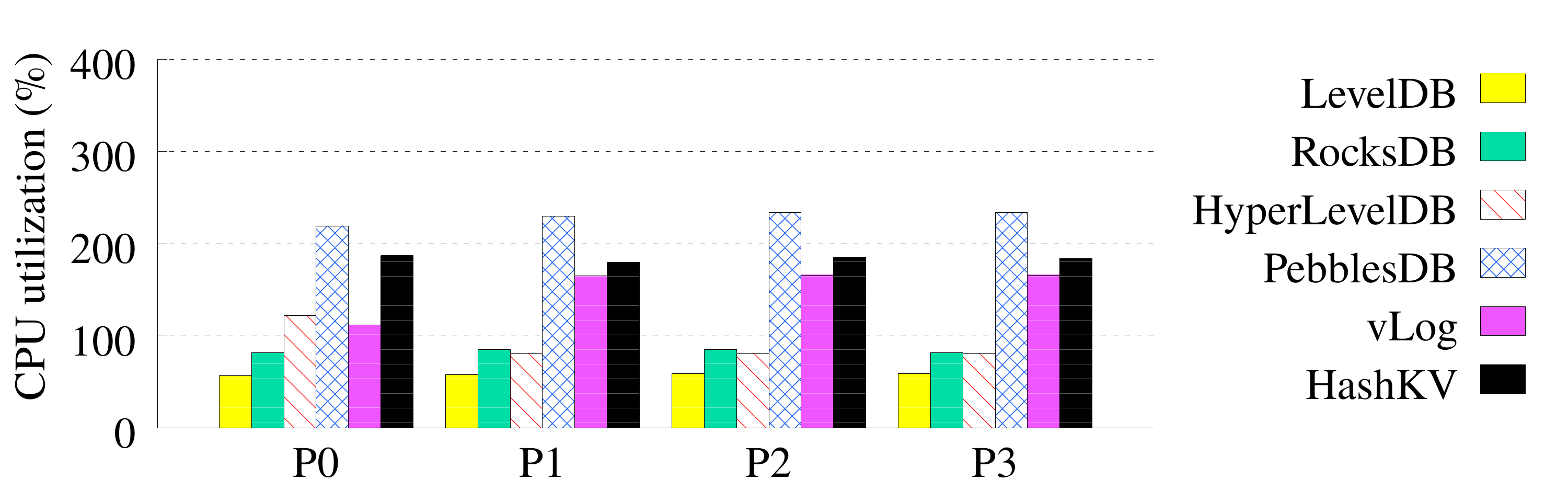}
}\\
\multicolumn{2}{c}{
\mbox{\small (e) CPU utilization}
}
\end{tabular}
\caption{Experiment~1: Performance comparison of KV stores under
update-intensive workloads.}
\label{fig:update}
\end{figure}

\paragraph{Experiment~1 (Load and update performance):} We evaluate LevelDB
(LDB), RocksDB (RDB), HyperLevelDB (HDB), PebblesDB (PDB), vLog, and \HASHKV
(HKV), under update-intensive workloads.  We first compare LevelDB, RocksDB,
vLog, and \HASHKV; later, we also include HyperLevelDB and PebblesDB into our
comparison. 

Figure~\ref{fig:update}(a) shows the performance of each phase.  For vLog and
\HASHKV, the throughput in the load phase is higher than those in the update
phases, as the latter is dominated by the GC overhead.  In the load phase, the
throughput of \HASHKV is 17.1$\times$ and 3.0$\times$ over LevelDB and RocksDB,
respectively.  \HASHKV's throughput is 7.9\% slower than vLog, due to random
writes introduced to distribute KV pairs via hashing.  In the update phases, the
throughput of \HASHKV is 6.3-7.9$\times$, 1.3-1.4$\times$, and 3.7-4.6$\times$
over LevelDB, RocksDB, and vLog, respectively. LevelDB has the lowest
throughput among all KV stores due to significant compaction overhead, while
vLog also suffers from high GC overhead.

Figures~\ref{fig:update}(b) and \ref{fig:update}(c) show the total write sizes
and the KV store sizes of different KV stores after all load and update
requests are issued.  \HASHKV reduces the total write sizes of LevelDB, RocksDB
and vLog by 71.5\%, 66.7\%, and 49.6\%, respectively.  Also, they have very
similar KV store sizes. 

For HyperLevelDB and PebblesDB, both of them have high load and update
throughput due to their low compaction overhead.  For example, PebblesDB
appends fragmented SSTables from the higher level to the lower level, without
rewriting SSTables at the lower level \cite{Raju17}.  Both HyperLevelDB and
PebblesDB achieve at least 2$\times$ throughput of \HASHKV, while incurring
lower write sizes than \HASHKV.  On the other hand, they incur significant
storage overhead, and their final KV store sizes are 2.2$\times$ and
1.7$\times$ over \HASHKV, respectively.  

We further analyze the storage overhead of LevelDB, RocksDB, HyperLevelDB,
PebblesDB, vLog, and \HASHKV under update-intensive workloads. In particular,
we elaborate the reasons on the significant storage overhead observed in
HyperLevelDB and PebblesDB. 

Figure~\ref{fig:update}(d) shows the KV store size of each KV store at the end
of each phase.  At the end of the load phase (Phase P0), the sizes of all KV
stores only differ by at most 4.6\%; in particular, the differences of the KV
store sizes among LevelDB, RocksDB, HyperLevelDB, and PebblesDB are only at
most 1.17\%, which aligns with the space amplification results under the
insertion-only workload in \cite{Raju17}.  Both vLog and \HASHKV have slightly
larger KV store sizes than others, mainly because the keys and metadata are
stored in both the LSM-tree and the value store due to KV separation 
(\S\ref{subsec:storage}).  From Phase~P0 to Phase~P3, the KV store sizes of
LevelDB, RocksDB, vLog, and \HASHKV increase by 42.1\%, 38.9\%, 28.9\%, and
27.0\%, respectively, while the KV store sizes of HyperLevelDB and PebblesDB
increase significantly and are 3.0$\times$ and 2.3$\times$ their sizes at the
end of Phase~P0, respectively.

The reasons of the significant increase in the KV store sizes of HyperLevelDB
and PebblesDB are two-fold.  First, both HyperLevelDB and PebblesDB compact
only selected ranges of keys to reduce write amplification.  Specifically,
HyperLevelDB selects the largest range of keys in the upper level that covers
the smallest range of keys in the lower level to perform compaction, and
places a limit on the total volume of KV pairs in each compaction.  This
reduces the amount of invalid KV pairs being reclaimed from the lower level
during compaction.  PebblesDB divides keys in each level into disjoint ranges
and compacts each range only when its size reaches a predefined threshold. To
enable fast compaction, PebblesDB only partitions KV pairs in the selected
ranges and directly inserts them into the lower level, without removing
invalid KV pairs in the lower level.  This also reduces the amount of invalid
KV pairs being reclaimed. 

Second, HyperLevelDB and PebblesDB trigger much fewer compaction operations
under the update-intensive workloads; for example, their numbers of compaction
operations are only 1.6\% and 0.02\% of that of LevelDB, respectively.  Such
infrequent compaction operations further delay the removal of invalid KV pairs
and hence lead to large KV store sizes.

We emphasize that this experiment only shows the baseline performance of
\HASHKV rather than its best performance. For example, if we allocate more
reserved space (Experiment~2) and issue larger-size KV pairs (Experiment~4),
\HASHKV achieves higher throughput.  Also, we can improve the performance of
\HASHKV by enabling hotness awareness (Experiment~6) and selective KV
separation (Experiment~7).  Furthermore, \HASHKV outperforms HyperLevelDB and
PebblesDB under the YCSB core workloads (Experiment~11).  In the following
experiments, except YCSB benchmarking and the impact of KV separation on
different KV stores, we mainly focus on LevelDB, RocksDB, vLog, and \HASHKV,
as they have comparable storage overhead.

Finally, Figure~\ref{fig:update}(e) shows the CPU utilization of each KV
store in different phases.  Here, we sample the CPU utilization (in
percentage) of each KV store every one second using \texttt{nmon} \cite{nmon},
and plot the median CPU usage.  Since the CPU has four cores, the maximum
achievable CPU utilization is 400\%.  We observe that in the load phase
(Phase~P0), \HASHKV has 75\% higher CPU utilization than vLog.  Our further
investigation finds that the high CPU utilization of \HASHKV is attributed to
the flushing of KV pairs from the write cache to different segments.  In the
update phases, vLog has higher CPU utilization than in its load phase, as its
GC overhead now becomes significant.  In contrast, \HASHKV maintains similar
CPU utilization in both load and update phases, and its CPU utilization is
15-18\% higher than vLog in each update phase.  PebblesDB has the highest CPU
utilization in all phases due to more aggressive compaction, as also reported
in \cite{Raju17}.


\begin{figure}[!t]
\centering
\begin{tabular}{cc}
\includegraphics[width=2.7in]{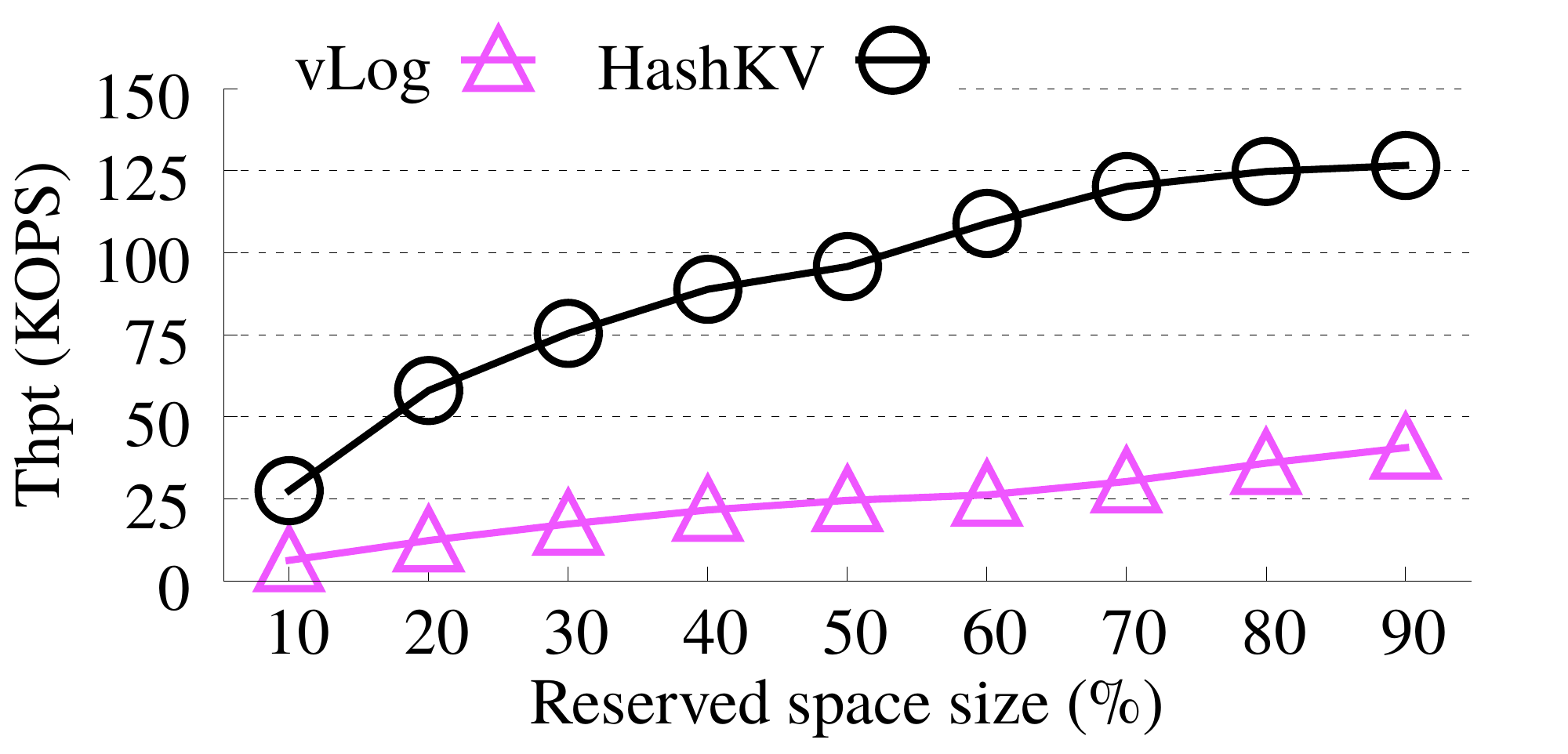} & 
\includegraphics[width=2.7in]{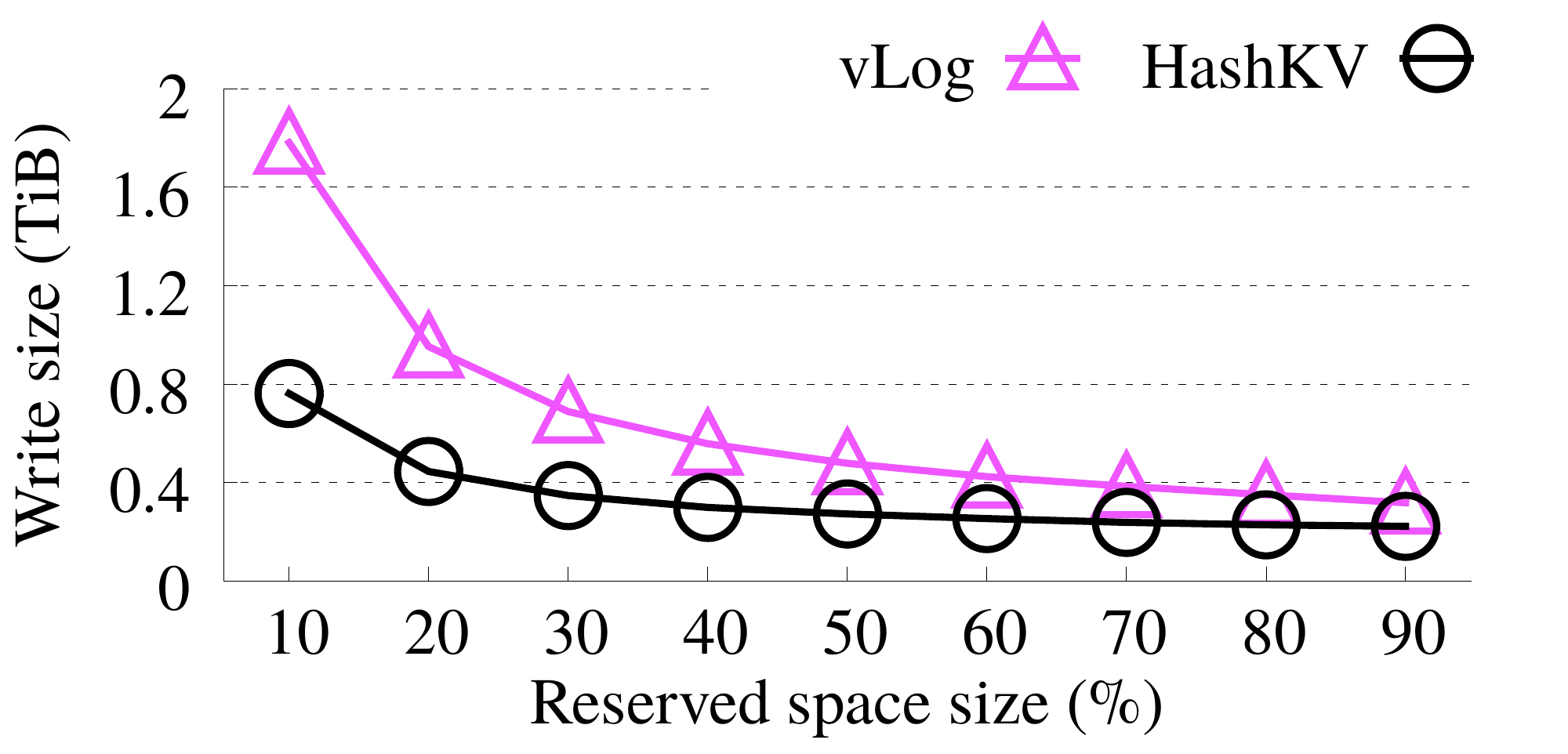}
\\
\mbox{\small (a) Throughput} &
\mbox{\small (b) Total write size} 
\\
\multicolumn{2}{c} {
\includegraphics[width=4.5in]{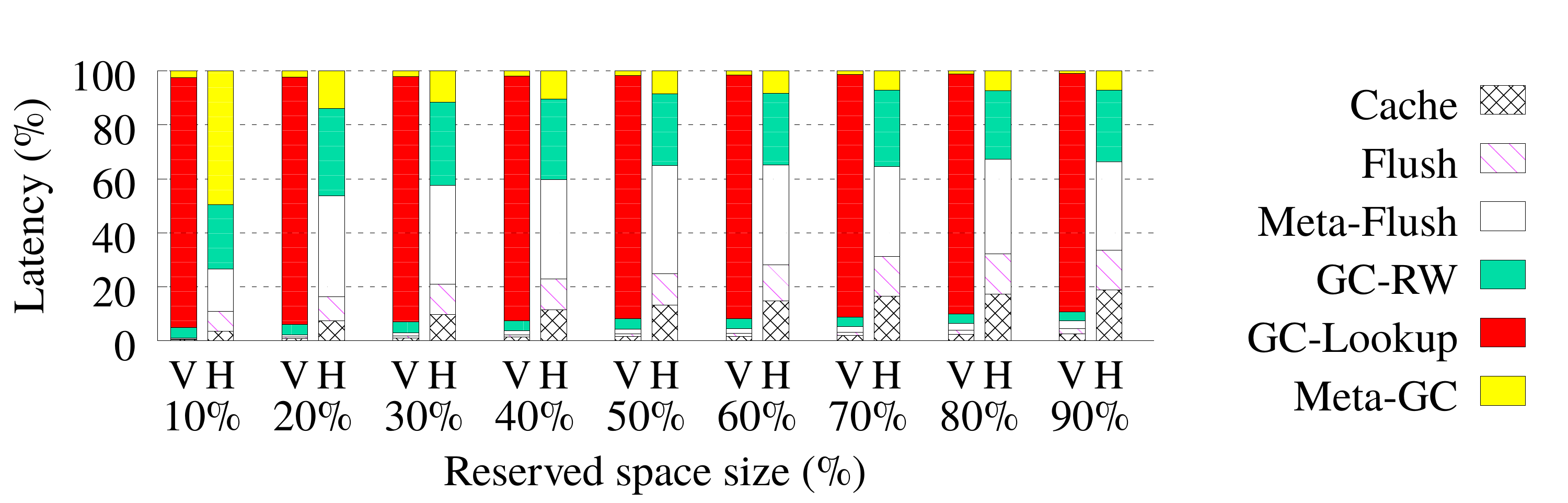}
}\\
\multicolumn{2}{c} {
\mbox{\small (c) Latency breakdown (`V'$=$ vLog; `H'$=$ \HASHKV)}
} 
\end{tabular}
\caption{Experiment~2: Impact of reserved space size.}
\label{fig:reserved}
\end{figure}

\paragraph{Experiment~2 (Impact of reserved space):}
We study the impact of reserved space size on the update performance of vLog
and \HASHKV. We vary the reserved space size from 10\% to 90\% (of 40\,GiB).
Figure~\ref{fig:reserved} shows the performance in Phase~P3,
including the update throughput, the total write size, and the latency
breakdown. Both vLog and \HASHKV benefit from the increase in reserved space.
Nevertheless, \HASHKV achieves 3.1-4.7$\times$ throughput of vLog
(Figure~\ref{fig:reserved}(a)) and reduces the write size of vLog by
30.1-57.3\% (Figure~\ref{fig:reserved}(b)) across different reserved space
sizes.

We further analyze the latency breakdown of the update requests in Phase~P3
for vLog and \HASHKV (Figure~\ref{fig:reserved}(c)).  The breakdown includes
the fractions of time spent on the major steps in an update request,
including: processing the write cache (``Cache''), flushing values from the
write cache (``Flush''), updating metadata in the LSM-tree during flush
(``Meta-Flush''), reading and writing values during GC (``GC-RW''), querying
the LSM-tree during GC (``GC-Lookup''), and updating metadata in the LSM-tree
during GC (``Meta-GC'').  We see that the high GC overhead of vLog is mainly
attributed to the queries of the LSM-tree for checking the validity of KV
pairs (i.e., ``GC-Lookup''), and such queries account for over 80\% of the
overall update latency.  On the other hand, \HASHKV eliminates this LSM-tree
query overhead in GC.  Furthermore, we observe that \HASHKV spends less time
on updating metadata during GC (i.e., ``Meta-GC'') in the LSM-tree with the
increasing reserved space size due to less frequent GC operations.

\begin{table}[t]
\centering
\begin{tabular}{c|@{\ }c|@{\ }c|@{\ }c|@{\ }c|@{\ }c|@{\ }c|@{\ }c|@{\ }c|@{\ }c}
\hline
Reserved space size		& 10\%    & 20\%  & 30\%  & 40\%  & 50\%  & 60\%  & 70\%  & 80\%  & 90\%  \\
\hline \hline
\textbf{vLog}         & 160.70  & 81.17 & 57.57 & 46.19 & 40.60 & 38.04 & 32.86 & 27.72 & 24.46 \\
\hline
\textbf{\HASHKV}      & 36.21   & 17.05 & 13.07 & 11.07 & 10.25 & 8.99  & 8.14  & 7.83  & 7.72  \\
\hline
\end{tabular}
\vspace{6pt}
\caption{Experiment~2: Average update latencies (in microseconds) of vLog and
\HASHKV.}
\label{table:reserved}
\end{table}

Table~\ref{table:reserved} shows the average update latencies of vLog and
\HASHKV in Phase~P3.  The update latencies of both vLog and \HASHKV decrease
when the reserved space size increases, due to less frequent GC operations.
The update latency of \HASHKV is generally lower than that of vLog for
different reserved space sizes.

\begin{figure}[!t]
\centering
\begin{tabular}{cc}
\includegraphics[width=2.1in]{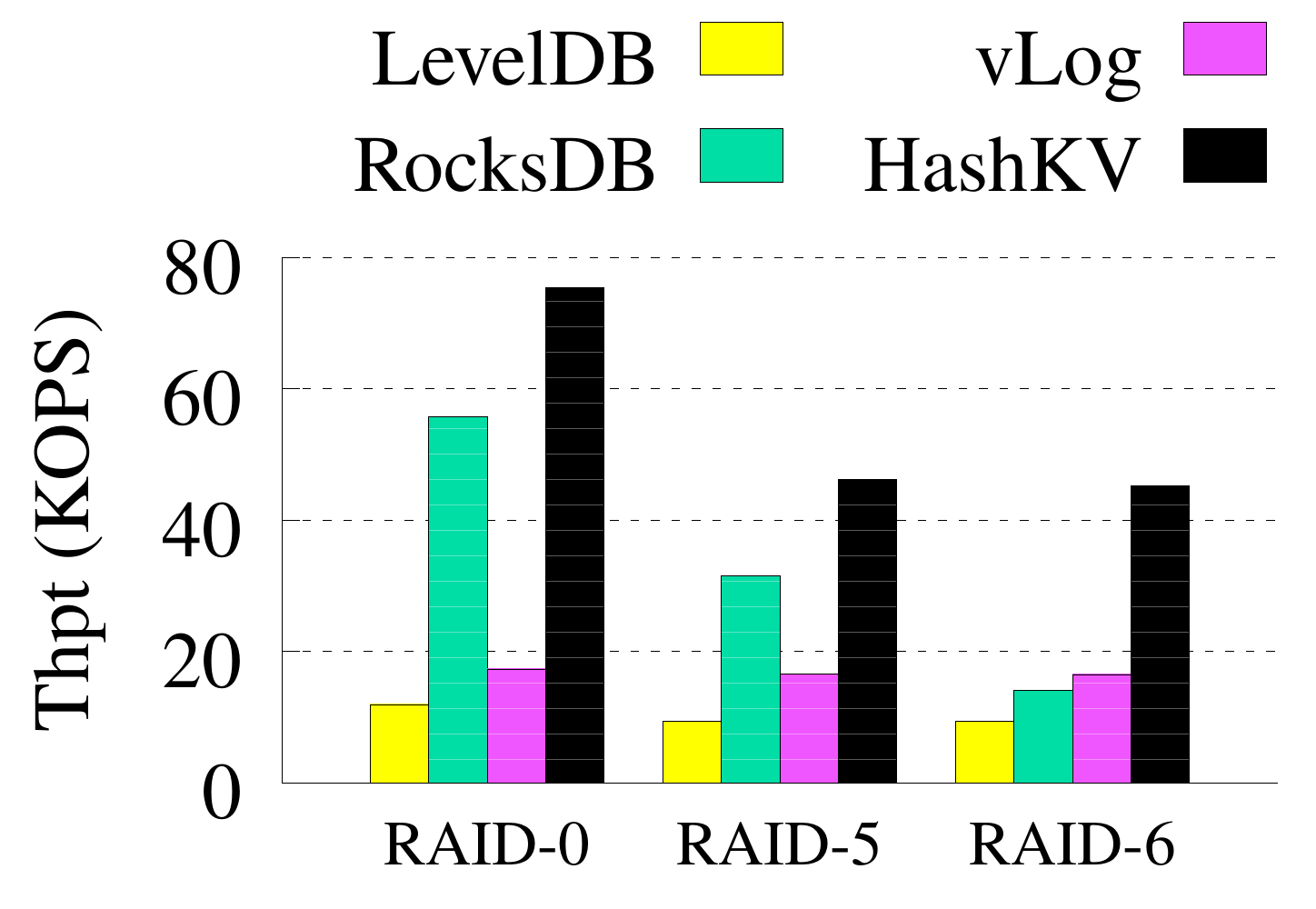} &
\includegraphics[width=2.1in]{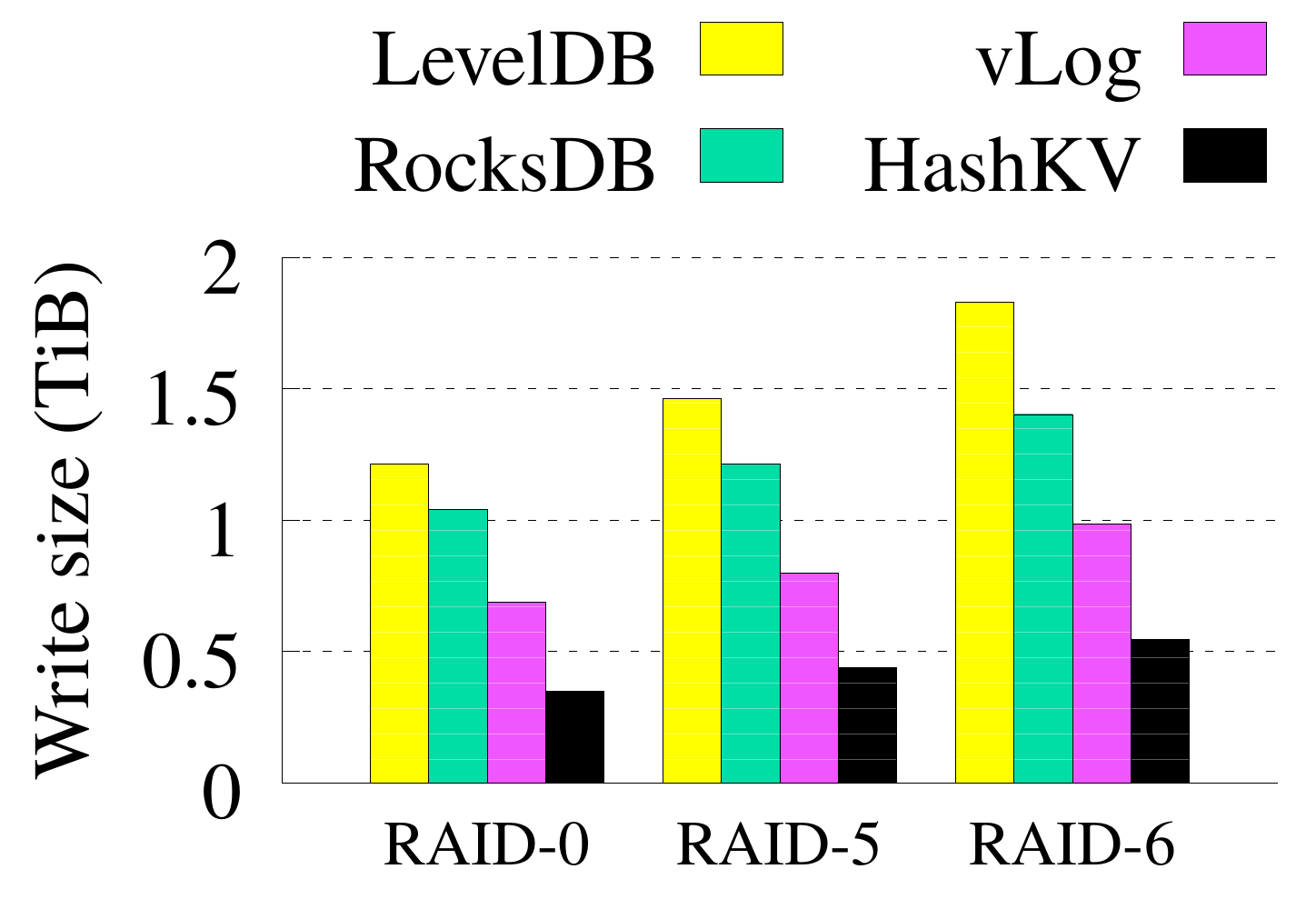} 
\\
{(a) Throughput} &
{(b) Total write size}
\end{tabular}
\caption{Experiment~3: Different RAID configurations.}
\label{fig:raid}
\end{figure}

\paragraph{Experiment~3 (Impact of parity-based RAID):}
We evaluate the impact of the fault tolerance configuration of RAID on the
update performance of LevelDB, RocksDB, vLog, and \HASHKV. We configure the
RAID volume to run two parity-based RAID schemes, RAID-5 (single-device fault
tolerance) and RAID-6 (double-device fault tolerance). We include the
results under RAID-0 for comparison.
Figure~\ref{fig:raid} shows the throughput in Phase P3 and the total write
size. RocksDB and \HASHKV are more sensitive to RAID configurations (larger
drops in throughput), since their performance is write-dominated. Nevertheless,
the throughput of \HASHKV is higher than other KV stores under parity-based
RAID schemes, e.g., 4.8$\times$, 3.2$\times$, and 2.7$\times$ over
LevelDB, RocksDB, and vLog, respectively, under RAID-6. The write sizes of KV
stores under RAID-5 and RAID-6 increase by around 20\% and 50\%, respectively,
compared to RAID-0, which match the amount of redundancy of the corresponding
parity-based RAID schemes.

\subsection{Performance under Different Workloads}
\label{subsec:exp_diff_workloads}

We now study the update and range scan performance of \HASHKV for different KV
pair sizes. 

\begin{figure}[t]
\centering
\begin{tabular}{cc}
\includegraphics[width=2.6in]{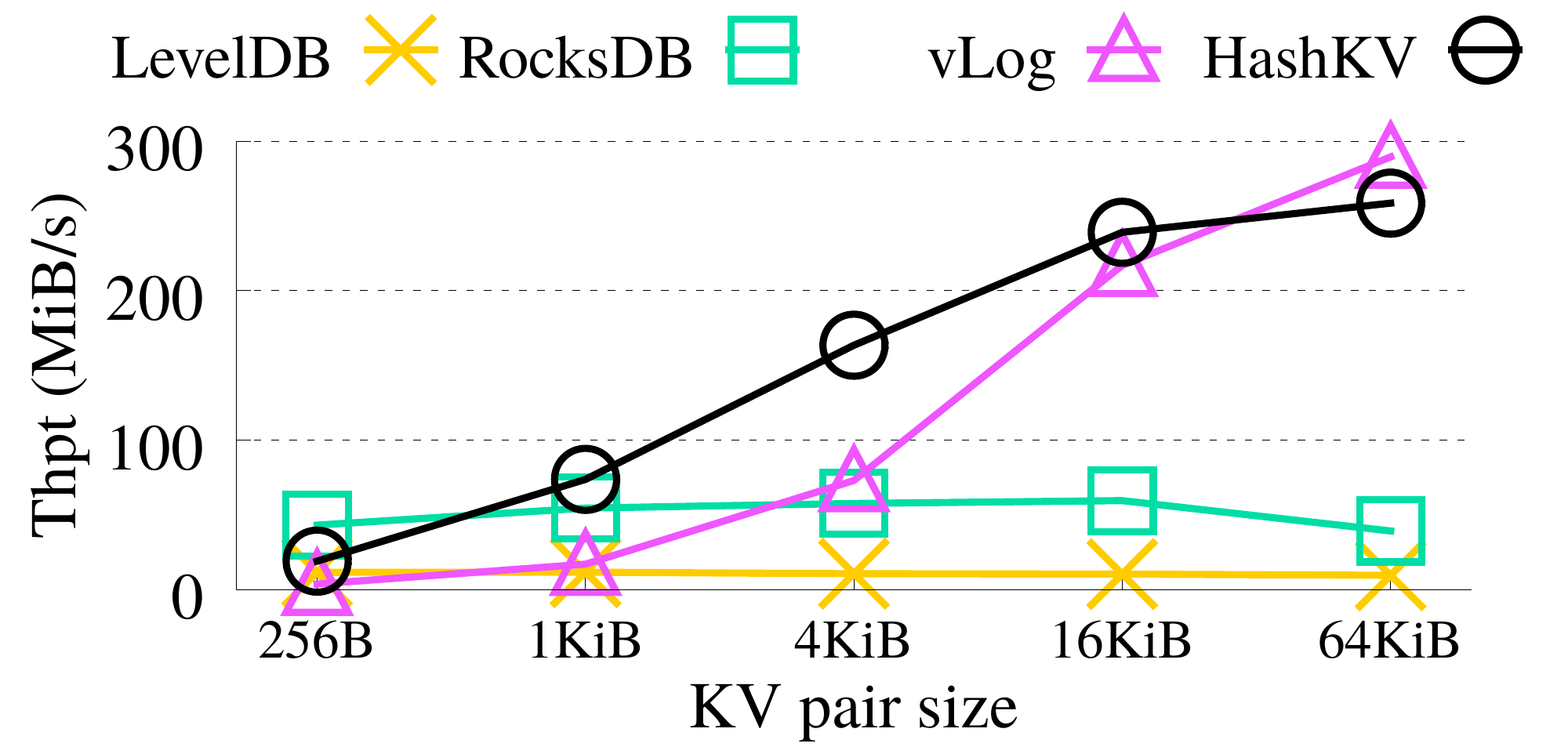} &
\includegraphics[width=2.6in]{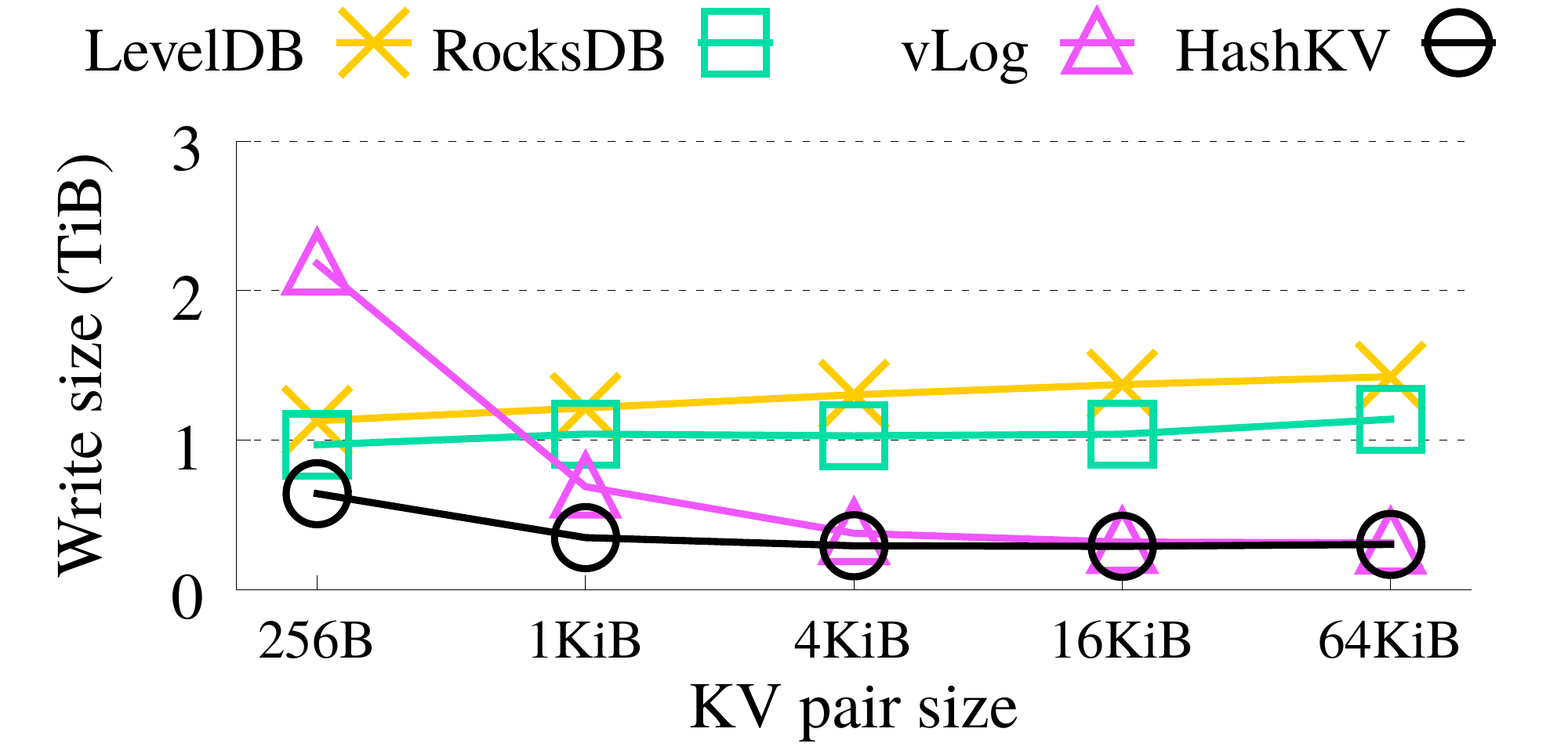} \\
\mbox{\small (a) Throughput} &
\mbox{\small (b) Total write size}
\end{tabular}
\caption{Experiment~4: Update performance versus the KV pair size.}
\label{fig:kvsize}
\end{figure}

\paragraph{Experiment~4 (Impact of KV pair size):}
We study the impact of KV pair sizes on the update performance of KV stores. We
vary the KV pair size from 256\,B to 64\,KiB.  Specifically, we increase the
KV pair size by increasing the value size and keeping the key size fixed at
24\,B.  We also reduce the number of KV pairs loaded or updated, so as to keep
the total size of KV pairs fixed at 40\,GiB.  Figure~\ref{fig:kvsize}
shows the update performance of KV stores in Phase P3 versus the KV pair size. 
The throughput of LevelDB and RocksDB remains similar across most KV pair
sizes, while the throughput of vLog and \HASHKV increases as the KV pair size
increases. Both vLog and \HASHKV have lower throughput than 
RocksDB when the KV pair size is 256\,B, since the overhead of writing small
values to the value store is more significant. Nevertheless, \HASHKV can
benefit from selective KV separation (Experiment~7).  As the KV pair size
increases, \HASHKV also sees increasing throughput. For example, \HASHKV
achieves 15.5$\times$ and 2.8$\times$ throughput over LevelDB and RocksDB,
respectively, for 4-KiB KV pairs. \HASHKV achieves 2.2-5.1$\times$ throughput
over vLog for KV pair sizes between 256\,B and 4\,KiB.  The performance gap
between vLog and \HASHKV narrows as the KV pair size increases, since the size
of the LSM-tree decreases with fewer KV pairs. Thus, the queries to the
LSM-tree of vLog are less expensive.  For 64-KiB KV pairs, \HASHKV has 10.7\%
less throughput than vLog.

When the KV pair size increases, the total write sizes of LevelDB and RocksDB
increase due to the increasing compaction overhead, while those of \HASHKV and
vLog decrease due to fewer KV pairs in the LSM-tree.  Overall, \HASHKV reduces
the total write sizes of LevelDB, RocksDB, and vLog by 43.2-78.8\%,
33.8-73.5\%, and 3.5-70.6\%, respectively.

\begin{figure}[!t]
\centering
\begin{tabular}{cc}
\includegraphics[width=2.6in]{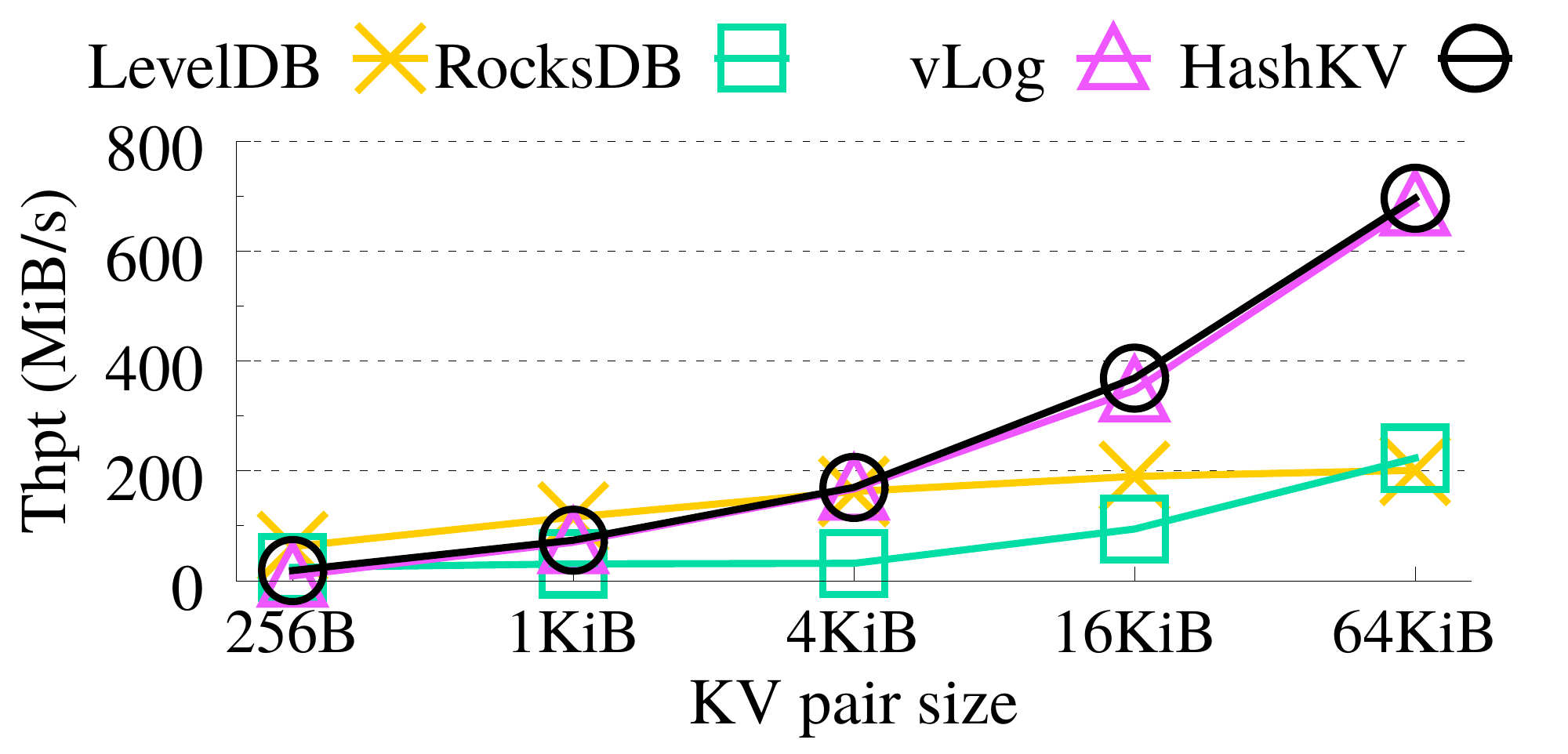} &
\includegraphics[width=2.6in]{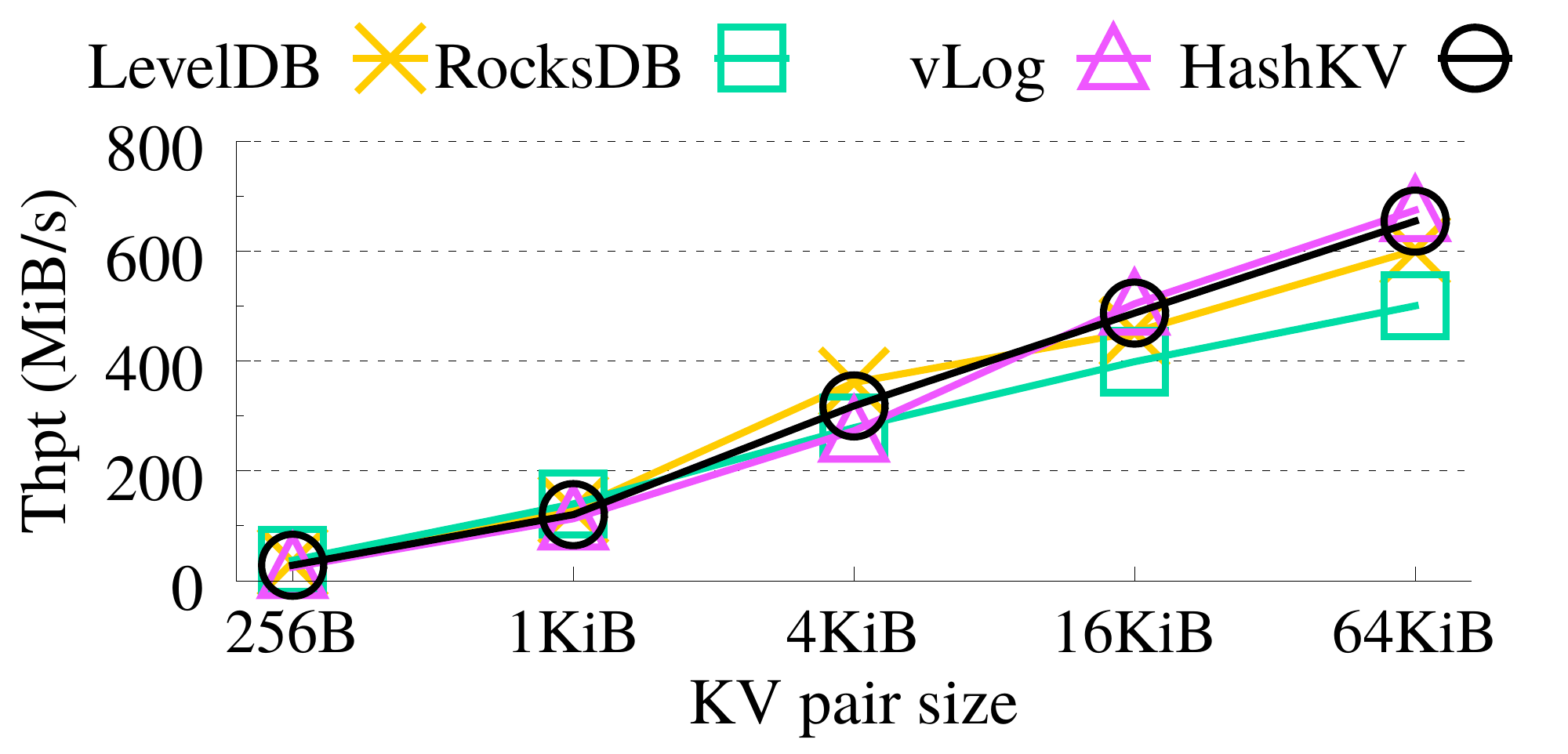} \\
\mbox{\small (a) After the load phase} &
\mbox{\small (b) After the update phases}
\end{tabular}
\caption{Experiment~5: Range scan performance versus the KV pair size.}
\label{fig:scan}
\end{figure}

\paragraph{Experiment~5 (Range scans):} We compare the range scan performance
of KV stores for different KV pair sizes.  Specifically, we first load 40\,GiB
of fixed-size KV pairs, and then issue scan requests whose start keys follow a
Zipf distribution with a Zipfian constant of 0.99. Each scan request reads
1\,MiB of KV pairs, and the total scan size is 4\,GiB.
Figure~\ref{fig:scan}(a) shows the results. \HASHKV has similar scan
performance to vLog across KV pair sizes. However, \HASHKV has 70.0\% and
36.3\% lower scan throughput than LevelDB for 256-B and 1-KiB KV pairs,
respectively, mainly because \HASHKV needs to issue reads to both the LSM-tree
and the value store and there is also high overhead of retrieving small values
from the value store via random reads.  Nevertheless, for KV pairs of 4\,KiB
or larger, \HASHKV outperforms LevelDB, e.g., by 94.2\% for 4-KiB KV pairs.
The lower scan performance for small KV pairs is also consistent with that of
WiscKey (see Figure~12 in \cite{Lu16}).  Note that the read-ahead mechanism
(\S\ref{subsec:range}) is critical to enabling \HASHKV to achieve high
range scan performance.  For example, the range scan throughput of \HASHKV
increases by 81.0\% for 256-B KV pairs compared to without read-ahead (not
shown in figures).  

We further study the range scan performance of KV stores after
update-intensive workloads. Specifically, we load 40\,GiB of fixed-size KV
pairs and run three phases of 40\,GiB of updates as in
\S\ref{subsec:exp_perf}.  We then issue the same range scan workload as above.
In particular, before issuing scan requests, we perform a manual LSM-tree
compaction on all KV pairs. 
Figure~\ref{fig:scan}(b) shows the results.  All KV stores achieve similar
scan throughput across KV pair sizes.  When compared to the scan performance
after the load phase (Figure~\ref{fig:scan}(a)), 
\HASHKV preserves its high scan performance after
updates, while both LevelDB and RocksDB see improved scan performance.  The
performance gains of LevelDB and RocksDB are attributed to the manual
compaction before we issue scans. The manual compaction compacts all KV pairs
to the same level, and ensures that the key ranges of all SSTables are
disjoint. This saves the overhead of searching through SSTables with
overlapped key ranges in order to determine the next smallest key during
scans.

\subsection{\HASHKV Features}
\label{subsec:exp_opt}

We study the two optimization techniques of \HASHKV, hotness awareness and
selective KV separation, as well as the GC and crash consistency mechanisms of
\HASHKV. We mainly report the throughput in Phase P3 and the total write size
using the same update-intensive workloads in
\S\ref{subsec:exp_perf}.  In Experiments~6-7, we configure 20\% of reserved
space to show that the optimized performance of smaller reserved space can
match the unoptimized performance of larger reserved space.

\begin{figure}[t]
\centering
\begin{tabular}{cc}
\includegraphics[width=2.1in]{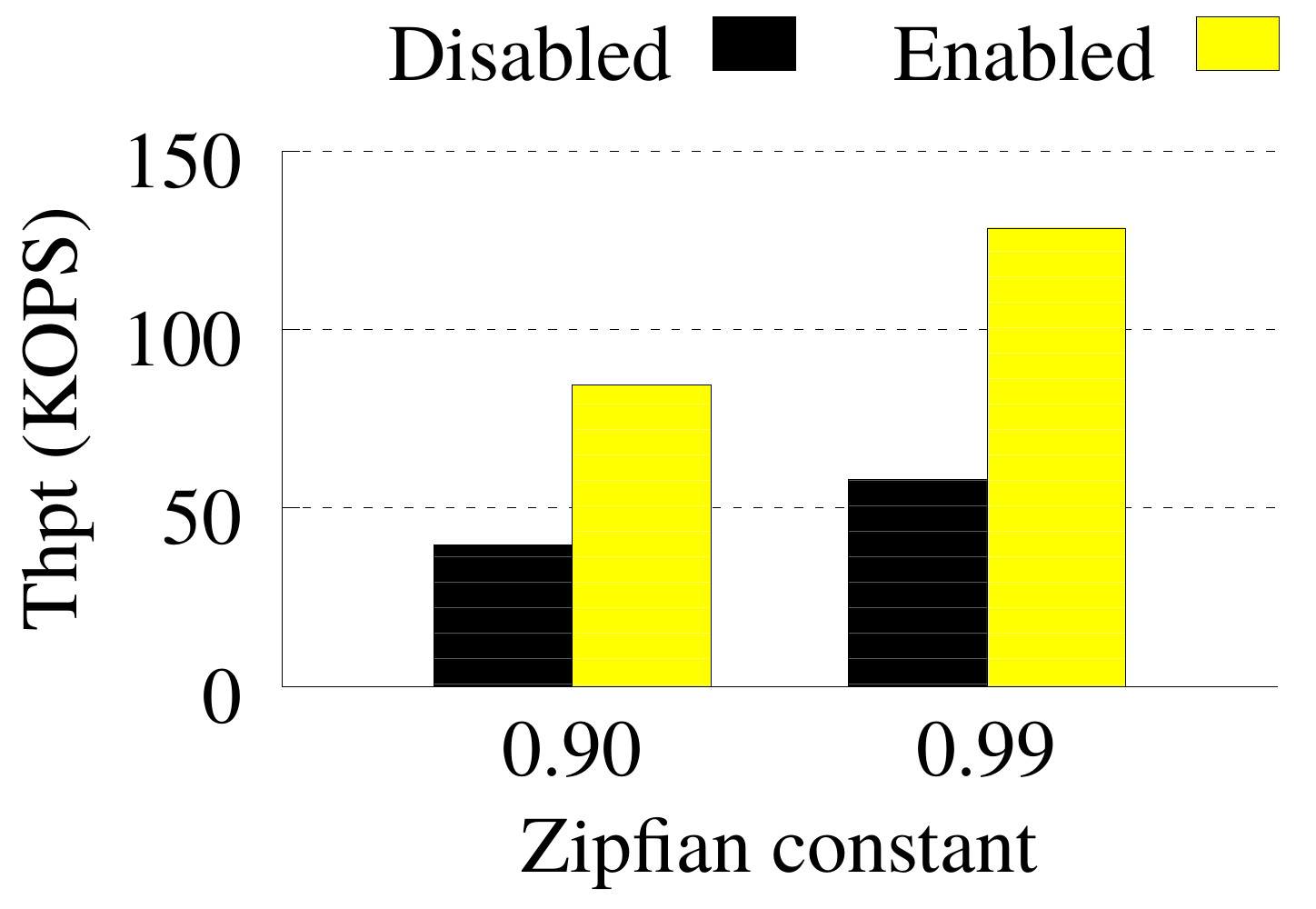} &
\includegraphics[width=2.1in]{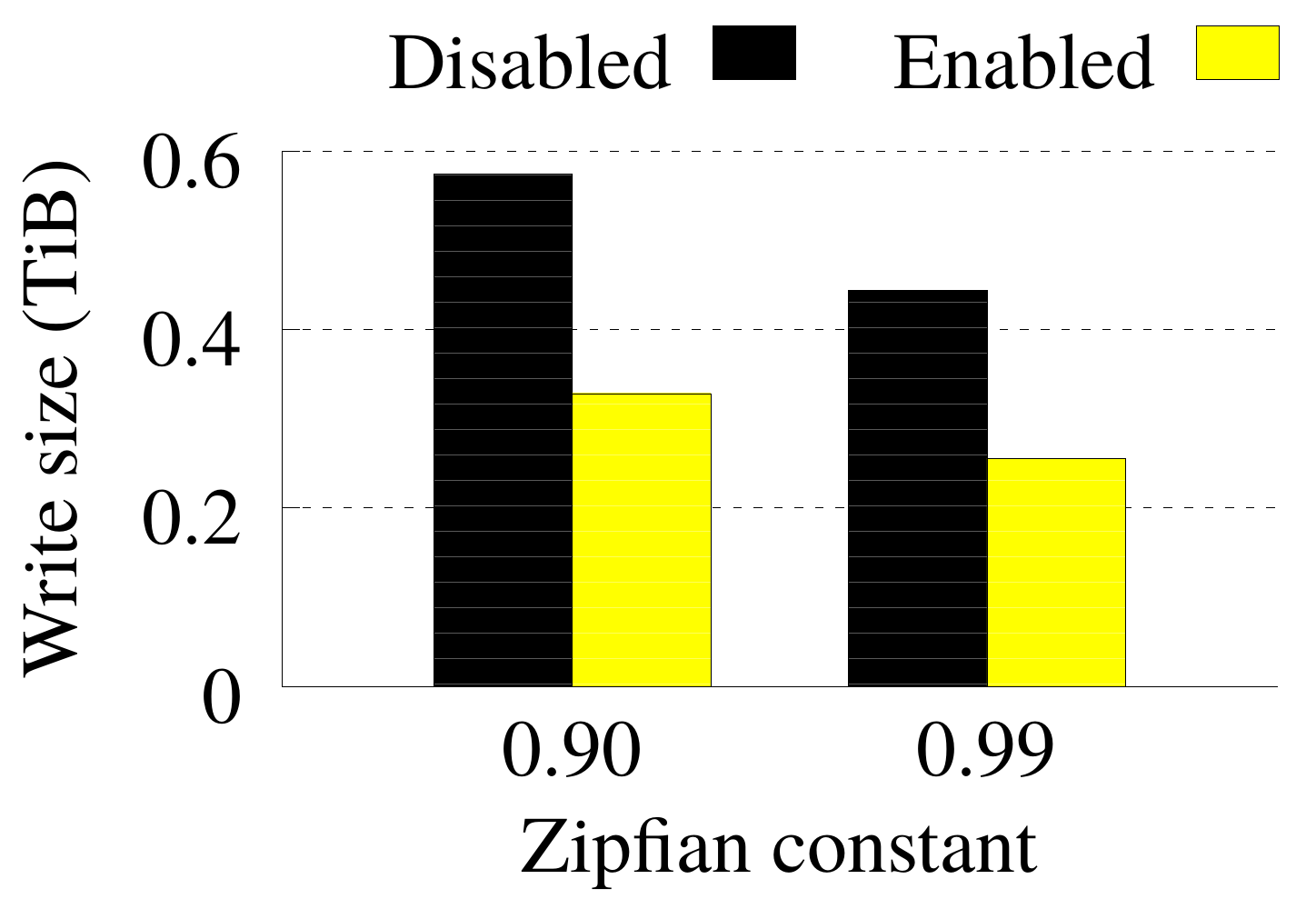} \\
\mbox{\small (a) Throughput} &
\mbox{\small (b) Total write size}
\end{tabular}
\caption{Experiment~6: Hotness awareness.}
\label{fig:hotcold}
\end{figure}

\paragraph{Experiment~6 (Hotness awareness):}
We evaluate the impact of hotness awareness on the update performance of
\HASHKV. We consider two Zipfian constants, 0.9 and 0.99, to capture different
skewness in workloads.
Figure~\ref{fig:hotcold} shows the results when hotness awareness is disabled
and enabled. When hotness awareness is enabled, the update throughput increases
by 113.1\% and 121.3\%, while the write size reduces by 42.8\% and 42.5\%, for
Zipfian constants 0.9 and 0.99, respectively.

\begin{figure}[!t]
\centering
\begin{tabular}{cc}
\includegraphics[width=2.1in]{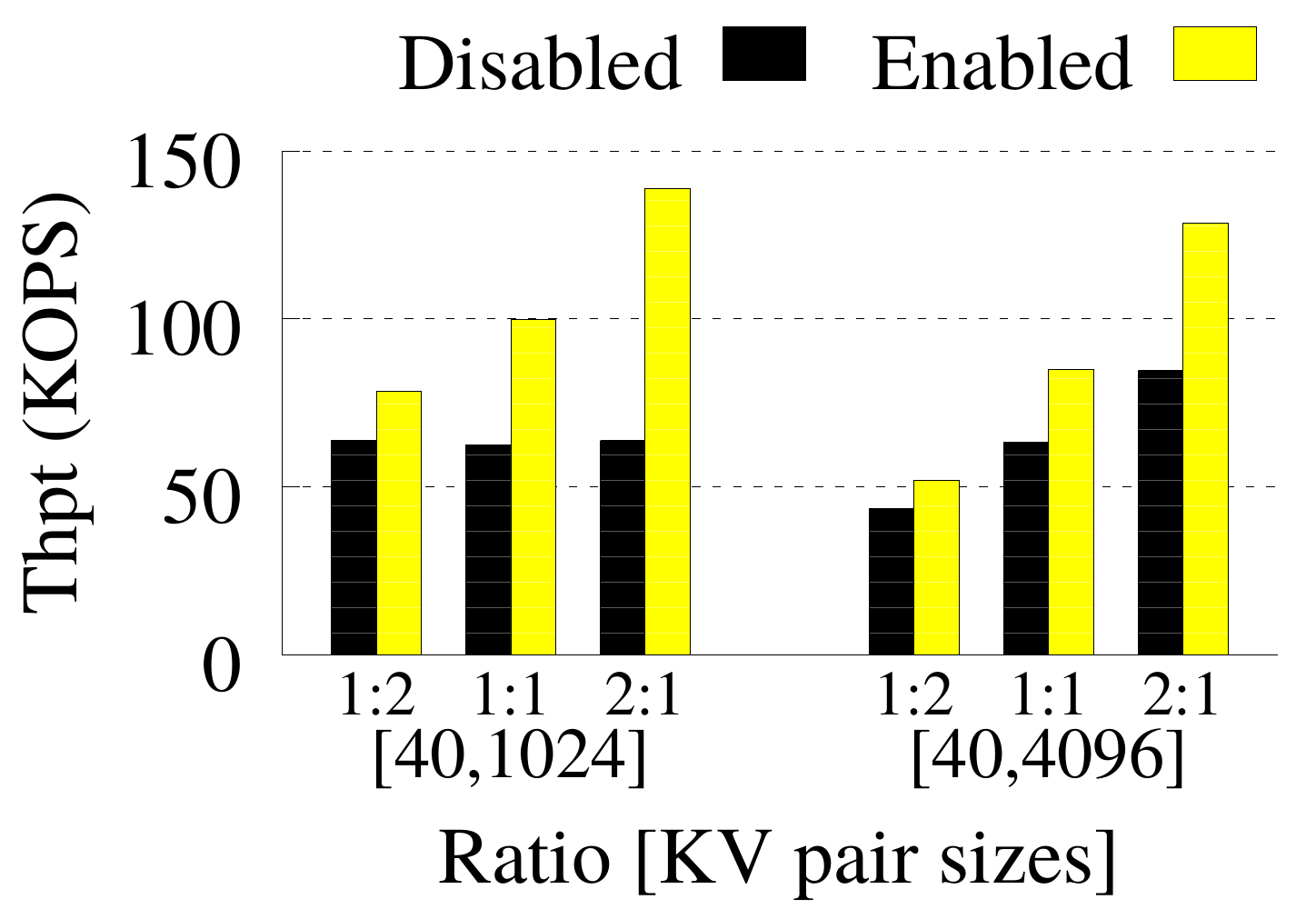} &
\includegraphics[width=2.1in]{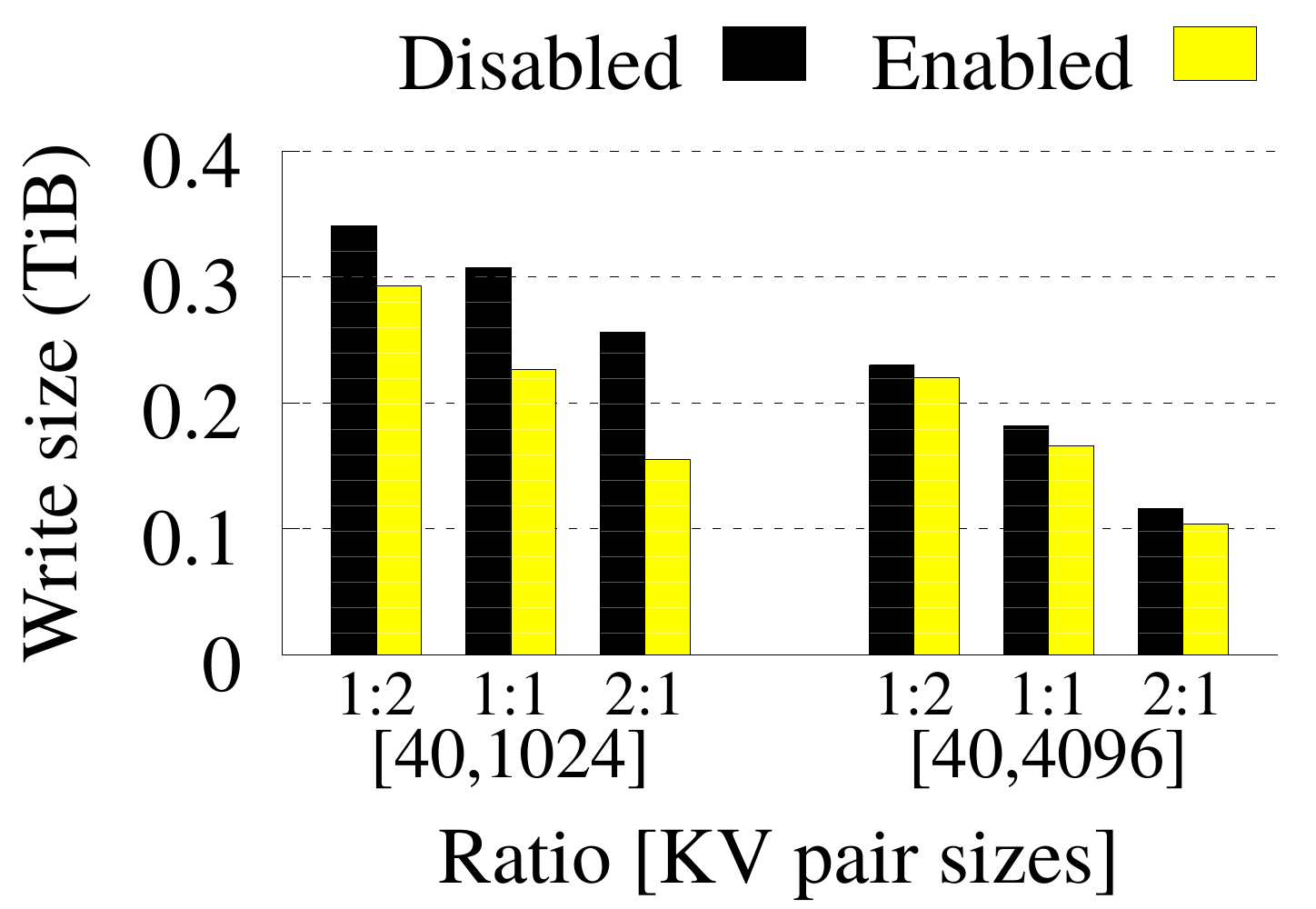} \\
\mbox{\small (a) Throughput} &
\mbox{\small (b) Total write size}
\end{tabular}
\caption{Experiment~7: Selective KV separation.}
\label{fig:kvsep}
\end{figure}

\paragraph{Experiment~7 (Selective KV separation):}
We evaluate the impact of selective KV separation on the update performance of
\HASHKV.  We consider three ratios of small-to-large KV pairs, including 1:2,
1:1, and 2:1. We set the small KV pair size as 40\,B, and the large KV
pair size as 1\,KiB or 4\,KiB.  Recall that we set the threshold of selective
KV separation as 192\,B by default (\S\ref{subsec:selective_kv_sep}), so
when selective KV separation is enabled, the small KV pairs are stored
entirely in the LSM-tree, while the large KV pairs are stored via KV
separation.  
Figure~\ref{fig:kvsep} shows the results when selective KV separation is
disabled or enabled. When selective KV separation is enabled, the throughput
increases by 23.2-118.0\% and 19.2-52.1\% when the large KV pair size is
1\,KiB and 4\,KiB, respectively.  We observe higher performance gain for
workloads with a higher ratio of small KV pairs, due to the high update
overhead of small KV pairs stored under KV separation.  Also, selective KV
separation reduces the total write size by 14.1-39.6\% and 4.1-10.7\% when the
large KV pair size is 1\,KiB and 4\,KiB, respectively.

\begin{figure}[!t]
\centering
\begin{tabular}{c@{\ }c}
\includegraphics[width=3.35in]{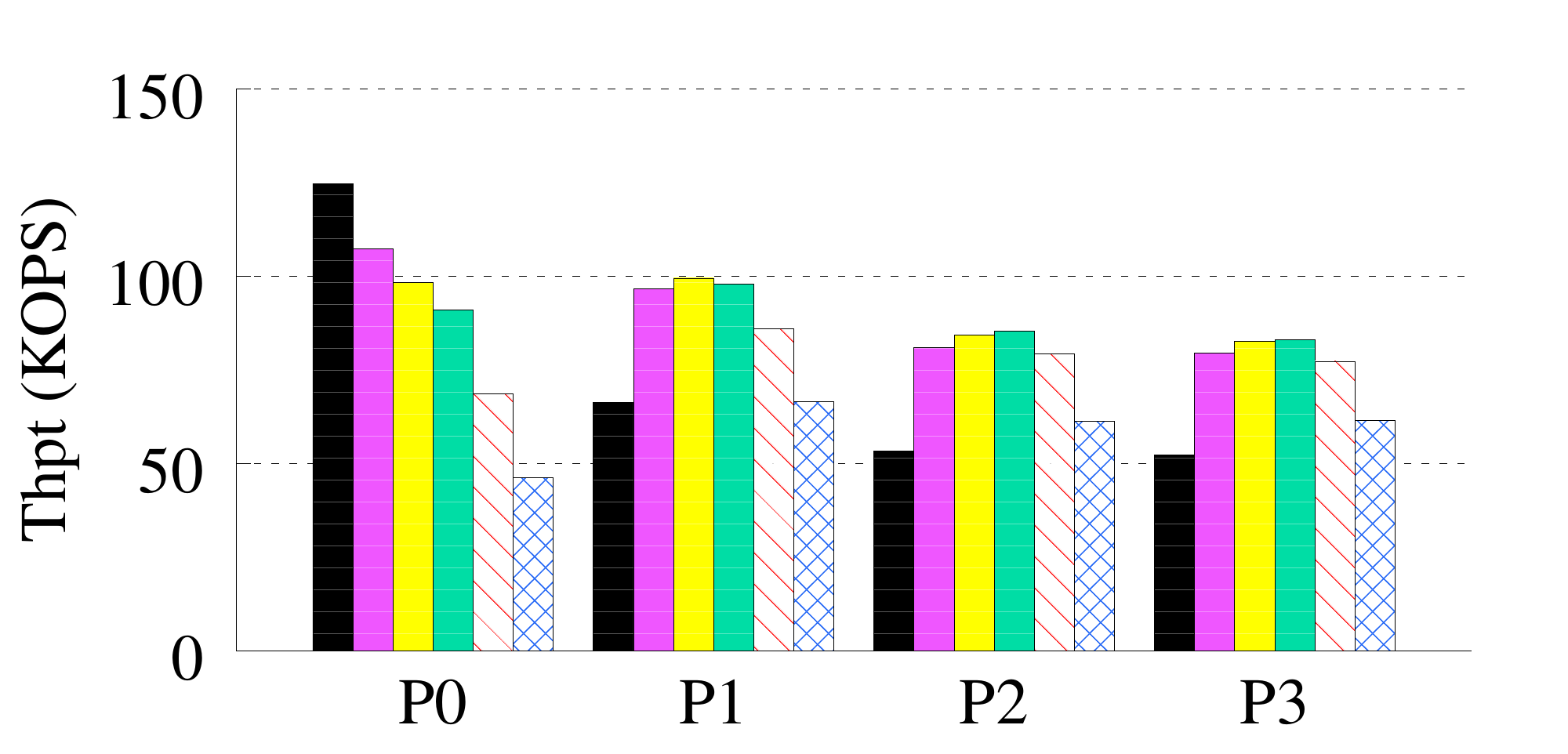} &
\includegraphics[width=2.15in]{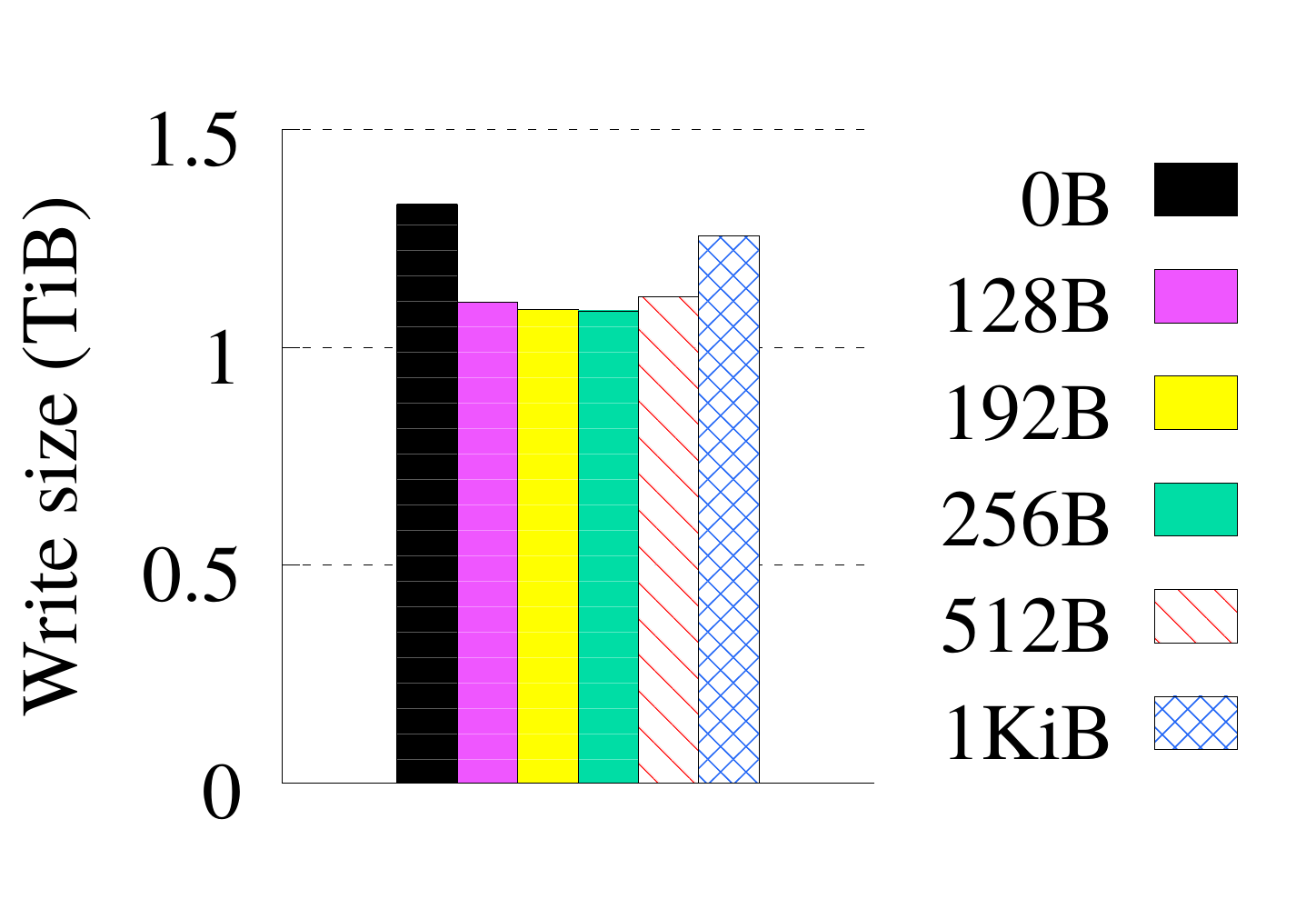} \\
\mbox{\small (a) Throughput} &
\mbox{\small (b) Total write size} 
\end{tabular}
\caption{Experiment~8: Threshold selection of selective KV separation in
\HASHKV under update-intensive workloads.}  
\label{fig:kvsep_120g}
\end{figure}


\paragraph{Experiment~8 (Threshold selection of selective KV separation):} We
study the impact of the KV pair size threshold on the performance of selective
KV separation.  We consider different thresholds; when the threshold is 0\,B,
it implies that KV separation is always used.  We modify the update-intensive 
workloads in \S\ref{subsec:exp_perf} by generating KV pairs of different sizes
ranging from 40\,B to 8\,KiB based on a Zipf distribution with a Zipfian
constant of 0.99 (note that we have also evaluated the case where the KV pairs
have sizes ranging from 40\,B to 16\,KiB and observed similar results).  To
ensure that different KV pair sizes are sufficiently covered, we increase the
volume of KV pairs being inserted or updated to 120\,GiB in each phase (i.e.,
3$\times$ of the data volume in the original update-intensive workloads in
\S\ref{subsec:exp_perf}), and configure 120\,GiB of storage space with 30\% of
reserved space. 

Figures~\ref{fig:kvsep_120g}(a) shows the throughput of \HASHKV under the
update-intensive workloads. In the load phase (Phase~P0), the throughput
decreases when the threshold increases, since \HASHKV loads more KV pairs into
LevelDB and triggers more compaction operations.  In the update phases
(Phases~P1-P3), the throughput of \HASHKV first increases and then decreases
with the increasing threshold.  Note that \HASHKV reaches the maximum update
throughput when the threshold is in the range from 128\,B to 256\,B, in which
the update throughput is at least 40\% higher compared to the threshold 0\,B
(i.e., KV separation is always used).  Also, the throughput is similar for the
thresholds of 128\,B, 192\,B, and 256\,B.  We further evaluate the impact of
the threshold selection via YCSB benchmarking in \S\ref{subsec:exp_ycsb}. 
		
Figure~\ref{fig:kvsep_120g}(b) shows the total write size of \HASHKV under the
update-intensive workloads.  When the threshold is in the range from 128\,B to
512\,B, \HASHKV reduces the write size by 16.9-18.4\%.  


\paragraph{Experiment~9 (Impact of GC approaches):} We study the
impact of different GC approaches on the performance of \HASHKV 
(\S\ref{subsec:gc}).  We compare three variants of GC approaches with our
default greedy approach: (i) CBA, (ii) GRA with $d=5$, and (iii) the random
approach (i.e., GRA with $d$ set to the total number of segment groups).  We
evaluate the performance of \HASHKV under the update-intensive workloads, and
consider two Zipfian constants, 0.9 and 0.99, as in Experiment~7.

\begin{figure}[!t]
\centering
\begin{tabular}{cc}
\includegraphics[width=2.1in]{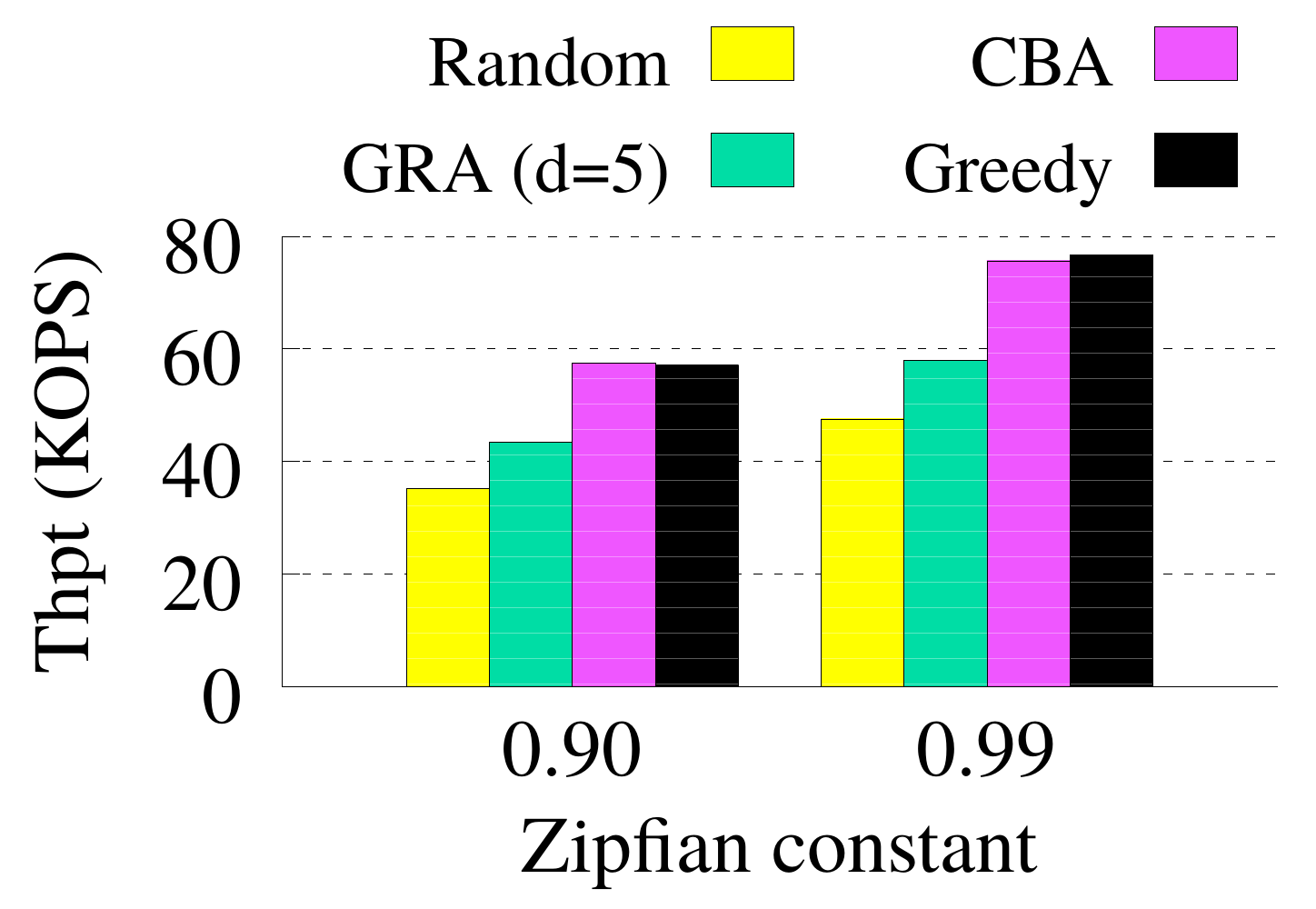} &
\includegraphics[width=2.1in]{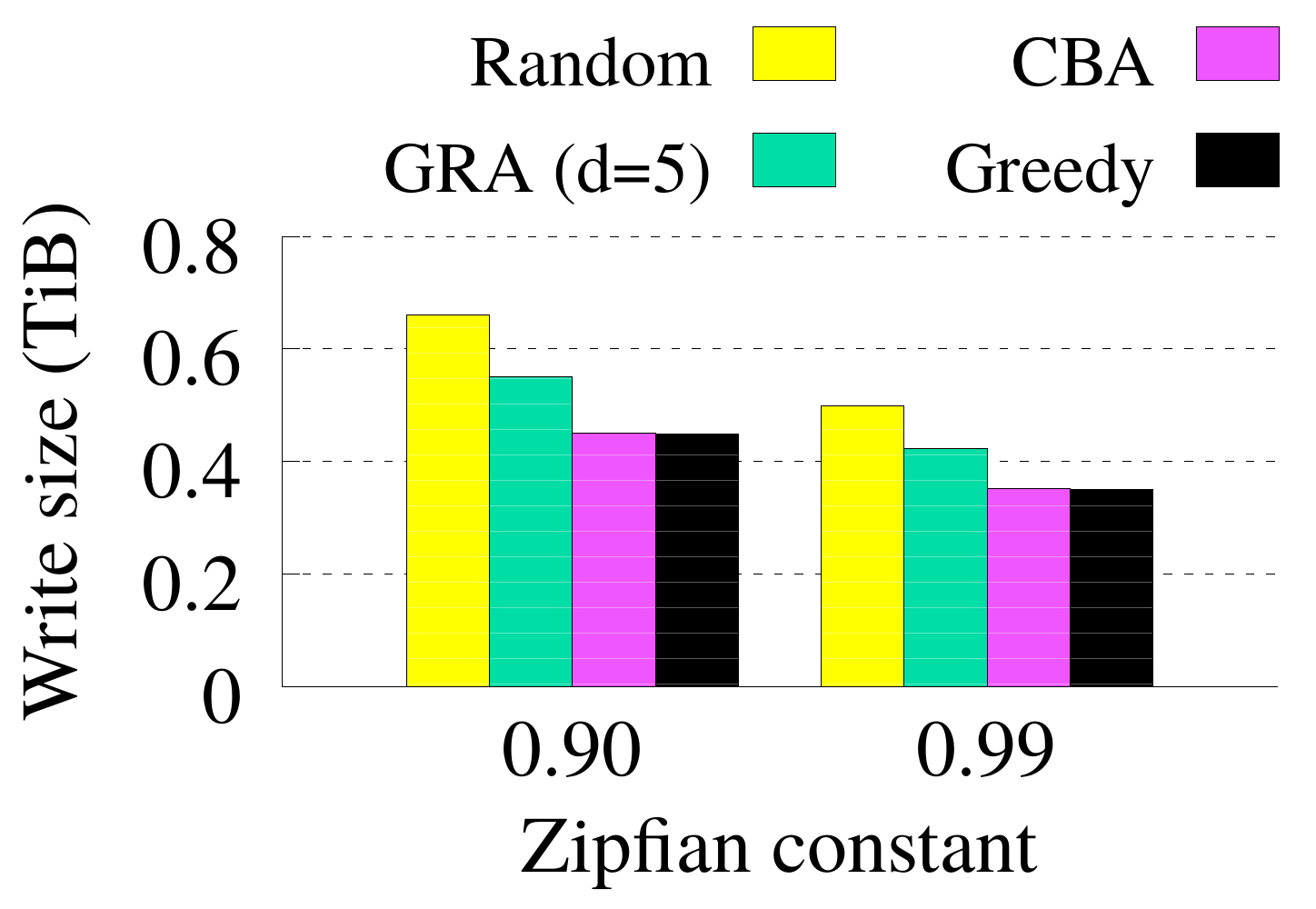} \\
\mbox{\small (a) Throughput} &
\mbox{\small (b) Total write size} 
\end{tabular}
\caption{Experiment~9: Impact of GC approaches.}
\label{fig:gc_policy}
\end{figure}

Figure~\ref{fig:gc_policy}(a) shows the update throughput of \HASHKV using 
different GC approaches under the update-intensive workloads. The update
throughput of \HASHKV using the greedy approach is similar to that using CBA,
and we find that the priority of choosing which segment group for GC is mainly
determined by the amount of free space that can be reclaimed
rather than the segment age.  The update throughput of \HASHKV using the
greedy approach is 61.3-62.2\% and 31.5-32.1\% higher than that using the
Random and GRA approaches, respectively.

Figure~\ref{fig:gc_policy}(b) shows the total write size of \HASHKV using
different GC approaches under the update-intensive workloads. Again, both the
greedy and CBA approaches have similar write sizes.  The greedy approach
reduces the write size of using the random and GRA approaches by 30.1-32.0\%
and 17.5-18.4\%, respectively.  

Our results confirm that the default greedy approach in \HASHKV can achieve
high update throughput and reduce the total write size. 


\paragraph{Experiment~10 (Crash consistency):}
We study the impact of the crash consistency mechanism on the performance of
\HASHKV. Table~\ref{table:consistency} shows the results.  When the crash
consistency mechanism is enabled, the update throughput of \HASHKV in Phase~P3
reduces by 6.5\% and the total write size increases by 4.2\%, which shows that
the impact of crash consistency mechanism remains limited. Note that we verify
the correctness of the crash consistency mechanism by crashing \HASHKV via
code injection and unexpected terminations during runtime.

\begin{table}[t]
\centering
\begin{tabular}{c|@{\ }c|@{\ }c}
\hline
							  & \textbf{Disabled} & \textbf{Enabled} \\
\hline \hline
\textbf{Throughput (KOPS)}     & 58.0              & 54.3  \\
\hline
\textbf{Total write size (GiB)} & 454.6             & 473.7 \\
\hline
\end{tabular}
\vspace{7pt}
\caption{Experiment~10: Performance of \HASHKV with crash consistency
disabled and enabled.}
\label{table:consistency}
\end{table}

\subsection{Performance under YCSB Core Workloads}
\label{subsec:exp_ycsb}

We study the performance of \HASHKV under the default YCSB core workloads
\cite{Cooper10} (Table~\ref{table:ycsb}).  We do not consider Workload~E,
which is about range scans, and we will specifically study the effects of range
scans for different KV pair sizes in Experiment~13 (see also Experiment~5).
Here, we focus on the aged KV stores that have executed a large number of
updates.  Specifically, before running each YCSB workload, we first run
Phases~P0-P2 on each KV store; in this case, both vLog and \HASHKV have
started to trigger GC in their value stores.  We then run each YCSB workload
based on the storage layout after issuing the update-intensive workloads, and
fix the KV pair size as 1\,KiB by default.  In addition, we set the MemTable
size, level0-slowdown, and level0-stop of RocksDB to 4\,MiB, 8, and 12,
respectively, so as to match the default parameters of LevelDB, HyperLevelDB,
and PebblesDB.

\begin{figure}[!t]
\centering
\begin{tabular}{c}
\includegraphics[width=4.4in]{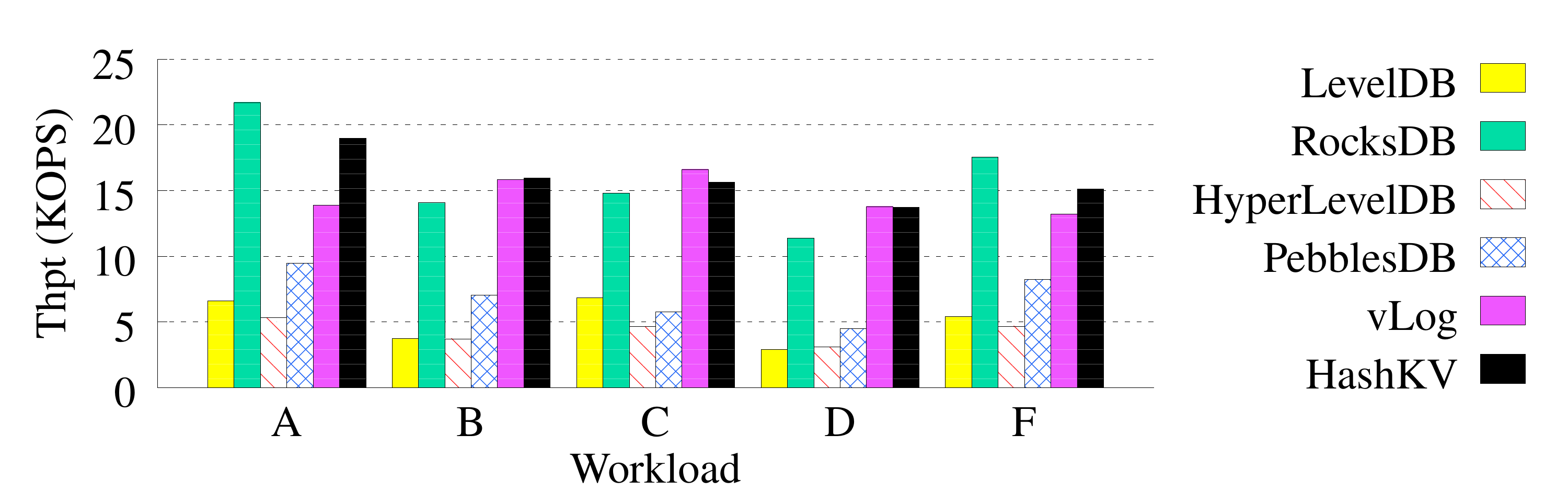} \\
\mbox{\small (a) Aggregate throughput} \\
\includegraphics[width=4.4in]{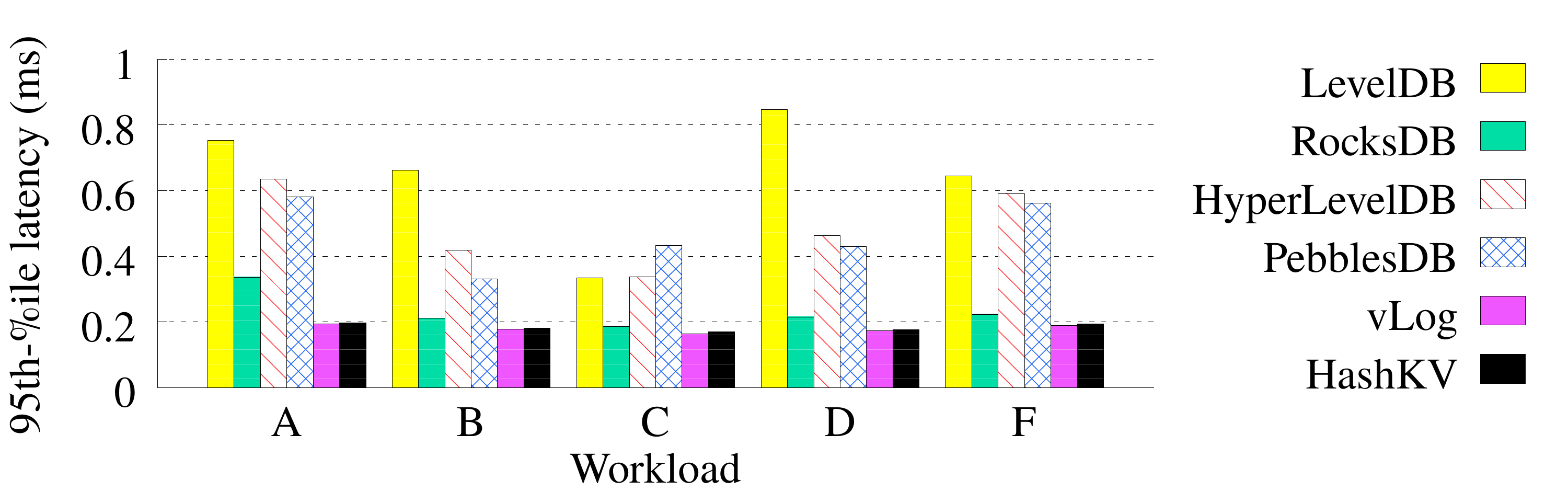} \\
\mbox{\small (b) 95th percentile read latency}
\end{tabular}
\caption{Experiment~11: YCSB benchmarking.}
\label{fig:ycsb}
\end{figure}

\paragraph{Experiment~11 (YCSB benchmarking):}
We first present the performance results of LevelDB, RocksDB, HyperLevelDB,
PebblesDB, vLog, and \HASHKV under the default YCSB core workloads.
Figures~\ref{fig:ycsb}(a) and \ref{fig:ycsb}(b) show the aggregate throughput
and the 95th percentile read latency of each KV store, respectively, under
each YCSB workload. 

We start with considering Workload~A and Workload~F, both of which contain
around 50\% of reads.  The throughput of \HASHKV is 2.8-2.9$\times$,
3.2-3.6$\times$, 1.8-2.0$\times$, and 1.1-1.4$\times$ over LevelDB,
HyperLevelDB, PebblesDB, and vLog, respectively. 
We observe that LevelDB, HyperLevelDB, and PebblesDB also have higher read
latencies, which also lead to less overall throughput.  Both vLog and \HASHKV
have similar read latencies, yet vLog has lower throughput than \HASHKV due to
the GC overhead.  However, the throughput of \HASHKV is 12.5-13.9\% lower than
RocksDB, mainly because RocksDB no longer stalls writes for flushing the
MemTable under the workloads with less intensive updates and it can better
serve reads and updates via multi-threading optimization \cite{rocksdbmt}.
Nevertheless, we show that \HASHKV can achieve higher throughput when it uses
RocksDB, instead of LevelDB, for key and metadata management (Experiment~13).

We next consider Workload~B, Workload~C, and Workload~D, all of which are
read-intensive.  \HASHKV, vLog, and RocksDB have similar read latencies and
hence similar throughput.   \HASHKV achieves 2.3-4.8$\times$, 3.2-4.4$\times$,
and 2.3-3.1$\times$ throughput over LevelDB, HyperLevelDB, and PebblesDB,
respectively.

\begin{table}[!t]
\centering
\begin{tabular}{l|@{\ }l}
\hline
Workload & Portions of Requests\\
\hline \hline
A (Update-heavy) & 50\% updates, 50\% reads \\
\hline
B (Read-mostly) & 5\% updates, 95\% reads \\
\hline
C (Read-only) & 100\% reads \\
\hline
D (Read-latest) & 5\% inserts, 95\% reads \\
\hline
F (Read-modify-write) & 50\% read-modify-write, 50\% reads \\
\hline
\end{tabular}
\vspace{3pt}
\caption{YCSB core workloads \cite{Cooper10}.}
\label{table:ycsb}
\end{table}

\begin{figure}[!t]
\centering
\begin{tabular}{c}
\includegraphics[width=4.5in]{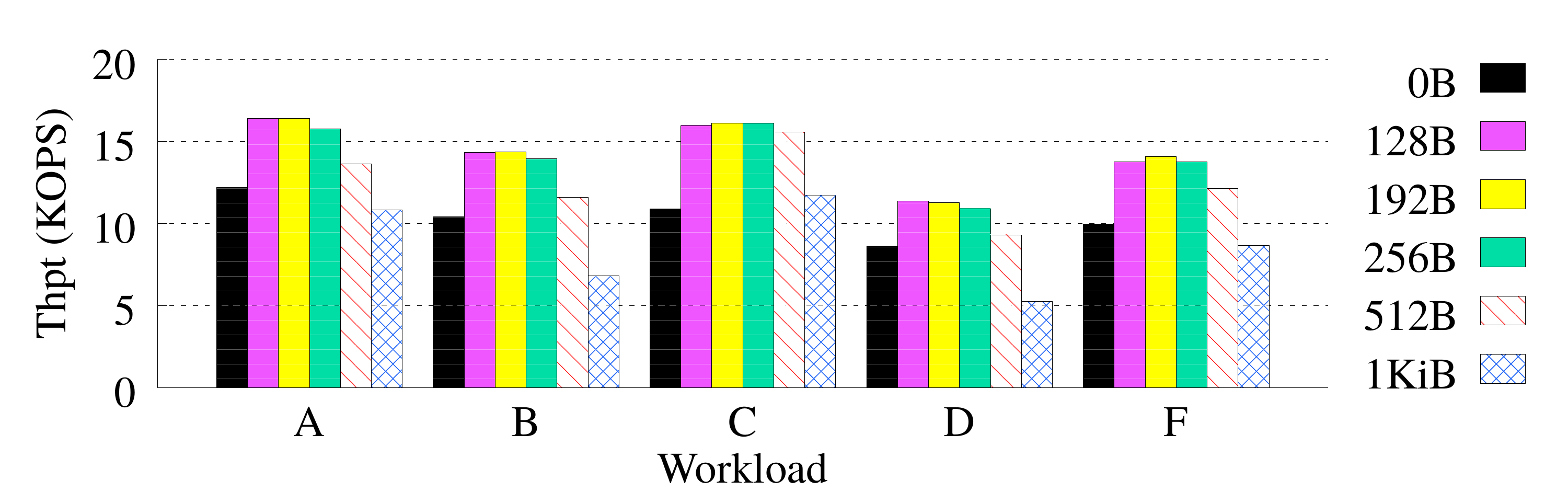} \\
\mbox{\small (a) Aggregate throughput} \\
\includegraphics[width=4.5in]{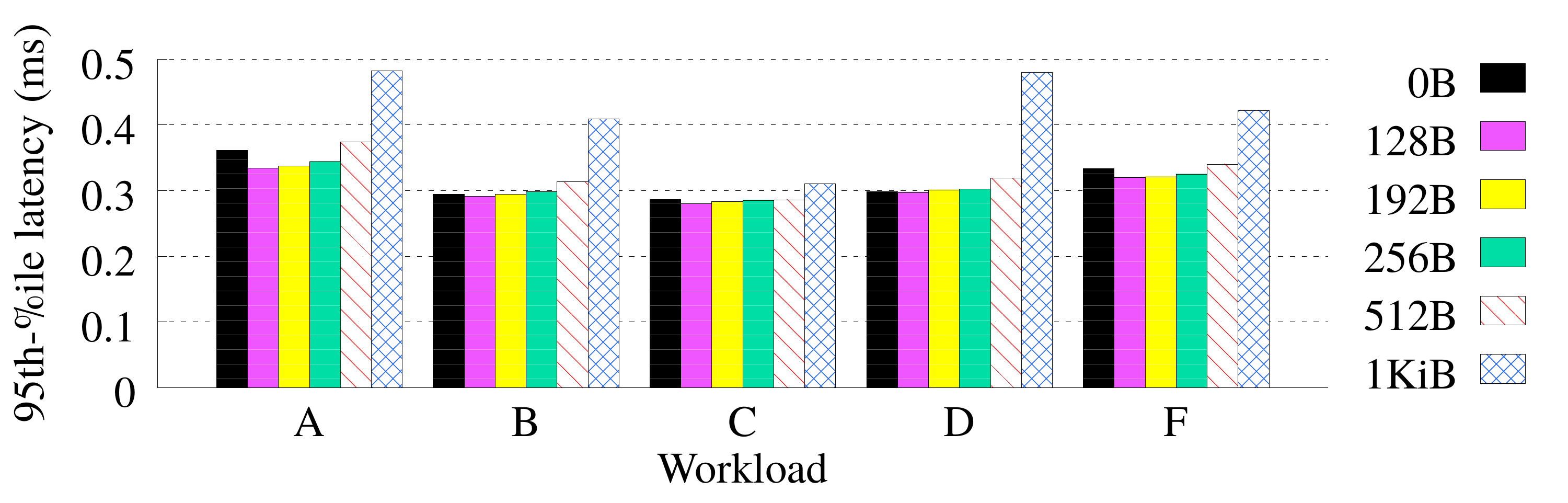} \\
\mbox{\small (b) 95th percentile read latency} 
\end{tabular}
\caption{Experiment~12: Threshold selection of selective KV separation under
YCSB core workloads.}
\label{fig:kvsep_ycsb_120g}
\end{figure}

\paragraph{Experiment 12 (Threshold selection of selective KV
separation):} We study the impact of the KV pair size threshold on the
performance of \HASHKV under YCSB core workloads.  We use the same
experiment setting as in Experiment~8 (i.e., we vary the KV pair size from
40\,B to 8\,KiB following the Zipf distribution with a Zipfian constant of
0.99 and configure 120\,GiB of storage space with 30\% of reserved space). 
Figure~\ref{fig:kvsep_ycsb_120g}(a) shows the aggregate throughput of \HASHKV
for different KV pair size thresholds. For all workloads, \HASHKV achieves the
maximum throughput when the threshold is in the range of 128\,B and 256\,B,
and improves the throughput by 26.5-48.3\% compared to the threshold 0\,B
(i.e., KV separation is always used).
Figure~\ref{fig:kvsep_ycsb_120g}(b) shows the 95th percentile read latency of
\HASHKV under the YCSB core workloads.  Again, the read latency is the lowest
when the threshold is in the range of 128\,B and 256\,B.  From Experiment~8
and this experiment, we see that our threshold selection approach in
\S\ref{subsec:selective_kv_sep} remains robust across different workloads. 

\paragraph{Experiment~13 (Deployment of KV separation on
different KV stores):}  We study the performance of two KV separation
designs, \HASHKV and vLog, when we deploy KV separation on different KV
stores.  Specifically, we configure \HASHKV and vLog to use RocksDB,
HyperLevelDB, and PebblesDB, instead of LevelDB, as the LSM-tree
implementation for key and metadata management.  For clarity, we append the
suffixes ``-RDB", ``-HDB", and ``-PDB" to \HASHKV and vLog to denote their
deployment on RocksDB, HyperLevelDB, and PebblesDB, respectively.

\begin{figure}[!t]
\centering
\begin{tabular}{ccc}
\includegraphics[width=1.7in]{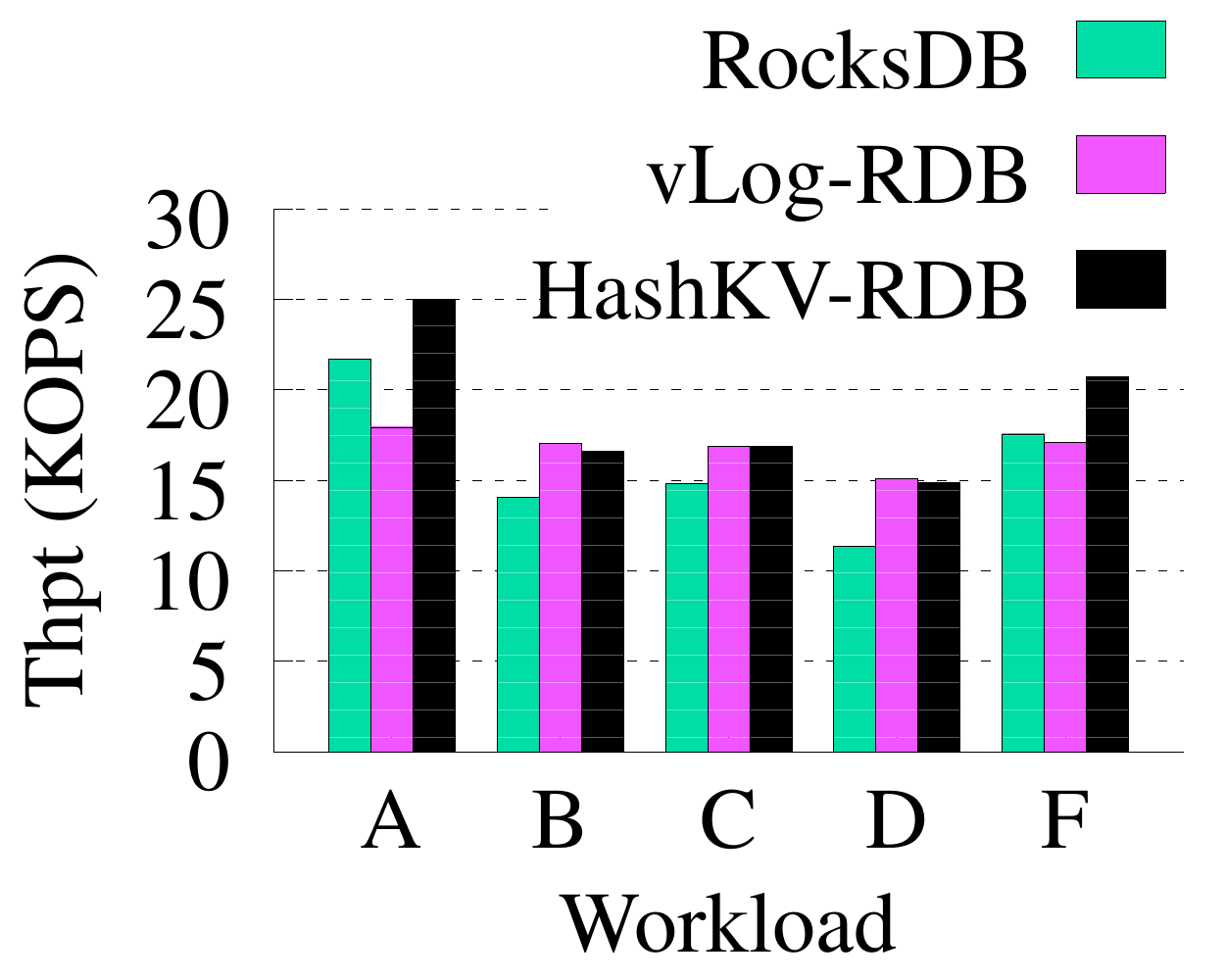} &
\includegraphics[width=1.7in]{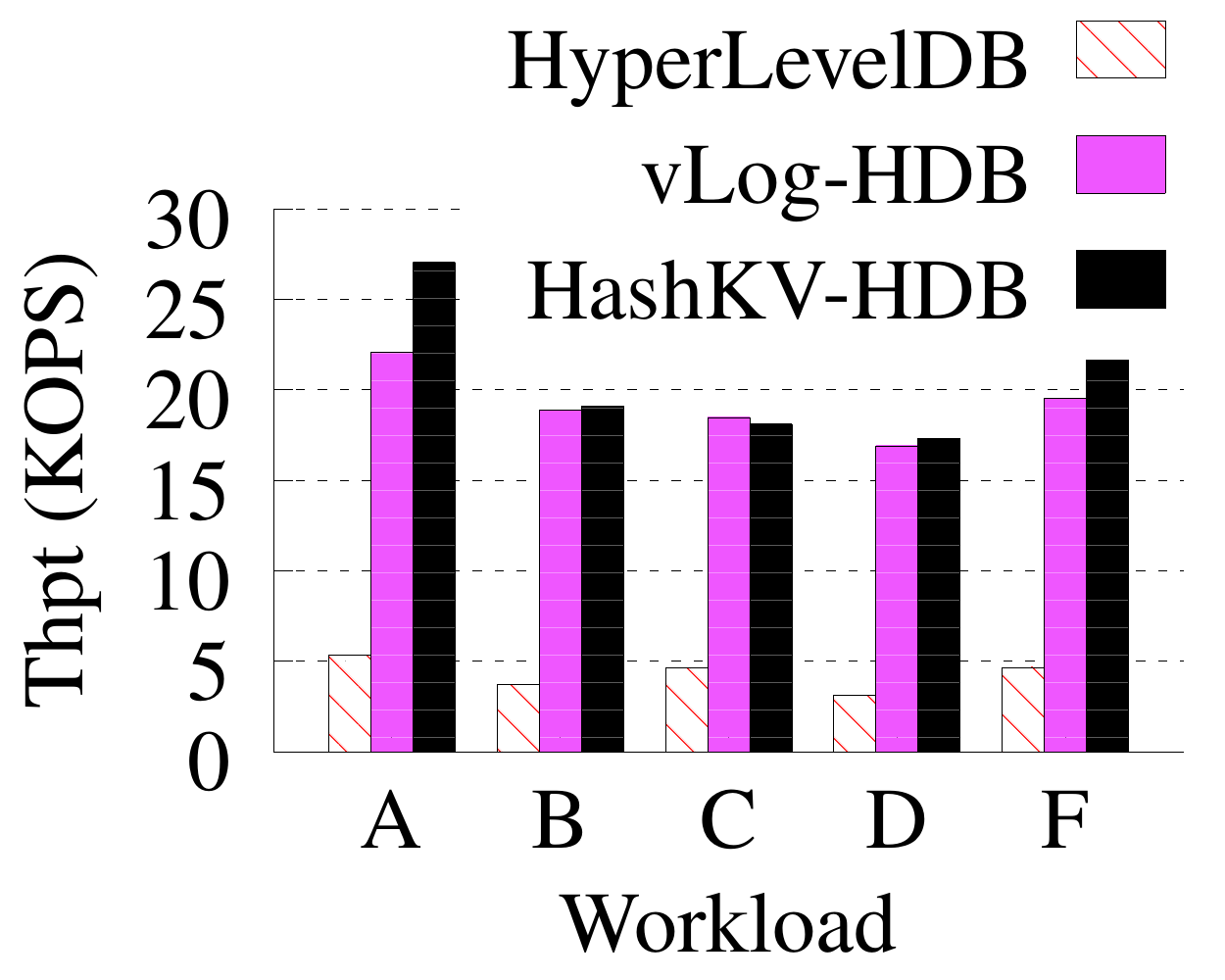} &
\includegraphics[width=1.7in]{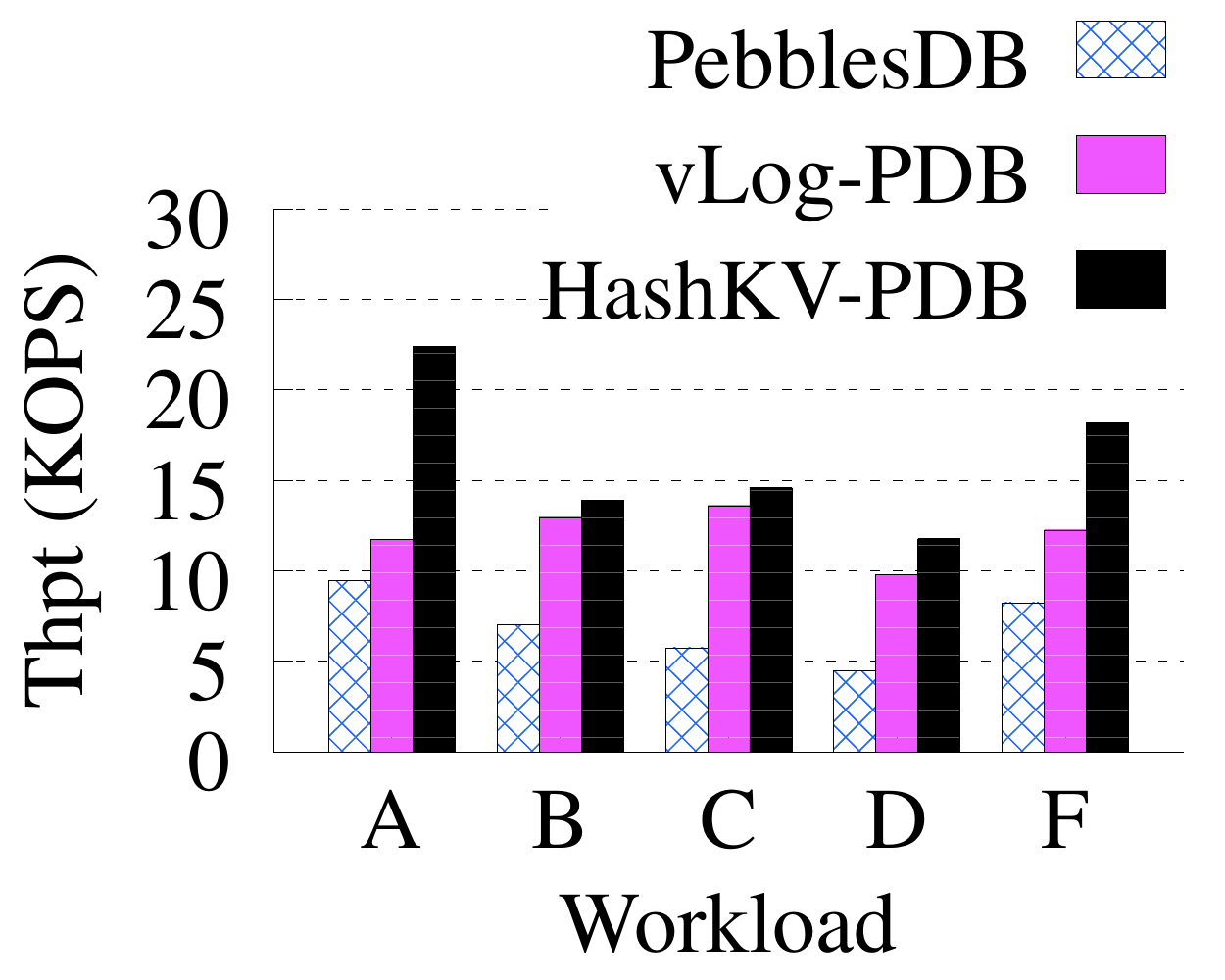} \\
\mbox{\small (a) RocksDB} &
\mbox{\small (b) HyperLevelDB} &
\mbox{\small (c) PebblesDB} 
\end{tabular}
\caption{Experiment~13: Deployment of KV separation on different KV stores
under YCSB core workloads.}
\label{fig:diffkvs_ycsb}
\end{figure}

Figure~\ref{fig:diffkvs_ycsb} shows the deployment of \HASHKV and vLog on
RocksDB, HyperLevelDB, and PebblesDB, under YCSB core workloads.  We first
compare the performance of \HASHKV and vLog to the KV stores without KV
separation.  Both \HASHKV and vLog increase the aggregate throughput of each
KV store across different YCSB core workloads via KV separation.  For example,
under Workload~A and Workload~F (i.e., the update-intensive workloads), the
throughput of \HASHKV-RDB, \HASHKV-HDB, and \HASHKV-PDB is 1.2$\times$,
4.6-5.0$\times$, and 2.2-2.4$\times$ over RocksDB, HyperLevelDB, and
PebblesDB, respectively.  Also, \HASHKV still outperforms vLog: the
throughput of \HASHKV-RDB, \HASHKV-HDB, and \HASHKV-PDB is 21.3-39.4\%,
10.7-22.5\%, and 48.5-91.0\% higher than vLog-RDB, vLog-HDB, and vLog-PDB,
respectively.  Under Workload~B, Workload~C, and Workload~D (i.e., the
read-intensive workloads), the throughput of \HASHKV-RDB, \HASHKV-HDB, and
\HASHKV-PDB is 1.1-1.3$\times$, 3.9-5.6$\times$, and 2.0-2.6$\times$ over
RocksDB, HyperLevelDB, and PebblesDB, respectively. Both \HASHKV and vLog
achieve similar throughput under the read-intensive workloads when they are
deployed over the same KV store.

\begin{figure}[!t]
\centering
\begin{tabular}{ccc}
\includegraphics[width=1.7in]{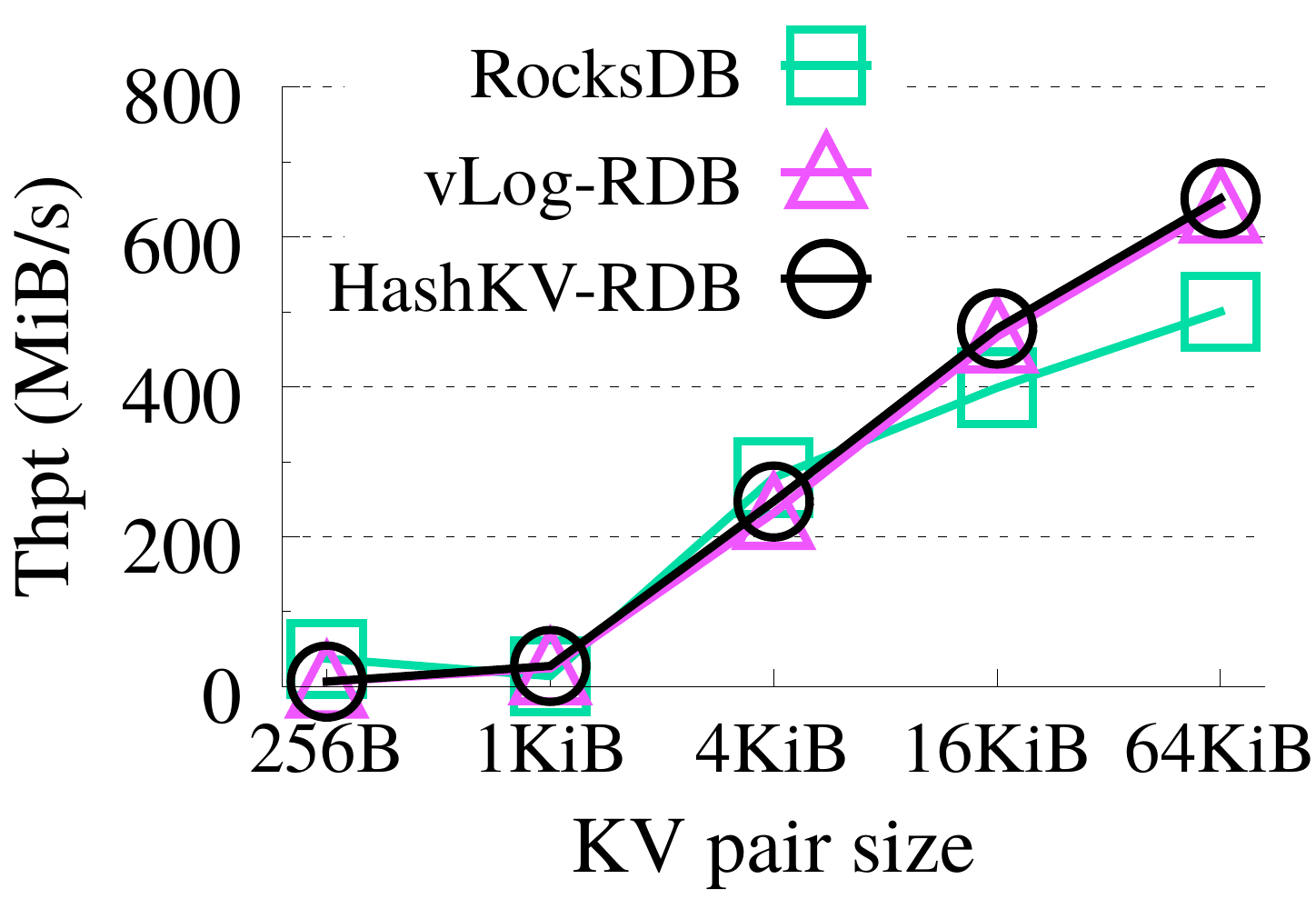} &
\includegraphics[width=1.7in]{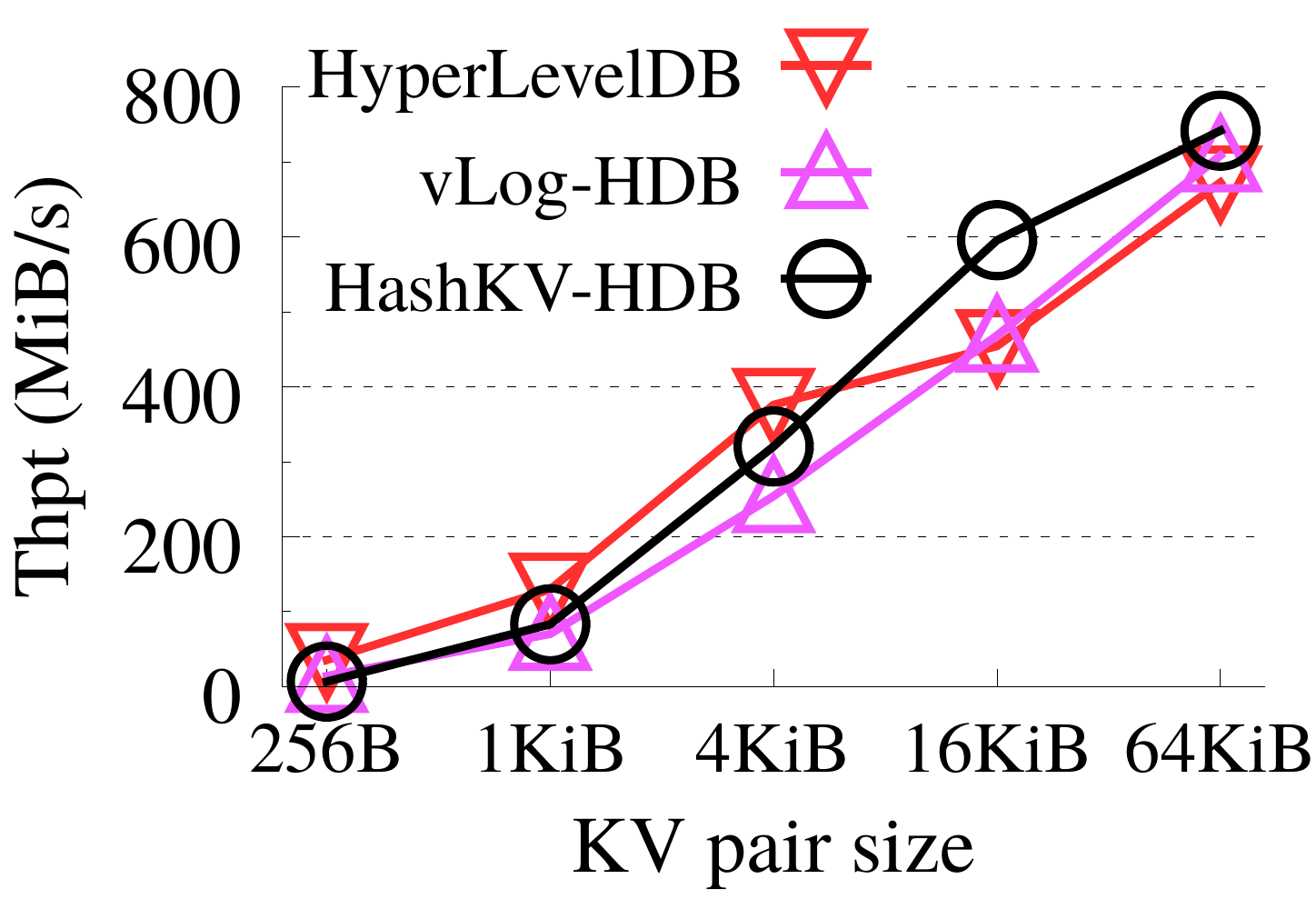} &
\includegraphics[width=1.7in]{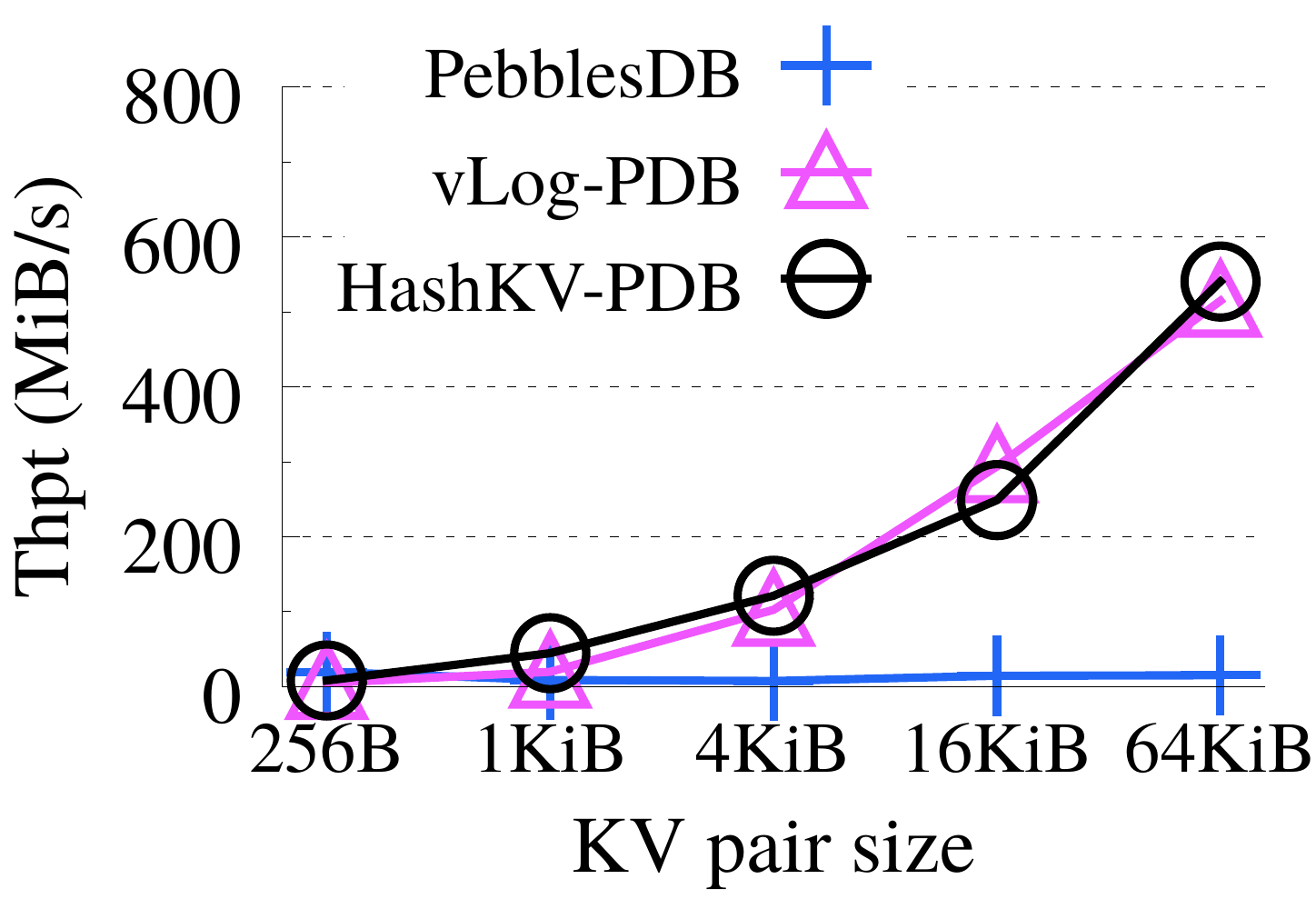} \\
\mbox{\small (a) RocksDB} &
\mbox{\small (b) HyperLevelDB} &
\mbox{\small (c) PebblesDB} 
\end{tabular}
\caption{Experiment~13: Impact of KV separation on different KV stores under
range scan workloads.}
\label{fig:diffkvs_scan}
\end{figure}

We further compare the range scan performance of \HASHKV and vLog to the KV
stores without KV separation.  We consider the same range scan workload as in
Experiment~5 (\S\ref{subsec:exp_diff_workloads}).
Figure~\ref{fig:diffkvs_scan} shows the range scan performance of \HASHKV and
vLog when they are deployed on RocksDB, HyperLevelDB, and PebblesDB.  Overall,
\HASHKV has similar range scan performance to vLog for most KV pair sizes in
all cases.  Also, both of them achieve similar or higher range scan
performance than RocksDB for larger KV pair sizes due to KV separation.  

Note that the range scan throughput of PebblesDB is significantly lower than
that of RocksDB and HyperLevelDB.  The reason is that PebblesDB triggers a
number of compaction operations during range scans, and such compaction
overhead significantly degrades the range scan performance especially for
large KV pairs.  Also, PebblesDB's fragmented LSM-tree design causes the KV
pairs to be scattered across multiple levels, and each range scan needs to
examine multiple levels to retrieve all KV pairs (see \cite{Raju17} for
details).  Nevertheless, KV separation (under \HASHKV or vLog) significantly
improves the range scan performance. 

\begin{figure}[!t]
\centering
\begin{tabular}{c@{\ \ \ }c}
\includegraphics[width=2.1in]{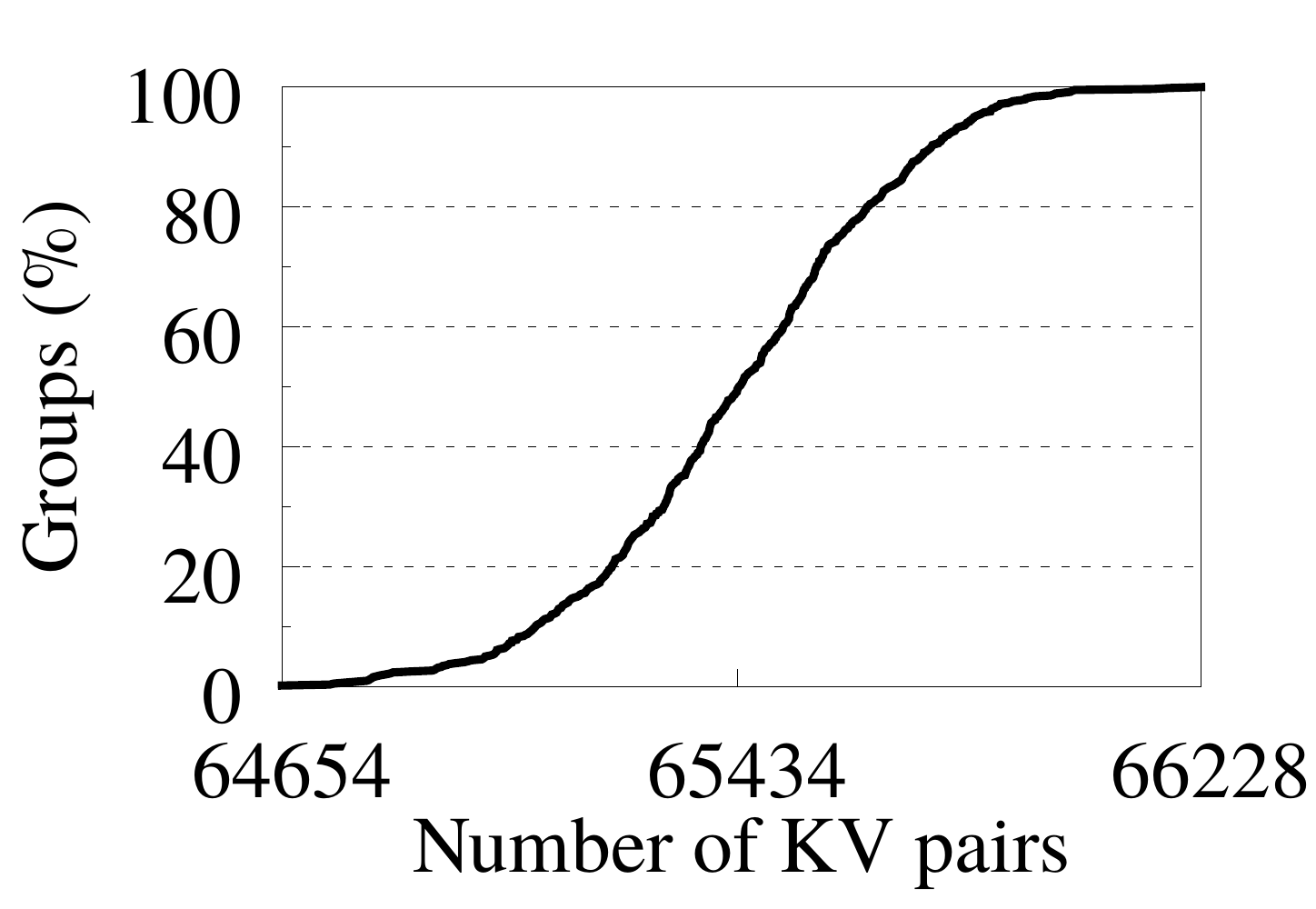} &
\includegraphics[width=2.1in]{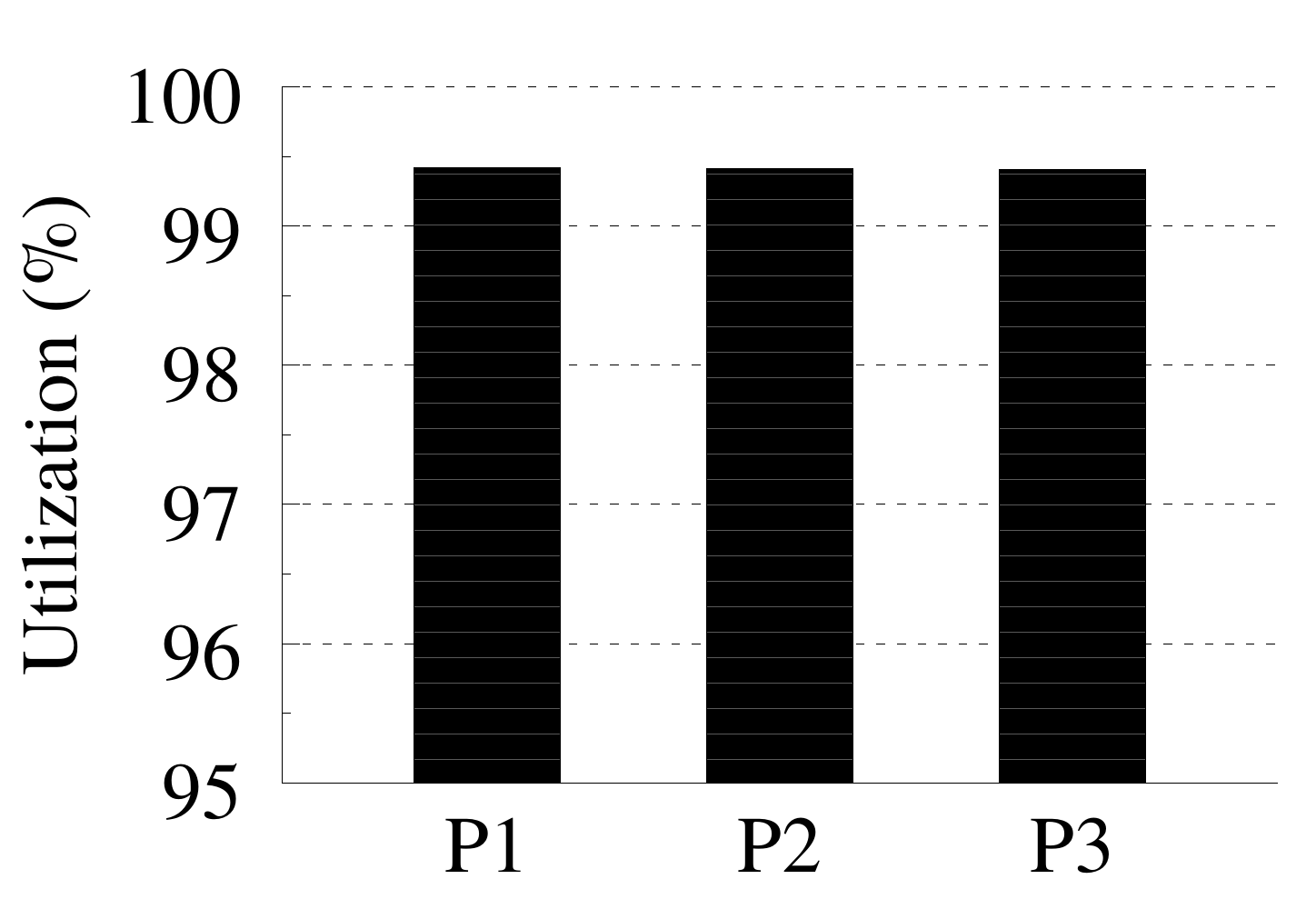} \\
\parbox[t]{2.4in}{(a) Cumulative distribution of KV pairs
among segment groups after Phase~P0} &
\parbox[t]{2.4in}{(b) Storage space utilization after each of
Phases~P1-P3} 
\end{tabular}
\caption{Experiment~14: Storage distribution of KV pairs in \HASHKV.}
\label{fig:utilize}
\end{figure}

\subsection{Storage Distribution of KV Pairs}
\label{subsec:exp_utilization}

\paragraph{Experiment~14 (Storage distribution of KV pairs)}:
We study the storage distribution of KV pairs in \HASHKV under
the update-intensive workloads in \S\ref{subsec:exp_perf}. Since \HASHKV
distributes KV pairs via hashing, it is possible that the KV pairs are
unevenly distributed across segment groups and some segment groups are not
fully utilized.  However, if there are a sufficiently large number of keys and
the hash function can produce uniformly distributed outputs, we argue that the
KV pairs are indeed evenly distributed across segment groups.
Figure~\ref{fig:utilize}(a) plots the cumulative distribution of the number of
KV pairs across segment groups at the end of the load phase (Phase~P0).  We
see that the number of KV pairs in each segment group varies between 64K and
66K, and the difference is within 2.5\% only.  
	 
Also, we argue that update-intensive workloads have limited impact on storage
space utilization (i.e., the fraction of valid and invalid data being stored
over the entire storage space), even though some segment groups may allocate
new log segments after receiving extensive updates (\S\ref{subsec:storage}).
Figure~\ref{fig:utilize}(b) shows the utilization of
the storage space at the end of each update phase (i.e., Phases~P1-P3).
\HASHKV achieves a high utilization of 99.4\% across the update phases.

\subsection{Parameter Choices}
\label{subsec:exp_param}

We further study the impact of parameters, including the main segment size,
the log segment size, and the write cache size on the update performance of
\HASHKV under the update-intensive workloads in \S\ref{subsec:exp_perf}.
We vary one parameter in each test, and use the default values for
other parameters. We report the update throughput in Phase~P3 and the total
write size. Here, we focus on 20\% and 50\% of reserved space.


\begin{figure}[!t]
\centering
\begin{tabular}{ccc}
\includegraphics[width=1.6in]{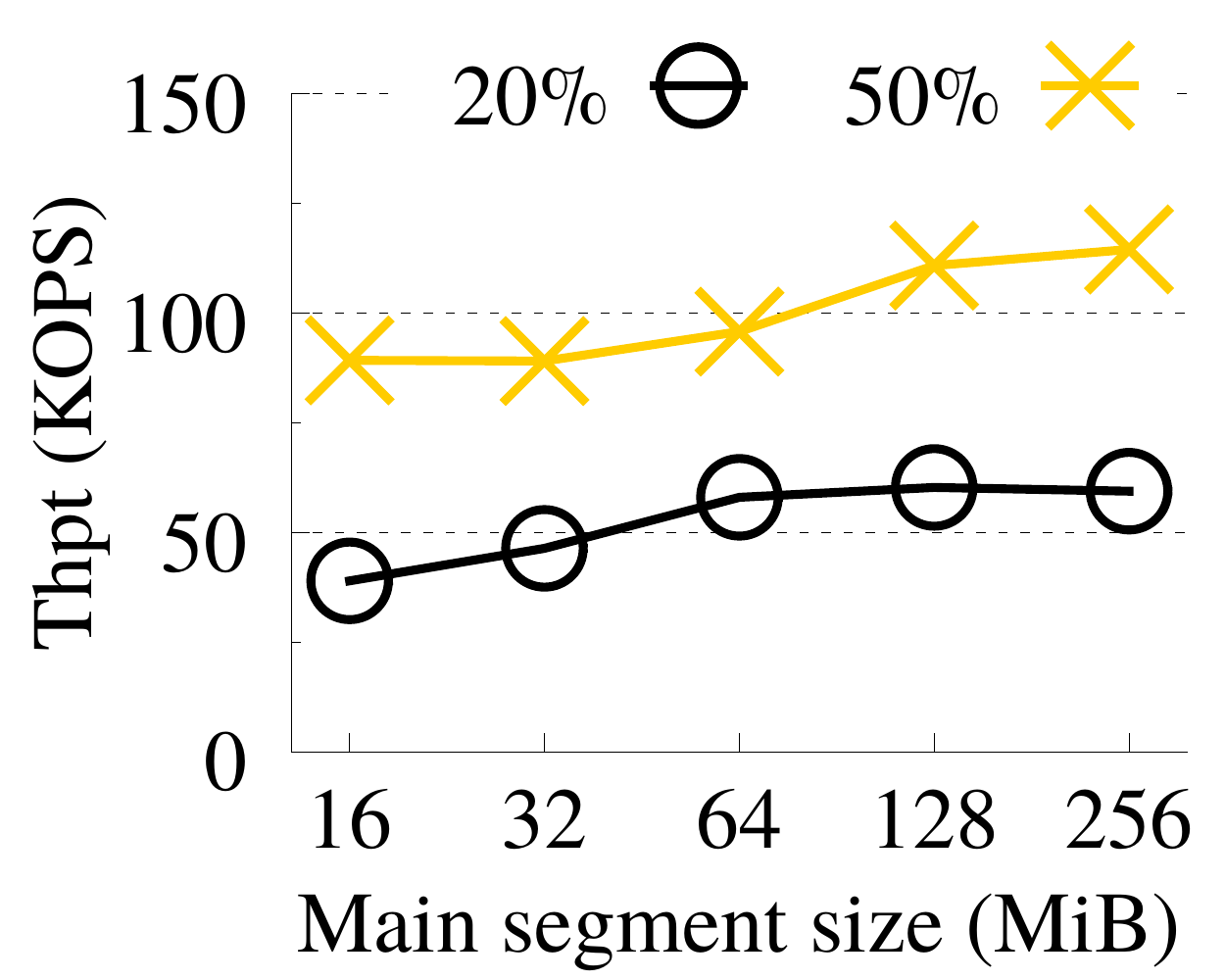}
&
\hspace{-5pt}
\includegraphics[width=1.6in]{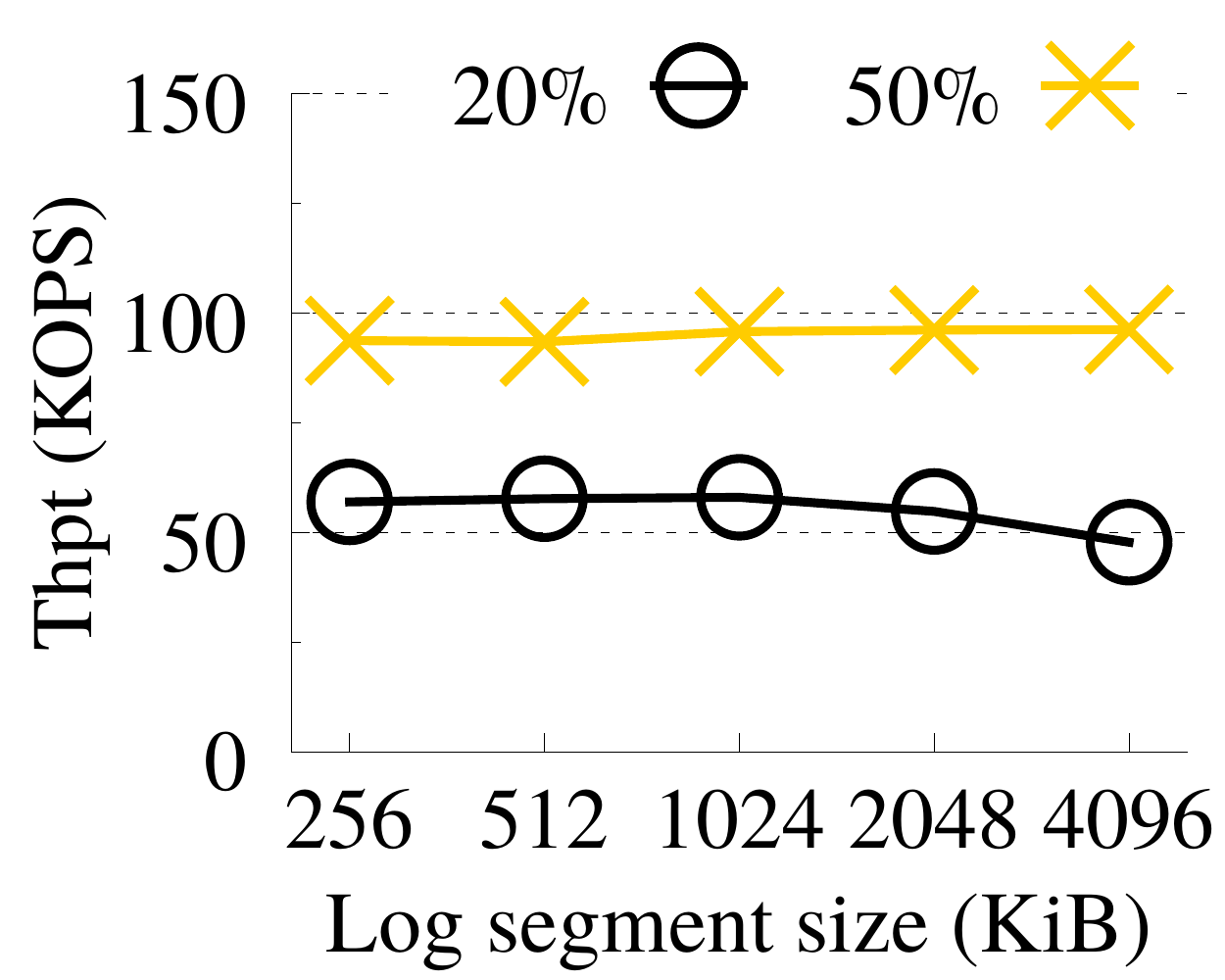}
&
\includegraphics[width=1.6in]{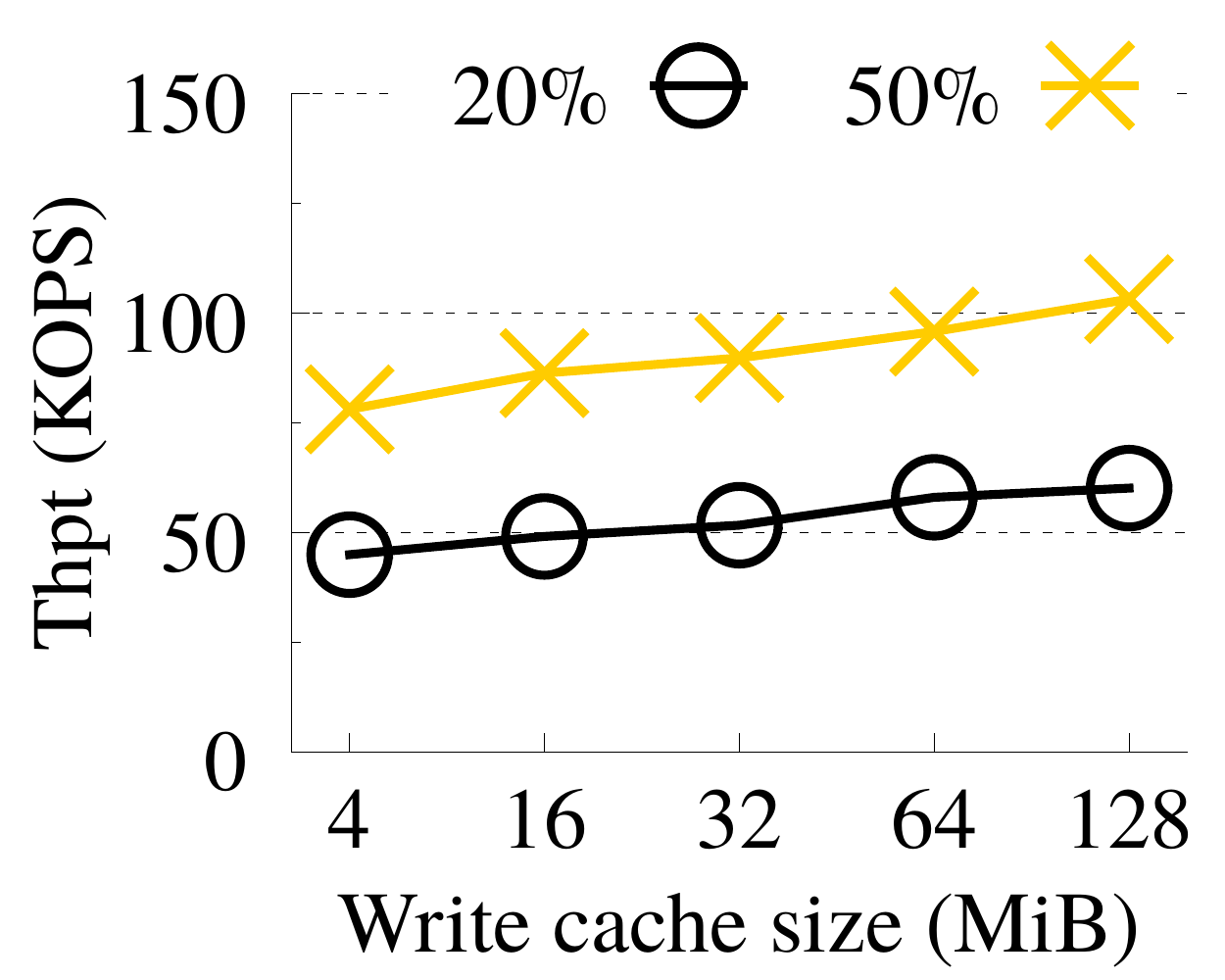}\\
\mbox{\small (a)} & \hspace{5pt} \mbox{\small (b)} & \hspace{5pt} 
\mbox{\small (c)} \\
\includegraphics[width=1.6in]{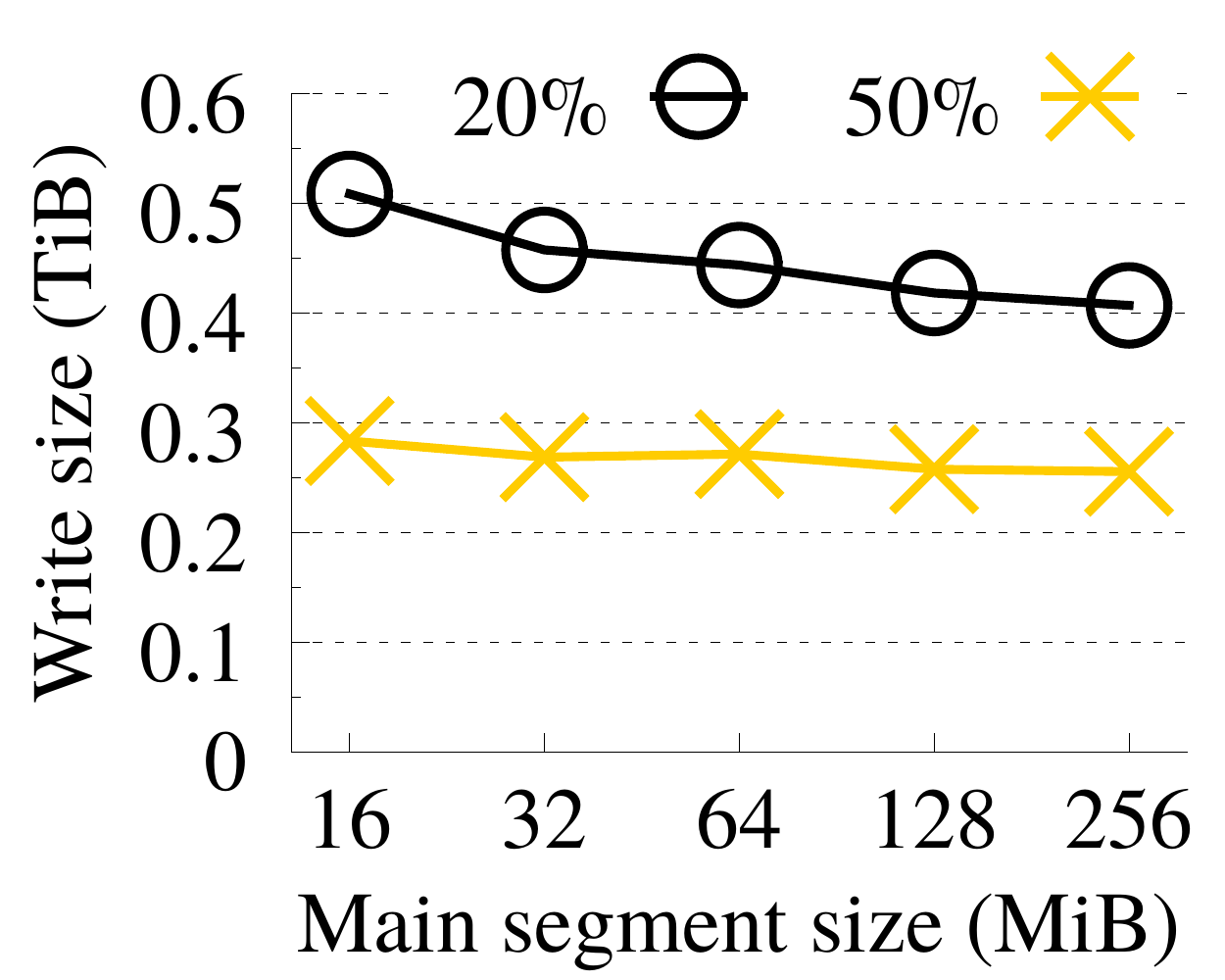}
&
\hspace{-5pt}
\includegraphics[width=1.6in]{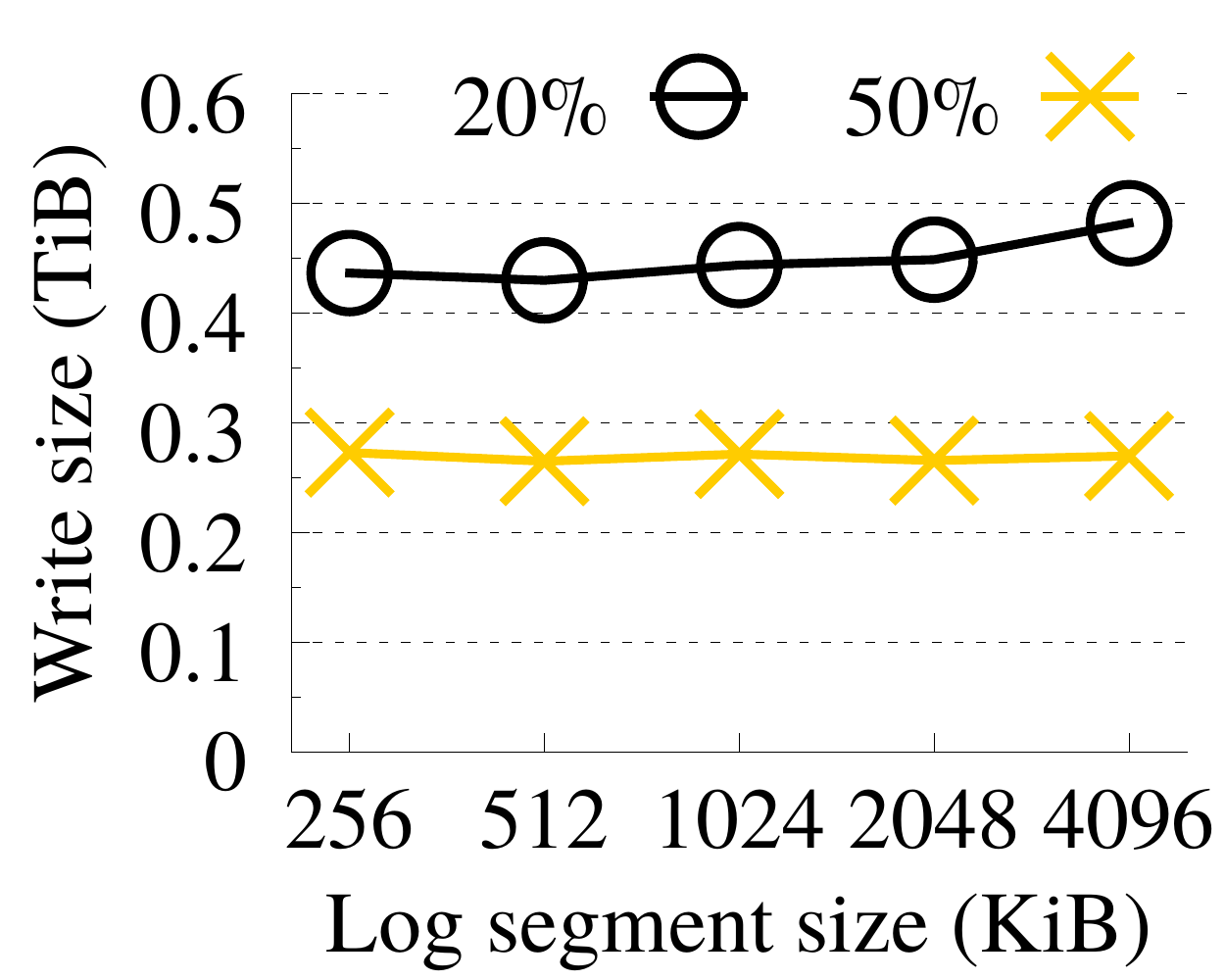}
&
\includegraphics[width=1.6in]{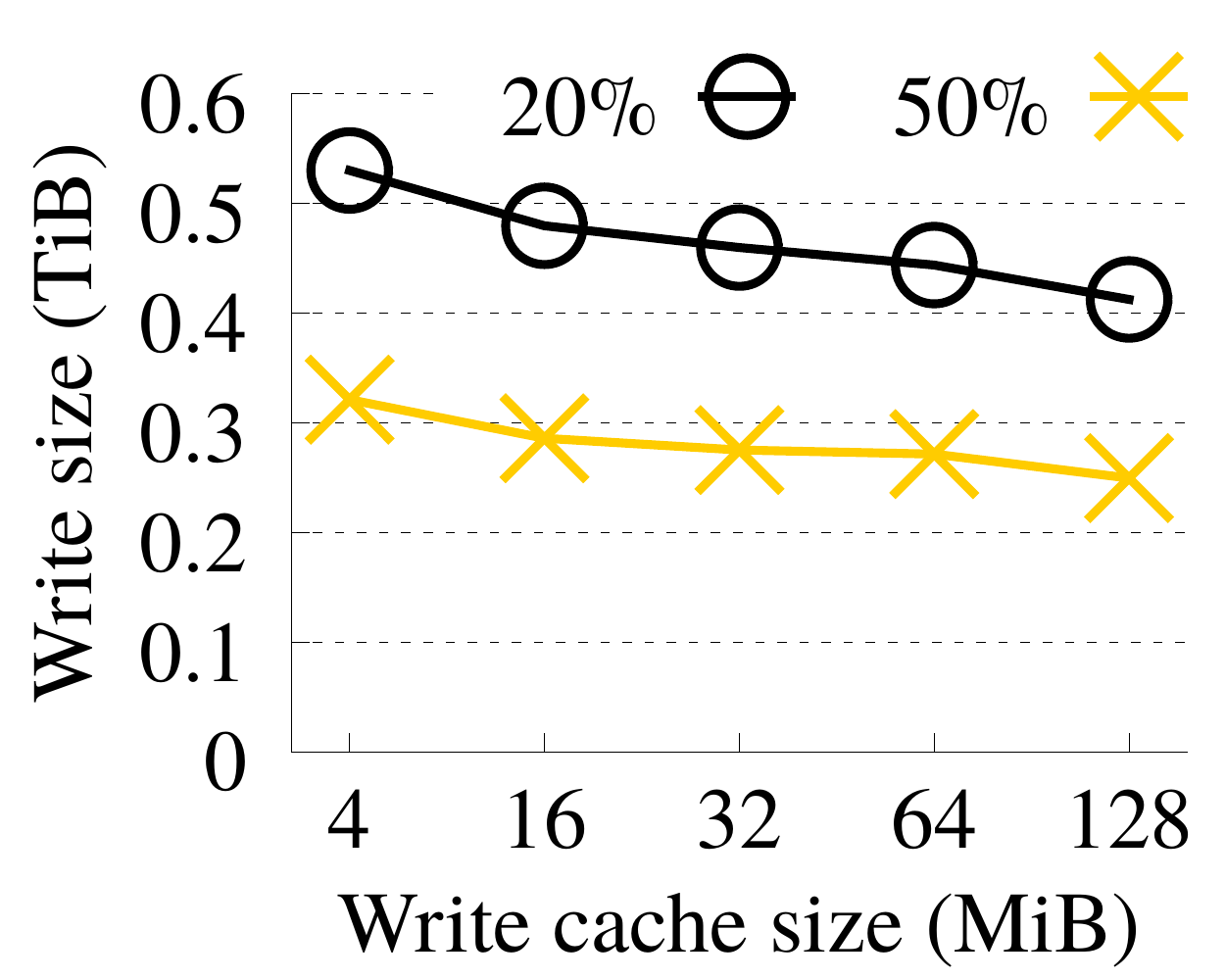} \\
\mbox{\small (d)} & \hspace{5pt} \mbox{\small (e)} & \hspace{5pt} 
\mbox{\small (f)}
\end{tabular}
\caption{Experiment~15: Throughput and total write size of \HASHKV versus
the main segment size ((a) and (d)), the log segment size ((b) and (e)), and
the write cache size ((c) and (f)).}
\label{fig:parm}
\end{figure}

\paragraph{Experiment~15 (Impact of main segment size, log segment size,
and write cache size):} We first consider the main segment size. 
Figures~\ref{fig:parm}(a) and \ref{fig:parm}(d) show the results versus
the main segment size.  When the main segment size increases, the throughput
of \HASHKV increases, while the total write size decreases. The reason is that
there are fewer segment groups for larger main segments, so each segment group
receives more updates in general.  Each GC operation can now reclaim more space
from more updates, so the performance improves.  We see that the update
performance of \HASHKV is more sensitive to the main segment size under
limited reserved space.  For example, the throughput increases by 52.5\% under
20\% of reserved space, but 28.3\% under 50\% of reserved space, when the main
segment size increases from 16\,MiB to 256\,MiB.

We next consider the log segment size.  Figures~\ref{fig:parm}(b) and
\ref{fig:parm}(e) show the results versus the log segment size.  We see that
when the log segment size increases from 256\,KiB to 4\,MiB, the throughput of
\HASHKV drops by 16.1\%, while the write size increases by 10.4\% under 20\%
of reserved space.  The reason is that the utilization of log segments
decreases as the log segment size increases.  Thus, each GC operation reclaims
less free space, and the performance drops.  However, when the reserved space
size increases to 50\%, we do not see significant performance differences, and
both the throughput and the write size remain almost unchanged across
different log segment sizes. 

We finally consider the write cache size. 
Figures~\ref{fig:parm}(c) and \ref{fig:parm}(f) show the results versus
the write cache size.  As expected, the throughput of \HASHKV increases and
the total write size drops as the write cache size increases, since a larger
write cache can absorb more updates.  For example, under 20\% of reserved
space, the throughput of \HASHKV increases by 29.1\% and the total write size
reduces by 16.3\% when the write cache size increases from 4\,MiB to 64\,MiB.  

\section{Related Work}
\label{sec:related}

\noindent
{\bf General KV stores:} Many KV store designs are proposed for different
types of storage backends, such as DRAM
\cite{Fitzpatrick04,redis,Fan13,Lim14}, commodity flash-based SSDs 
\cite{Debnath10,Debnath11,Lim11,Lu16}, open-channel SSDs \cite{Shen17}, and
emerging non-volatile memories \cite{Marmol15,Xia17}.    
The above KV stores and \HASHKV are designed for a single server. They can
serve as building blocks of a distributed KV store (e.g., \cite{Nishtala13}). 

\paragraph{LSM-tree-based KV stores:} As stated in \S\ref{sec:intro}, 
the LSM-tree \cite{Oneil96} is a major building block for most today's KV
stores that target workloads with high volumes of inserts or updates.  
Many studies extend the LSM-tree design for improved compaction performance;
we refer readers to the survey \cite{Luo18} on state-of-the-art
LSM-tree-based KV stores.
To name a few, bLSM \cite{Sears12} proposes a new merge scheduler to prevent
the compaction operations from blocking write requests, and uses Bloom
filters for efficient indexing.  VT-Tree \cite{Shetty13} stitches already
sorted blocks of SSTables to allow lightweight compaction overhead, at the
expense of incurring fragmentation.  LSM-trie \cite{Wu15} maintains a trie
structure and organizes KV pairs by hash-based buckets within each SSTable.
It also organizes large Bloom filters in clustered disk blocks for efficient
I/O access.  LWC-store \cite{Yao17} decouples data and metadata management in
compaction by merging and sorting only the metadata in SSTables. 
SkipStore \cite{Yue17} pushes KV pairs across non-adjacent levels to reduce
the number of levels traversed during compaction.  
TRIAD \cite{Balmau17} combines different techniques to reduce compaction
overhead, and addresses data skewness by keeping hot data in memory while
flushing only cold data into disk (note that our write cache also buffers 
recently written KV pairs and allows them to be directly updated in-place). 	
PebblesDB \cite{Raju17} relaxes the restriction of keeping disjoint key ranges
in each level, and pushes partial SSTables across levels to limit compaction
overhead. 
LSbM \cite{Teng18} keeps frequently accessed KV pairs in a compaction buffer
to avoid cache invalidation caused by compaction, thereby improving read
performance.  Note that the aforementioned LSM-tree-based KV stores do not
address KV separation.  Their performance is still limited by the compaction
and lookup overheads due to the storage of values in the LSM-tree (see
Experiment~13 in \S\ref{subsec:exp_ycsb}).

\paragraph{KV separation:}  WiscKey \cite{Lu16} employs KV separation to
remove value compaction in the LSM-tree (see \S\ref{subsec:kvsep}). Atlas
\cite{Lai15} also applies KV separation in cloud storage, in
which keys and metadata are stored in an LSM-tree that is replicated,
while values are separately stored and erasure-coded for low-redundancy fault
tolerance.  Cocytus \cite{Zhang16} is an in-memory KV store that separates
keys and values for replication and erasure coding, respectively.  
Tucana \cite{Papagiannis16} uses a B$^\epsilon$-tree for
indexing, so as to reduce the I/O amplifications compared to the LSM-tree.
Similar to WiscKey, Tucana employs KV separation and stores the values 
in an append-only log.  While \HASHKV also builds on KV separation, it takes
one step further to address efficient GC management in value storage via
hash-based data grouping. 

\paragraph{GC in log-structured storage:} Many efficient GC designs have been
proposed for log-structured storage, including log-structured file systems
\cite{Rosenblum92,Matthews97,Rumble14} and SSDs \cite{Min12,Lee13}, yet
applying them directly to LSM-tree-based KV storage is challenging 
(\S\ref{subsec:kvsep}).  For KV storage, Berkeley DB \cite{berkeleydb}
maintains no-overwrite log files, and selects the log file with the fewest
active records for GC.  BadgerDB \cite{badgerdb}, a WiscKey-based
implementation in Golang, divides the value store into regions. It leverages
the statistics obtained from LSM-tree compaction to identify the regions with
the most free space for GC.  In contrast, \HASHKV specifically aims for
efficient GC in LSM-tree-based KV separation via hash-based data grouping,
which eliminates the need of querying the LSM-tree during GC.  We also show
how hash-based data grouping can be extended with hotness awareness. 

\paragraph{Hash-based data organization:}  Distributed storage systems (e.g.,
\cite{Weil06,DeCandia07,Maccormick09}) use hash-based data placement to avoid
centralized metadata lookups.   NVMKV \cite{Marmol15} also uses hashing to map
KV pairs in physical address space.  However, it assumes sparse address space
to limit the overhead of resolving hash collisions, and incurs internal
fragmentation for small-sized KV pairs.  In contrast, \HASHKV does not cause
internal fragmentation as it packs KV pairs in each main/log segment in a
log-structured manner. It also supports dynamic reserved space allocation when
the main segments become full. 

\section{Conclusion}
\label{sec:conclusion}

This paper presents \HASHKV, which enables efficient updates in KV stores
under update-intensive workloads.  Its novelty lies in leveraging hash-based
data grouping for deterministic data organization so as to mitigate GC
overhead.  We also enhance \HASHKV with several extensions including dynamic
reserved space allocation, hotness awareness, and selective KV separation.
Testbed experiments show that \HASHKV achieves high update throughput and
reduces the total write size.  We further demonstrate that \HASHKV can build
on different LSM-tree-based KV stores to improve their respective performance.

\bibliographystyle{abbrv}
\bibliography{ref}

\begin{thebibliography}{10}

\bibitem{Agrawal08}
N.~Agrawal, V.~Prabhakaran, T.~Wobber, J.~D. Davis, M.~Manasse, and
  R.~Panigrahy.
\newblock {Design Tradeoffs for SSD Performance}.
\newblock In {\em Proceedings of the 2008 USENIX Annual Technical Conference
  (ATC'08)}, pages 57--70, 2008.

\bibitem{Atikoglu12}
B.~Atikoglu, Y.~Xu, E.~Frachtenberg, S.~Jiang, and M.~Paleczny.
\newblock {Workload Analysis of a Large-Scale Key-Value Store}.
\newblock In {\em Proceedings of the 12th ACM SIGMETRICS/PERFORMANCE Joint
  International Conference on Measurement and Modeling of Computer Systems
  (SIGMETRICS'12)}, pages 53--64, 2012.

\bibitem{Balmau17}
O.~Balmau, D.~Didona, R.~Guerraoui, W.~Zwaenepoel, H.~Yuan, A.~Arora, K.~Gupta,
  and P.~Konka.
\newblock {TRIAD: Creating Synergies Between Memory, Disk and Log in Log
  Structured Key-Value Stores}.
\newblock In {\em Proceedings of the 2017 USENIX Annual Technical Conference
  (ATC'17)}, pages 363--375, 2017.

\bibitem{Beaver10}
D.~Beaver, S.~Kumar, H.~C. Li, J.~Sobel, and P.~Vajgel.
\newblock {Finding a Needle in Haystack: Facebook's Photo Storage}.
\newblock In {\em Proceedings of the 9th USENIX Symposium on Operating Systems
  Design and Implementation (OSDI'10)}, pages 47--60, 2010.

\bibitem{chan18}
H.~H.~W. Chan, Y.~Li, P.~P.~C. Lee, and Y.~Xu.
\newblock {HashKV: Enabling Efficient Updates in KV Storage via Hashing}.
\newblock In {\em Proceedings of the 2018 USENIX Annual Technical Conference
  (ATC'18)}, pages 1007--1019, 2018.

\bibitem{Chang06}
F.~Chang, J.~Dean, S.~Ghemawat, W.~C. Hsieh, D.~A. Wallach, M.~Burrows,
  T.~Chandra, A.~Fikes, and R.~E. Gruber.
\newblock {Bigtable: A Distributed Storage System for Structured Data}.
\newblock In {\em Proceedings of the 7th USENIX Symposium on Operating Systems
  Design and Implementation (OSDI'06)}, pages 15--15, 2006.

\bibitem{Chen17}
Y.~L. Chen, S.~Mu, J.~Li, C.~Huang, J.~Li, A.~Ogus, and D.~Phillips.
\newblock {Giza: Erasure Coding Objects across Global Data Centers}.
\newblock In {\em Proceedings of the 2017 USENIX Annual Technical Conference
  (ATC'17)}, pages 539--551, 2017.

\bibitem{Cooper10}
B.~F. Cooper, A.~Silberstein, E.~Tam, R.~Ramakrishnan, and R.~Sears.
\newblock {Benchmarking Cloud Serving Systems with YCSB}.
\newblock In {\em Proceedings of the 1st ACM Symposium on Cloud Computing
  (SoCC'10)}, pages 143--154, 2010.

\bibitem{Debnath10}
B.~Debnath, S.~Sengupta, and J.~Li.
\newblock {FlashStore: High Throughput Persistent Key-value Store}.
\newblock {\em Proceedings of the VLDB Endowment}, 3(1-2):1414--1425, Sep 2010.

\bibitem{Debnath11}
B.~Debnath, S.~Sengupta, and J.~Li.
\newblock {SkimpyStash: RAM Space Skimpy Key-value Store on Flash-based
  Storage}.
\newblock In {\em Proceedings of the 2011 ACM SIGMOD International Conference
  on Management of Data (SIGMOD'11)}, pages 25--36, 2011.

\bibitem{DeCandia07}
G.~DeCandia, D.~Hastorun, M.~Jampani, G.~Kakulapati, A.~Lakshman, A.~Pilchin,
  S.~Sivasubramanian, P.~Vosshall, and W.~Vogels.
\newblock {Dynamo: Amazon's Highly Available Key-value Store}.
\newblock In {\em Proceedings of the 21st ACM SIGOPS Symposium on Operating
  Systems Principles (SOSP'07)}, pages 205--220, 2007.

\bibitem{badgerdb}
{\relax Dgraph Labs}.
\newblock {BadgerDB}.
\newblock \url{https://github.com/dgraph-io/badger/}, Retrieved June 2019.

\bibitem{Hyperleveldb}
R.~Escriva.
\newblock {HyperLevelDB}.
\newblock \url{https://github.com/rescrv/HyperLevelDB/}, Retrieved in June
  2019.

\bibitem{RocksDB}
Facebook.
\newblock {RocksDB}.
\newblock \url{https://rocksdb.org}, Retrieved in June 2019.

\bibitem{rocksdbmt}
Facebook.
\newblock {RocksDB Features that are not in LevelDB}.
\newblock
  \url{https://github.com/facebook/rocksdb/wiki/Features-Not-in-LevelDB},
  Retrieved in June 2019.

\bibitem{Fan13}
B.~Fan, D.~G. Andersen, and M.~Kaminsky.
\newblock {MemC3: Compact and Concurrent MemCache with Dumber Caching and
  Smarter Hashing}.
\newblock In {\em Proceedings of the 10th USENIX Conference on Networked
  Systems Design and Implementation (NSDI'13)}, pages 371--384, 2013.

\bibitem{Fitzpatrick04}
B.~Fitzpatrick.
\newblock {Distributed Caching with Memcached}.
\newblock {\em Linux Journal}, 2004(124), Aug 2004.

\bibitem{LevelDB}
S.~Ghemawat and J.~Dean.
\newblock {LevelDB}.
\newblock \url{https://leveldb.org}, Retrieved in June 2019.

\bibitem{nmon}
N.~Griffiths.
\newblock {nmon for Linux}.
\newblock \url{http://nmon.sourceforge.net/}, Retrieved in June 2019.

\bibitem{Hsieh06}
J.-W. Hsieh, T.-W. Kuo, and L.-P. Chang.
\newblock {Efficient Identification of Hot Data for Flash Memory Storage
  Systems}.
\newblock {\em ACM Transactions on Storage}, 2(1):22--40, Feb 2006.

\bibitem{Kavalanekar08}
S.~Kavalanekar, B.~Worthington, Q.~Zhang, and V.~Sharda.
\newblock {Characterization of Storage Workload Traces from Production Windows
  Servers}.
\newblock In {\em Proceedings of the 2008 IEEE International Symposium on
  Workload Characterization (IISWC'08)}, pages 119--128, 2008.

\bibitem{Lai15}
C.~Lai, S.~Jiang, L.~Yang, S.~Lin, G.~Sun, Z.~Hou, C.~Cui, and J.~Cong.
\newblock {Atlas: Baidu's Key-value Storage System for Cloud Data}.
\newblock In {\em Proceedings of the 31st Symposium on Mass Storage Systems and
  Technologies (MSST'15)}, pages 1--14, 2015.

\bibitem{Lee13}
J.~Lee and J.-S. Kim.
\newblock {An Empirical Study of Hot/Cold Data Separation Policies in Solid
  State Drives (SSDs)}.
\newblock In {\em Proceedings of the 6th International Systems and Storage
  Conference (SYSTOR'13)}, page~12, 2013.

\bibitem{Li15}
Y.~Li, P.~P. Lee, J.~C. Lui, and Y.~Xu.
\newblock {Impact of Data Locality on Garbage Collection in SSDs: A General
  Analytical Study}.
\newblock In {\em Proceedings of the 6th ACM/SPEC International Conference on
  Performance Engineering (ICPE'15)}, pages 305--315, 2015.

\bibitem{Lim11}
H.~Lim, B.~Fan, D.~G. Andersen, and M.~Kaminsky.
\newblock {SILT: A Memory-Efficient, High-Performance Key-Value Store}.
\newblock In {\em Proceedings of the 23rd ACM Symposium on Operating Systems
  Principles (SOSP'11)}, pages 1--13, 2011.

\bibitem{Lim14}
H.~Lim, D.~Han, D.~G. Andersen, and M.~Kaminsky.
\newblock {MICA: A Holistic Approach to Fast In-Memory Key-Value Storage}.
\newblock In {\em Proceedings of the 11th USENIX Conference on Networked
  Systems Design and Implementation (NSDI'14)}, pages 429--444, 2014.

\bibitem{mdadm}
{Linux Raid Wiki}.
\newblock {RAID setup}.
\newblock \url{https://raid.wiki.kernel.org/index.php/RAID_setup}, Retrieved in
  June 2019.

\bibitem{Lu16}
L.~Lu, T.~S. Pillai, A.~C. Arpaci-Dusseau, and R.~H. Arpaci-Dusseau.
\newblock {WiscKey: Separating Keys from Values in SSD-Conscious Storage}.
\newblock {\em ACM Transactions on Storage}, 13(1):5, Mar 2017.

\bibitem{Luo18}
C.~Luo and M.~J. Carey.
\newblock {LSM-based Storage Techniques: A Survey}.
\newblock {\em CoRR}, abs/1812.07527, Dec 2018.
\newblock \url{http://arxiv.org/abs/1812.07527}.

\bibitem{Maccormick09}
J.~MacCormick, N.~Murphy, V.~Ramasubramanian, U.~Wieder, J.~Yang, and L.~Zhou.
\newblock {Kinesis: A New Approach to Replica Placement in Distributed Storage
  Systems}.
\newblock {\em ACM Transactions on Storage}, 4(4):11, 2009.

\bibitem{Marmol15}
L.~Marmol, S.~Sundararaman, N.~Talagala, and R.~Rangaswami.
\newblock {{NVMKV}: A Scalable, Lightweight, FTL-aware Key-Value Store}.
\newblock In {\em Proceedings of the 2015 USENIX Annual Technical Conference
  (ATC'15)}, pages 207--219, 2015.

\bibitem{Matthews97}
J.~N. Matthews, D.~Roselli, A.~M. Costello, R.~Y. Wang, and T.~E. Anderson.
\newblock {Improving the Performance of Log-Structured File Systems with
  Adaptive Methods}.
\newblock In {\em Proceedings of the 16th ACM Symposium on Operating Systems
  Principles (SOSP'97)}, pages 238--251, 1997.

\bibitem{Min12}
C.~Min, K.~Kim, H.~Cho, S.-W. Lee, and Y.~I. Eom.
\newblock {SFS: Random Write Considered Harmful in Solid State Drives}.
\newblock In {\em Proceedings of the 10th USENIX Conference on File and Storage
  Technologies (FAST'12)}, pages 12--12, 2012.

\bibitem{Nishtala13}
R.~Nishtala, H.~Fugal, S.~Grimm, M.~Kwiatkowski, H.~Lee, H.~C. Li, R.~McElroy,
  M.~Paleczny, D.~Peek, P.~Saab, D.~Stafford, T.~Tung, and enkateshwaran
  Venkataramani.
\newblock {Scaling Memcache at Facebook}.
\newblock In {\em Proceedings of the 10th USENIX Conference on Networked
  Systems Design and Implementation}, pages 385--398, 2013.

\bibitem{Oneil96}
P.~O'Neil, E.~Cheng, D.~Gawlick, and E.~O'Neil.
\newblock {The Log-Structured Merge-Tree (LSM-tree)}.
\newblock {\em Acta Informatica}, 33(4):351--385, 1996.

\bibitem{berkeleydb}
Oracle.
\newblock {Oracle Berkeley DB, Java Edition - Getting Started with Berkeley DB
  Java Edition - 12c Release 2 Library Version 12.2.7.5}, Oct 2017.

\bibitem{Papagiannis16}
A.~Papagiannis, G.~Saloustros, P.~Gonz\'alez-F\'erez, and A.~Bilas.
\newblock {Tucana: Design and Implementation of a Fast and Efficient Scale-up
  Key-value Store}.
\newblock In {\em Proceedings of the 2016 USENIX Annual Technical Conference
  (ATC'16)}, pages 537--550, 2016.

\bibitem{Raju17}
P.~Raju, R.~Kadekodi, V.~Chidambaram, and I.~Abraham.
\newblock {PebblesDB: Building Key-Value Stores using Fragmented Log-Structured
  Merge Trees}.
\newblock In {\em Proceedings of the 26th Symposium on Operating Systems
  Principles (SOSP'17)}, pages 497--514, 2017.

\bibitem{redis}
Redis.
\newblock \url{http://redis.io}, Retrieved in June 2019.

\bibitem{Rosenblum92}
M.~Rosenblum and J.~K. Ousterhout.
\newblock {The Design and Implementation of a Log-structured File System}.
\newblock {\em ACM Transactions on Computer Systems}, 10(1):26--52, Feb 1992.

\bibitem{Rumble14}
S.~M. Rumble, A.~Kejriwal, and J.~Ousterhout.
\newblock {Log-structured Memory for DRAM-based Storage}.
\newblock In {\em Proceedings of the 12th USENIX conference on File and Storage
  Technologies (FAST'16)}, pages 1--16, 2014.

\bibitem{Sears12}
R.~Sears and R.~Ramakrishnan.
\newblock {bLSM: A General Purpose Log Structured Merge Tree}.
\newblock In {\em Proceedings of the 2012 ACM SIGMOD International Conference
  on Management of Data (SIGMOD'12)}, pages 217--228, 2012.

\bibitem{Shen17}
Z.~Shen, F.~Chen, Y.~Jia, and Z.~Shao.
\newblock {DIDACache: A Deep Integration of Device and Application for Flash
  Based Key-Value Caching}.
\newblock {\em ACM Transactions on Storage}, 14(3):26, Nov 2018.

\bibitem{Shetty13}
P.~J. Shetty, R.~P. Spillane, R.~R. Malpani, B.~Andrews, J.~Seyster, and
  E.~Zadok.
\newblock {Building Workload-Independent Storage with VT-Trees}.
\newblock In {\em Proceedings of the 11th USENIX Conference on File and Storage
  Technologies (FAST'13)}, pages 17--30, 2013.

\bibitem{Teng18}
D.~Teng, L.~Guo, R.~Lee, F.~Chen, Y.~Zhang, S.~Ma, and X.~Zhang.
\newblock {A Low-cost Disk Solution Enabling LSM-tree to Achieve High
  Performance for Mixed Read/Write Workloads}.
\newblock {\em ACM Transactions on Storage}, 14(2):15:1--15:26, 2018.

\bibitem{threadpool}
Threadpool.
\newblock \url{http://threadpool.sourceforge.net/}, Retrieved in June 2019.

\bibitem{tpcc}
TPC.
\newblock {TPC-C is an On-Line Transaction Processing Benchmark}.
\newblock \url{http://www.tpc.org/tpcc/}, Retrieved in June 2019.

\bibitem{Weil06}
S.~A. Weil, S.~A. Brandt, E.~L. Miller, D.~D.~E. Long, and C.~Maltzahn.
\newblock {Ceph: A Scalable, High-Performance Distributed File System}.
\newblock In {\em Proceedings of the 7th USENIX Symposium on Operating Systems
  Design and Implementation (OSDI'06)}, pages 307--320, 2006.

\bibitem{Wu15}
X.~Wu, Y.~Xu, Z.~Shao, and S.~Jiang.
\newblock {LSM-trie: An LSM-tree-based Ultra-Large Key-Value Store for Small
  Data}.
\newblock In {\em Proceedings of the 2015 USENIX Annual Technical Conference
  (ATC'15)}, pages 71--82, 2015.

\bibitem{Xia17}
F.~Xia, D.~Jiang, J.~Xiong, and N.~Sun.
\newblock {HiKV: A Hybrid Index Key-Value Store for DRAM-NVM Memory Systems}.
\newblock In {\em Proceedings of the 2017 USENIX Annual Technical Conference
  (ATC'17)}, pages 349--362, 2017.

\bibitem{Yao17}
T.~Yao, J.~Wan, P.~Huang, X.~He, Q.~Gui, F.~Wu, and C.~Xie.
\newblock {A Light-weight Compaction Tree to Reduce I/O Amplification toward
  Efficient Key-Value Stores}.
\newblock In {\em Proceedings of the 33rd International Conference on Massive
  Storage Systems and Technology (MSST'17)}, 2017.

\bibitem{Yue17}
Y.~Yue, B.~He, Y.~Li, and W.~Wang.
\newblock {Building an Efficient Put-Intensive Key-Value Store with Skip-Tree}.
\newblock {\em IEEE Transactions on Parallel and Distributed Systems},
  28(4):961--973, Apr 2017.

\bibitem{Zhang16}
H.~Zhang, M.~Dong, and H.~Chen.
\newblock {Efficient and Available In-memory KV-Store with Hybrid Erasure
  Coding and Replication}.
\newblock {\em ACM Transactions on Storage}, 13(3):25, Oct 2017.

\end{thebibliography}

\end{document}